\documentclass[10pt,journal,comsoc]{IEEEtran}
\IEEEoverridecommandlockouts

\usepackage{cite}
\usepackage{amsmath,amssymb,amsfonts}
\usepackage{mathtools}
\usepackage{bm}
\usepackage{array}
\usepackage{makecell}

\usepackage{graphicx}
\usepackage{textcomp}
\usepackage{xcolor}
\usepackage{epstopdf}
\usepackage[caption=false]{subfig}

\usepackage{enumitem}

\usepackage[hidelinks]{hyperref}

\usepackage{algorithm}
\usepackage{algpseudocodex}

\usepackage{microtype}
\AtBeginDocument{%
	\setlength{\abovedisplayskip}{4pt}
	\setlength{\belowdisplayskip}{4pt}
	\setlength{\abovedisplayshortskip}{2pt}
	\setlength{\belowdisplayshortskip}{2pt}
}
\def\BibTeX{{\rm B\kern-.05em{\sc i\kern-.025em b}\kern-.08em
		T\kern-.1667em\lower.7ex\hbox{E}\kern-.125emX}}

\title{HermesHFL: Incentive-Aware Hierarchical Federated Unlearning for Dynamic LLM Fine-Tuning}

\author{
	Chenxi~Sun,~\IEEEmembership{Student Member,~IEEE},
	Minghui~Liwang,~\IEEEmembership{Senior Member,~IEEE},
	Wusi He,  
	Yuhan Su, Zhang Liu,\\ Sai Zou, \IEEEmembership{Senior Member,~IEEE}, 
	Wei Ni, \IEEEmembership{Fellow,~IEEE}, and  
	Seyyedali~Hosseinalipour,~\IEEEmembership{Senior Member,~IEEE} \\
	
	\vspace{-8mm}
}

\begin{document}	
	\maketitle
	\begin{abstract}
		Hierarchical federated learning (HFL) offers a promising paradigm for scalable fine-tuning of large language models (LLMs). However, in dynamic HFL environments, clients may withdraw, request the removal of their  contributions to model training, and subsequently seek to rejoin as their operational demands evolve. Accommodating this leave–unlearn–rejoin lifecycle in HFL, where local model updates are aggregated first at edge servers and subsequently at a cloud server,  poses two intertwined challenges: (i) selectively removing client contributions intertwined through successive layers of model aggregation, and (ii) coordinating client selection, edge association, and economic incentives.
		To address these underexplored challenges, we propose HermesHFL (\underline{h}i\underline{e}rarchical \underline{r}ejoinable \underline{m}achine l\underline{e}arning with \underline{s}elective unlearning for incentive-aware \underline{HFL}), a unified framework leveraging low-rank adaptation (LoRA) for parameter-efficient learning and unlearning in HFL settings.
		Under HermesHFL, we formulate a cross-layer optimization problem that couples continuous incentive allocation with discrete client selection and edge association decisions. To efficiently solve this problem, we develop Neogen (\underline{n}\underline{e}ural netw\underline{o}rk-\underline{g}uided dual-level \underline{e}volutionary optimizatio\underline{n}), a bilevel evolutionary optimizer that combines \underline{c}ovariance \underline{m}atrix \underline{a}daptation \underline{e}volution \underline{s}trategy (CMA-ES) for continuous incentive optimization with a \underline{c}ross-generational elitist selection, \underline{h}eterogeneous recombination, and \underline{c}ataclysmic mutation evolutionary algorithm (CHC)-based mechanism for discrete client participation and edge association decisions. We evaluate HermesHFL on GPT-2 considering two downstream tasks: news-topic classification (AGNews) and banking customer-service intent classification (Banking77). Experiments across varying network sizes and data heterogeneity levels demonstrate that HermesHFL enables effective unlearning with higher economic utility and significantly lower computational cost than full retraining, while maintaining comparable model accuracy.
	\end{abstract}
	\vspace{-1mm}
	% \vspace{5cm}
	\vspace{-3mm}
	\section{Introduction}
	\label{sec:introduction}
	\IEEEPARstart{R}{ecent} advances in large language models (LLMs)~\cite{vaswani2017attention} have motivated their adaptation for specialized applications, such as language services in mobile crowdsourcing~\cite{11534581}. However, the data required for such adaptation are often dispersed across clients and cannot be centralized due to data-governance constraints.
	Federated learning (FL)~\cite{pmlr-v54-mcmahan17a,fu2026differentially} offers a promising solution by enabling collaborative model training without sharing data. However, its communication and coordination overhead grows with the number of  clients, particularly in resource-intensive LLM fine-tuning.
	To alleviate this overhead, hierarchical federated learning (HFL)~\cite{9054634} introduces multi-tier model aggregation, offering a promising framework for collaborative LLM fine-tuning~\cite{11413826,11479917}. 
	
	Despite its potential for scalable LLM fine-tuning, HFL faces fundamental challenges in dynamic environments where client participation and data regulations evolve over time. For example, consider a regional media alliance in which news organizations collaboratively fine-tune an LLM for news-topic classification, or a banking consortium in which banks jointly adapt an LLM for customer-service intent classification. In both scenarios, participating organizations may temporarily withdraw due to operational or regulatory constraints and subsequently seek to rejoin once these constraints are lifted (see Appx.~\ref{sec:appendix_drift_scenarios} for additional examples). These challenges become particularly pronounced when departing organizations revoke authorization to use their data: while their withdrawal prevents future contributions, their historical contributions remain embedded in the fine-tuned LLM through successive layers of model aggregation. This necessitates model unlearning (i.e., removing the influence of previously used data from a trained model). However, fulfilling each unlearning request through model retraining can incur prohibitive computational and communication costs, calling for more efficient mechanisms that enable selective unlearning and subsequent client reintegration without restarting the collaborative fine-tuning process.
	
	% However, sustaining such collaboration in dynamic environments requires accommodating changes in client participation and data-authorization requirements, raising new challenges beyond conventional HFL.
	
	% These challenges are particularly relevant in cross-organizational LLM fine-tuning scenarios. For example, in a regional converged-media alliance, local news organizations (workers) collaboratively fine-tune a shared LLM for topic classification using locally stored articles, while regional servers (managers) and a cloud coordinator (president) perform hierarchical aggregation. Similarly, a banking customer-service consortium can employ the same architecture for intent classification without sharing sensitive customer dialogues. In such settings, organizations may temporarily suspend participation due to operational or compliance constraints and subsequently rejoin with largely unchanged local data (find more examples in Appx. \ref{sec:appendix_drift_scenarios}). Meanwhile, data-authorization withdrawals may require the removal of previously contributed information from the shared model, as simply disconnecting a client does not eliminate the influence of its historical contributions. Accommodating these changes through repeated fine-tuning or retraining can incur substantial computational and communication costs, particularly in hierarchical LLM fine-tuning systems. 
	
	\subsection{Research Questions and Motivation}
	\label{subsec:motivation}
	The above considerations motivate the following research questions (RQs) concerning efficient LLM fine-tuning and unlearning under dynamic client participation in HFL systems, the pursuit of which constitutes the central focus of this work.
	
	\noindent{\textit{RQ 1: How can client contributions be selectively unlearned in multi-tier HFL systems while preserving the fine-tuned LLM utility and avoiding costly  retraining?}}
	% Growing privacy and data-governance requirements, exemplified by the right to erasure under the EU General Data Protection Regulation (GDPR), have motivated increasing interest in federated unlearning (FUL) \cite{gdpr2016,zhou2025federated}. Although FL preserves data locality, information learned from client data remains embedded in shared model parameters through successive aggregation, making local data deletion insufficient to eliminate its historical influence. FUL addresses this limitation by enabling the removal of previously learned contributions upon data withdrawal or deletion requests.  
	% However, HFL introduces additional complexity, as client contributions become intertwined through multi-tier aggregation, making targeted unlearning difficult without disturbing knowledge retained from other clients. This challenge is further amplified in LLM fine-tuning, where extensive parameter dependencies complicate selective knowledge removal, while full retraining after each unlearning request incurs prohibitive computational costs. Consequently, the central challenge is to achieve effective and computationally efficient unlearning while preserving the utility and stability of the retained model.
	The right to erasure, as stipulated by the EU General Data Protection Regulation (GDPR), raises a fundamental concern in FL: although raw data remain local at each client, its influence persists in the shared model even after client departure. This has motivated federated unlearning (FUL)~\cite{gdpr2016,zhou2025federated}, which seeks to eliminate the influence of designated data from a collaboratively trained model without the need for retraining the model from scratch.  Realizing FUL in HFL settings, however, poses a distinct challenge: a client's contribution is first fused with those of its neighboring clients at an edge server and subsequently with those of geo-distributed clients at the cloud server. Selectively reversing this process without disrupting the knowledge acquired from other clients is particularly challenging for LLMs due to their extensive parameter dependencies. Thus, the key challenge lies in designing unlearning mechanisms for HFL that selectively suppress withdrawn clients' influence on the fine-tuned model while preserving model utility.

	% \noindent~ {\textit{RQ 2: How can client rejoining after unlearning be coordinated with hierarchical aggregation and incentive allocation while preserving unlearning effectiveness and training stability? }} In long-term HFL collaborations, client participation is inherently dynamic: clients may temporarily suspend participation due to operational or regulatory constraints and subsequently rejoin with updated data and resources. Such re-participation can sustain model utility and collaborative efficiency, but introduces challenges beyond conventional client admission. In particular, rejoining clients should contribute new updates without reintroducing previously unlearned information. Moreover, changes in client participation and edge association can alter hierarchical aggregation dynamics, potentially affecting training stability and incentive fairness. The key challenge, therefore, is to coordinate client reintegration, multi-tier aggregation, and incentive allocation while maintaining prior unlearning effectiveness and stable LLM fine-tuning.
	
	\noindent\textit{RQ 2: How can HFL accommodate client rejoining after unlearning without reversing prior unlearning efforts or disrupting collaborative fine-tuning?}
	Unlike conventional FUL, where departing clients are typically excluded from subsequent training, dynamic HFL environments may exhibit a leave-unlearn-rejoin (LUR) lifecycle, in which clients temporarily withdraw, request the removal of their historical contributions, and subsequently resume participation once the conditions prompting their withdrawal are resolved. However, client rejoining poses challenges beyond conventional client admission: returning clients must be reintegrated into the HFL system without restoring outdated model states or undoing prior unlearning operations. Moreover, changes in their local data, computational resources, and participation costs may render their previous edge associations and incentive agreements suboptimal, necessitating their joint re-optimization upon rejoining. Thus, client rejoining cannot be treated as a simple restoration of prior participation but requires a coordinated redesign of client admission, edge association, and incentive allocation that preserves the effectiveness of prior unlearning while sustaining stable LLM fine-tuning.
	
	% \noindent~ {\textit{RQ 3: How can cross-layer optimization and incentive design coordinate strategic client participation and unlearning while ensuring model utility, stability, and resource efficiency in HFL? }}HFL with unlearning and dynamic client participation introduces tightly coupled decisions across cloud, edge, and client tiers. Heterogeneous resources, participation costs, and privacy preferences may induce self-interested behaviors, potentially undermining collaborative efficiency and fairness. In particular, client selection and edge association determine both aggregation outcomes and individual contributions, while unlearning and rejoining further alter their marginal benefits and associated costs. Consequently, independently optimizing participation, unlearning, and incentives may lead to conflicting decisions and unsatisfying services. Addressing these interdependencies requires a unified cross-layer mechanism that jointly coordinates client participation, edge association, contribution evaluation, and incentive allocation under budget and resource constraints, where the central challenge is to align individual incentives with system-wide objectives while maintaining efficient and stable LLM fine-tuning.
	
	\noindent\textit{RQ 3: How can incentives and cross-layer decisions be jointly optimized to sustain efficient LLM fine-tuning under dynamic client participation and unlearning in HFL settings?}
	Beyond the technical challenges of unlearning and client reintegration, sustaining collaborative LLM fine-tuning in HFL settings poses an economic challenge stemming from clients' self-interested/selfish behavior and heterogeneous computational resources, participation costs, and privacy preferences. Specifically, while clients seek to maximize their individual utilities through the incentives offered, the system seeks to recruit clients whose data and resources can enhance the fine-tuned LLM utility with minimal participation and potential unlearning costs. These economic and technical considerations are further intertwined by the hierarchical architecture of HFL, where cloud-level budget allocation determines the recruitment capacity of edge servers, whose client selection and association decisions, in turn, shape the global model performance. Client withdrawal and rejoining further reshape these interdependencies by altering the costs and benefits of participation over time. The key challenge, therefore, is to jointly optimize budget allocation, client selection, edge association, and incentive contracts to reconcile individual economic interests with system-wide learning and unlearning objectives in HFL settings.

	\subsection{Novelty and Contributions}
	\label{subsec:contribution}
	To address the above RQs, we propose \underline{h}i\underline{e}rarchical \underline{r}ejoinable \underline{m}achine l\underline{e}arning with \underline{s}elective unlearning for incentive-aware \underline{HFL} (abbreviated as HermesHFL, depicted in Fig.~\ref{fig:my_figure}), a novel framework that explicitly accommodates the LUR lifecycle of clients and incorporates economic incentives into hierarchical federated fine-tuning of LLMs. Our main contributions can be  summarized as follows:
	
	\begin{figure}[!t]
		\vspace{-2mm}
		\centering
		\includegraphics[width=1\linewidth]{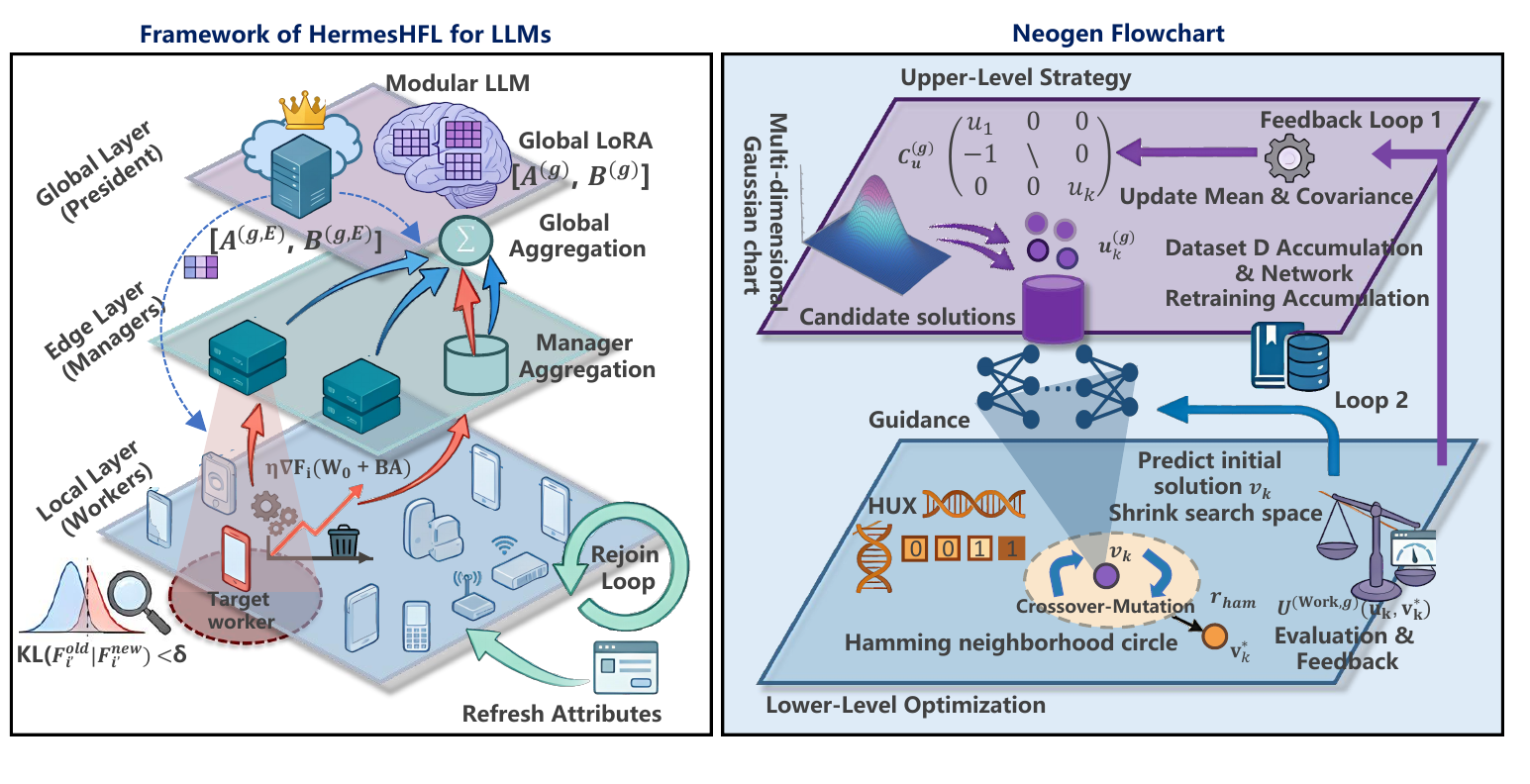}
		\vspace{-10mm}
		\caption{
			{Schematic illustration of the HermesHFL (left) and the Neogen optimization algorithm (right).}
			The left box illustrates the hierarchical LoRA-based federated training and aggregation process across workers, managers, and the president. 
			The right box presents the dual-level Neogen optimization procedure, where upper-level CMA-ES and lower-level CHC collaboratively optimize incentive strategies and worker decisions through iterative feedback.
		}
		\vspace{-5mm}
		\label{fig:my_figure}
	\end{figure}
	
	\begin{itemize}[leftmargin=3.5mm]
		\item \textit{LLM Adaptation in Dynamic HFL systems with LUR Lifecycle:} We develop a hierarchical federated unlearning (HFUL) framework for LLM fine-tuning that integrates selective unlearning, dynamic client participation, and economic incentives. To the best of our knowledge, this is among the first efforts to explicitly model the LUR lifecycle in incentive-aware HFL systems for LLM fine-tuning.
		\item \textit{President-Manager-Worker Architecture with Incentive-Aware Optimization:}
		We design the president-manager-worker architecture of HermesHFL, which jointly captures client selection, edge association, incentive allocation, selective unlearning, and rejoining under budget constraints.
		\item \textit{Dual-Level Optimization Guided by Neural Networks:}
		We develop Neogen (\underline{n}\underline{e}ural netw\underline{o}rk-\underline{g}uided dual-level \underline{e}volutionary optimizatio\underline{n}), depicted in Fig.~\ref{fig:my_figure}, which adopts a bilevel architecture, where the upper level optimizes incentive allocation and budget planning through a customized CMA-ES (\underline{c}ovariance \underline{m}atrix \underline{a}daptation \underline{e}volution \underline{s}trategy) with adaptive covariance control, while the lower level obtains client participation and edge association via a tailored CHC (\underline{c}ross-generational elitist selection, \underline{h}eterogeneous recombination, and \underline{c}ataclysmic mutation) evolutionary mechanism. To accelerate solution acquisition, we design a neural network that predicts high-quality lower-level solutions for each upper-level candidate, thereby guiding the CHC search toward promising regions of the discrete decision space.
		% This design enables stable and incentive-aware decision making while improving resource efficiency and convergence stability in dynamic LLM-oriented hierarchical federated unlearning (HFUL) systems.
		\item \textit{Experimental Validation:}
		We evaluate HermesHFL on GPT-2 using two representative tasks: news-topic classification and banking customer-service intent classification. Experiments across varying network sizes and degrees of data heterogeneity assess classification accuracy, unlearning effectiveness, economic utility, and computational efficiency.
	\end{itemize}

	\section{Literature Review and Preliminaries}
	\label{subsec:literature} 
	\textit{Related Work:} In the following, we review the relevant studies and highlight their key distinctions from our work.
	
	\noindent
	$\bullet$
	\textit{FUL: }FUL, e.g., class-discriminative pruning \cite{10.1145/3485447.3512222}, seeks to mitigate the influence of designated clients or data from trained models in FL systems. Existing FUL methods are primarily developed for flat FL (i.e., distributed learning with direct client-to-server communications) and rely on gradient ascent (GA), influence functions, differential privacy, or parameter resetting. For example, FedOSD \cite{pan2025federated} introduces unlearning cross-entropy loss with orthogonal steepest descent (OSD). FedRecovery \cite{zhang2023fedrecovery} removes client gradient residuals under differential privacy, producing models statistically close to retraining without full retraining. FRAMU \cite{shaik2024framu} combines attention-based unlearning with federated reinforcement learning (RL) to support continual model evolution. F2UL~\cite{10605088} promotes fairness in FUL through label-free model quality assessment and adaptive model allocation, mitigating the adverse effects of low-quality client data. For LLM-oriented unlearning, GRU\cite{wang2025gru} modifies gradient update strategies to reduce interference between the retained and removed knowledge during unlearning, whereas DSMGDA\cite{10889776} formulates unlearning as a multi-objective optimization to prevent gradient explosion. KNOT~\cite{10229075} reduces retraining costs through cluster-based model aggregation, confining unlearning operations to affected clusters. These studies are mostly designed for flat FL and further overlook the LUR lifecycle over dynamic clients in HFL. Moreover, they rarely integrate incentive design with unlearning, overlooking the economic considerations necessary for  LLM fine-tuning in HFL systems.
	
	\noindent
	$\bullet$
	\textit{Parameter-Efficient LLM Fine-tuning via FL:}
	Full training of LLMs is computationally prohibitive in resource-constrained FL environments, making parameter-efficient fine-tuning (PEFT)\cite{11364256} a dominant approach. Frameworks, such as FATE-LLM \cite{fan2023fate}, enable industrial-scale federated LLM fine-tuning by combining local PEFT updates with privacy-preserving learning mechanisms, reducing both computational and communication costs. FedALoRA \cite{11048575} combines the LoRA technique with personalized model aggregation to accommodate cross-domain non-independent and identically distributed (Non-IID) data across clients. FedBiOT~\cite{wu2024fedbiot} reduces client-side computational and communication overhead through bilevel optimization, where the server aligns a compressed LLM (emulator) with the original model, while clients fine-tune lightweight adapters using the emulator and their local data.
	FLORA \cite{wang2024flora} addresses model aggregation biases in heterogeneous LoRA configurations across clients through stacked aggregation strategies. LoRA-FAIR \cite{bian2025lora} introduces correction terms to mitigate server-side aggregation bias and inconsistent model initialization across clients. PAC+ \cite{11355763} integrates parallel adapters, activation caching and distributed collaborative training to achieve efficient LLM fine-tuning with reduced memory footprint. MIRA \cite{10877919} enables personalized LLM optimization, while HRL-FLLM \cite{11387738} adaptively optimizes LLM fine-tuning in edge networks to reduce computing/communication overhead and training latency while ensuring model convergence. FedARA \cite{11363369} employs truncated singular value decomposition (SVD) adaptation, dynamic rank allocation, and rank-based module pruning to optimize model performance, communication efficiency, and resource consumption on edge devices. These existing works on PEFT-based FL predominantly focus on LLM fine-tuning in flat FL architectures, overlooking both the complexities of multi-tier aggregation in HFL and the need for selective unlearning under LUR lifecycle.
	
	\noindent
	$\bullet$
	\textit{Incentive Design Mechanisms for FUL:}
	Sustaining client participation in FL/FUL systems calls for economic incentive mechanisms that compensate self-interested clients for their contributions and associated costs. An early study \cite{yu2020sustainable} introduced context-aware dynamic budget allocation to attract high-quality participants for FL, while mitigating reward imbalance and communication latency disparities. Another work \cite{10.1145/3589334.3645462} integrated clustering-based FUL with incentives, where unlearning tasks are executed within client clusters while client participation is dynamically optimized using evolutionary game theory and deep RL. Studies \cite{11049912} and \cite{ding2025incentivized} applied principal-agent theory and reverse game design to enforce verifiable and incentive-compatible client behavior for machine unlearning in distributed network settings. Unfortunately, these studies have not explored designing incentive mechanisms for HFUL in dynamic environments under LUR lifecycle, an effort carried out for the first time in this work.
	
	\begin{table}[!t]
		\centering
		\caption{Comparison of HermesHFL with representative studies.}
		\label{tab:literature_comparison}
		\resizebox{\columnwidth}{!}{%
			\setlength{\tabcolsep}{3pt}
			\begin{tabular}{@{}p{3.4cm}>{\centering\arraybackslash}p{0.6cm}>{\centering\arraybackslash}p{0.8cm}>{\centering\arraybackslash}p{1.55cm}>{\centering\arraybackslash}p{2.0cm}>{\centering\arraybackslash}p{1.5cm}>{\centering\arraybackslash}p{1.7cm}@{}}
				%\footnotesize
				\hline
				Method & HFL & LLM\slash PEFT & Client unlearning & Dynamic participation\slash rejoining & Incentives & Mobile/edge focus \\
				\hline
				FedRecovery~\cite{zhang2023fedrecovery} & $\times$ & $\times$ & $\checkmark$ & -- & -- & -- \\
				FRAMU~\cite{shaik2024framu} & -- & $\times$ & $\checkmark$ & -- & -- & -- \\
				F2UL~\cite{10605088} & $\times$ & $\times$ & $\checkmark$ & -- & -- & -- \\
				KNOT~\cite{10229075} & -- & $\times$ & $\checkmark$ & -- & -- & -- \\
				Server-initiated FUL~\cite{wang2024server} & -- & $\times$ & $\checkmark$ & -- & -- & -- \\
				Distilled-data FUL~\cite{11399903} & -- & $\times$ & $\checkmark$ & -- & -- & -- \\
				GRU~\cite{wang2025gru} & -- & $\checkmark$ & -- & -- & -- & -- \\
				Hier-FUN~\cite{ma2024hier} & $\checkmark$ & $\times$ & $\checkmark$ & -- & -- & $\checkmark$ \\
				Incentivized FUL~\cite{ding2025incentivized} & -- & $\times$ & $\checkmark$ & -- & $\checkmark$ & -- \\
				The price of forgetting~\cite{11049912} & -- & $\times$ & -- & -- & $\checkmark$ & -- \\
				Incentive and dynamic \\ selection~\cite{10.1145/3589334.3645462} & -- & $\times$ & $\checkmark$ & $\checkmark$ & $\checkmark$ & -- \\
				FATE-LLM~\cite{fan2023fate} & -- & $\checkmark$ & $\times$ & -- & -- & -- \\
				FedBiOT~\cite{wu2024fedbiot} & -- & $\checkmark$ & $\times$ & -- & -- & -- \\
				FLORA~\cite{wang2024flora} & -- & $\checkmark$ & $\times$ & -- & -- & -- \\
				LoRA-FAIR~\cite{bian2025lora} & -- & $\checkmark$ & $\times$ & -- & -- & -- \\
				MIRA~\cite{10877919} & -- & $\checkmark$ & $\times$ & -- & -- & -- \\
				PAC+~\cite{11355763} & -- & $\checkmark$ & $\times$ & -- & -- & -- \\
				FedALoRA~\cite{11048575} & -- & $\checkmark$ & $\times$ & -- & -- & -- \\
				FedARA~\cite{11363369} & -- & $\checkmark$ & $\times$ & -- & -- & -- \\
				HRL-FLLM~\cite{11387738} & -- & $\checkmark$ & $\times$ & -- & -- & $\checkmark$ \\
				Hierarchical federated \\ foundation models~\cite{11413826} & $\checkmark$ & $\checkmark$ & -- & -- & -- & $\checkmark$ \\
				Intermittent-participation HFL~\cite{11215887} & $\checkmark$ & -- & -- & $\checkmark$ & -- & $\checkmark$ \\
				% \rowcolor{gray!10}
				HermesHFL (Ours) & $\checkmark$ & $\checkmark$ & $\checkmark$ & $\checkmark$ & $\checkmark$ & $\checkmark$ \\
				\hline
			\end{tabular}%
		}
		\par\vspace{2pt}
		\begin{minipage}{\columnwidth}
			\footnotesize
			\noindent\textit{Note:} $\checkmark$ feature supported, $\times$ feature does not match, and ``--'' feature not clearly specified. ``Dynamic participation/rejoining'' combines client selection, intermittent participation, leave, and rejoining because these properties describe the same client-lifecycle dimension. The entries summarize the stated scope of each work and are not intended to imply that an unlisted feature is impossible.
		\end{minipage}
		\vspace{-5mm}
	\end{table}

	In summary, as reflected in Table~\ref{tab:literature_comparison}, existing studies on FUL predominantly focus on flat FL architectures, those on federated LLM adaptation primarily address fine-tuning rather than unlearning, and those on incentive design for FUL overlook the client LUR lifecycle in hierarchical settings. Our work bridges these gaps by bringing hierarchical federated LLM fine-tuning, selective unlearning, dynamic client participation, and incentive design into a unified framework.

	% \noindent
	\textit{Preliminaries:} Below, we briefly review the three technical bases of HermesHFL: LoRA-based PEFT, HFL, and FUL, building upon which we introduce the system-specific notation and decision variables for our system of interest in Section~\ref{sec:overview_modeling}.
	
	\noindent
	$\bullet$ \textit{LoRA:} For a pretrained weight matrix $\mathbb{W}_0 \in \mathbb{R}^{d\times k}$ of an LLM, LoRA represents the model update $\Delta\mathbb{W}$ as follows:
	\begin{equation}
		\mathbb{W}_0 + \Delta\mathbb{W} = \mathbb{W}_0 + \boldsymbol{B}\boldsymbol{A},
		\label{eq:lora_core}
	\end{equation}
	where $\boldsymbol{B} \in \mathbb{R}^{d\times r}$ and $\boldsymbol{A} \in \mathbb{R}^{r\times k}$ are trainable LoRA factors, and $r$ is the adaptation rank. By keeping $\mathbb{W}_0$ frozen and choosing $r \ll \min\{d,k\}$, LoRA reduces the number of trainable parameters from $d\times k$ to $r(d+k)$, thereby lowering the  overhead of model fine-tuning and communication.
	
	\noindent
	$\bullet$ \textit{HFL:} Unlike conventional FL, where clients directly exchange model updates with a central server, HFL introduces an intermediate layer of edge servers that perform regional model aggregation before forwarding their aggregated models to the cloud server. Let $g$, $e$, and $l$ index the global aggregation rounds, edge-level aggregation rounds, and local training iterations, respectively. Each global round comprises $E$ edge-level aggregations, each preceded by $L$ local training iterations. Let $\boldsymbol{W}$ and $\boldsymbol{M}$ denote the sets of clients and edge servers, respectively, and $\widehat{\boldsymbol{W}}_j^{(g)}\subseteq\boldsymbol{W}$ denote the set of participating clients associated with edge server $m_j$ during global round $g$. Upon completing $L$ local training iterations, edge server $m_j$ aggregates the model parameters of its associated clients as:
	\begin{equation}
		\omega_j^{\mathsf{midd},(g,e+1)}
		=
		\frac{\sum_{w_i\in\widehat{\boldsymbol{W}}_j^{(g)}}
			d_i^{\mathsf{size},(g)}
			\omega_i^{\mathsf{client},(g,e,L)}}
		{D_j^{(g)}},
		\label{eq:edge_aggregation}
	\end{equation}
	where $d_i^{\mathsf{size},(g)}$ denotes the local dataset size of client $w_i$ and $D_j^{(g)}=\sum_{w_i\in\widehat{\boldsymbol{W}}_j^{(g)}}d_i^{\mathsf{size},(g)}$ denotes the aggregate dataset size under edge server $m_j$. The aggregated model initializes the next local training stage. After $E$ edge-level aggregations, the cloud server obtains the global model for the next round as:
	\begin{equation}
		\omega^{\mathsf{glob},(g+1)}
		=
		\frac{\sum_{m_j\in\boldsymbol{M}}
			D_j^{(g)}\omega_j^{\mathsf{midd},(g,E)}}
		{\sum_{m_j\in\boldsymbol{M}}D_j^{(g)}}.
		\label{eq:global_aggregation}
	\end{equation}
	
	\noindent
	$\bullet$ \textit{FUL:} Let $\mathcal{D}$ denote the collective training dataset distributed across FL clients, partitioned as $\mathcal{D}=\mathcal{D}_{\mathsf{rtn}}\cup\mathcal{D}_{\mathsf{fgt}}$, where $\mathcal{D}_{\mathsf{rtn}}$ and $\mathcal{D}_{\mathsf{fgt}}$ denote the data to be retained and forgotten, respectively. The goal of FUL is to remove the influence of $\mathcal{D}_{\mathsf{fgt}}$ from the trained model without retraining it from scratch. Ideally, the resulting model $\omega^{-}$ should approximate the model obtained by training exclusively on $\mathcal{D}_{\mathsf{rtn}}$, i.e., $\omega^{-}\approx\mathbb{A}(\mathcal{D}_{\mathsf{rtn}})$, where $\mathbb{A}(\cdot)$ denotes the training algorithm. Fig.~\ref{fig:unlearning_mechanism} (Sec.~\ref{sec:overview_modeling}) illustrates the unlearning mechanism and the LUR lifecycle.

	\section{Overview and Basic Modeling}
	\label{sec:overview_modeling}
	
	\subsection{Overview}
	\label{subsec:system_overview}
	
	We consider an HFUL system with economic incentives and dynamic client LUR behaviors to support LLM fine-tuning and unlearning. The system consists of three main entities/players: \textit{(i)} a cloud server responsible for global aggregation, referred to as the \emph{president}; \textit{(ii)} a set of geographically distributed edge aggregators, referred to as \emph{managers}, represented by $\boldsymbol{M} = \{m_1, ..., m_j, ..., m_{|\boldsymbol{M}|}\}$; and \textit{(iii)} a set of distributed clients participating in local model training, referred to as \emph{workers}, represented by $\boldsymbol{W} = \{w_1, ..., w_i, ..., w_{|\boldsymbol{W}|}\}$. To facilitate LLM fine-tuning and unlearning over such a market, we propose HermesHFL, a cross-layer optimization framework integrating participation control, aggregation policies, incentives, and selective unlearning. In HermesHFL, client joining, withdrawal, unlearning, and rejoining behaviors induce tightly coupled system dynamics that simultaneously affect worker-level decisions, manager-side aggregation, and global coordination at the president. At a high level, HermesHFL operates through the following three steps:
	
	\noindent
	$\bullet$\textit{ Step 1. Global Model Downloading and Budget Distribution: }
	Prior to the \(g\)-th global aggregation round, the president distributes the latest global model to all managers, while allocating the total incentive budget across managers to attract high-quality workers for collaborative fine-tuning.
	
	\noindent
	$\bullet$\textit{ Step 2. Manager-Worker Contracting, Local Fine-Tuning, and Edge Aggregation: }Managers establish contracts with workers, specifying incentive payments, target training quality, and potential unlearning penalties based on system and worker characteristics. Subsequently, workers perform local LLM fine-tuning using LoRA and upload the resulting LoRA parameters to their associated managers for edge-level aggregation. The aggregated parameters are then disseminated back to the associated workers to initialize the next local fine-tuning stage. This process of local fine-tuning and edge-level aggregation is repeated for a prescribed number of rounds, after which the managers forward their final aggregated parameters to the president for global aggregation, completing one global round.
	
	\noindent
	$\bullet$\textit{ Step 3. Unlearning Process: }
	Upon completion of the $g$-th global aggregation round, participating workers receive their contractual payments, and some may subsequently request unlearning due to privacy concerns or market exit. During subsequent rounds, these workers perform gradient ascent (GA) instead of conventional gradient descent, with their unlearning directions refined through OSD projection~\cite{pan2025federated} (see Sec.~\ref{subsubsec:unlearning_process}). They also pay the contractual unlearning penalties, which contribute to their associated managers' budgets for the $(g+1)$-th global round. The progress of unlearning is assessed using the Kullback--Leibler divergence (KLD) between the model's predictive distributions on the requesting workers' data before and after unlearning. Once the prescribed unlearning criterion is satisfied, the workers become eligible to rejoin the market with updated attributes, including their computing capacity, communication capability, privacy cost, and reputation score. The latter is estimated using a sliding window of historical unlearning requests. This procedure continues until a predefined number of global rounds is reached. 
	% Fig.~\ref{fig:unlearning_mechanism} summarizes the mechanism and the measured LUR lifecycle.
	
	\begin{figure}[!t]
		\vspace{-5mm}
		\centering
		\includegraphics[width=.85\linewidth]{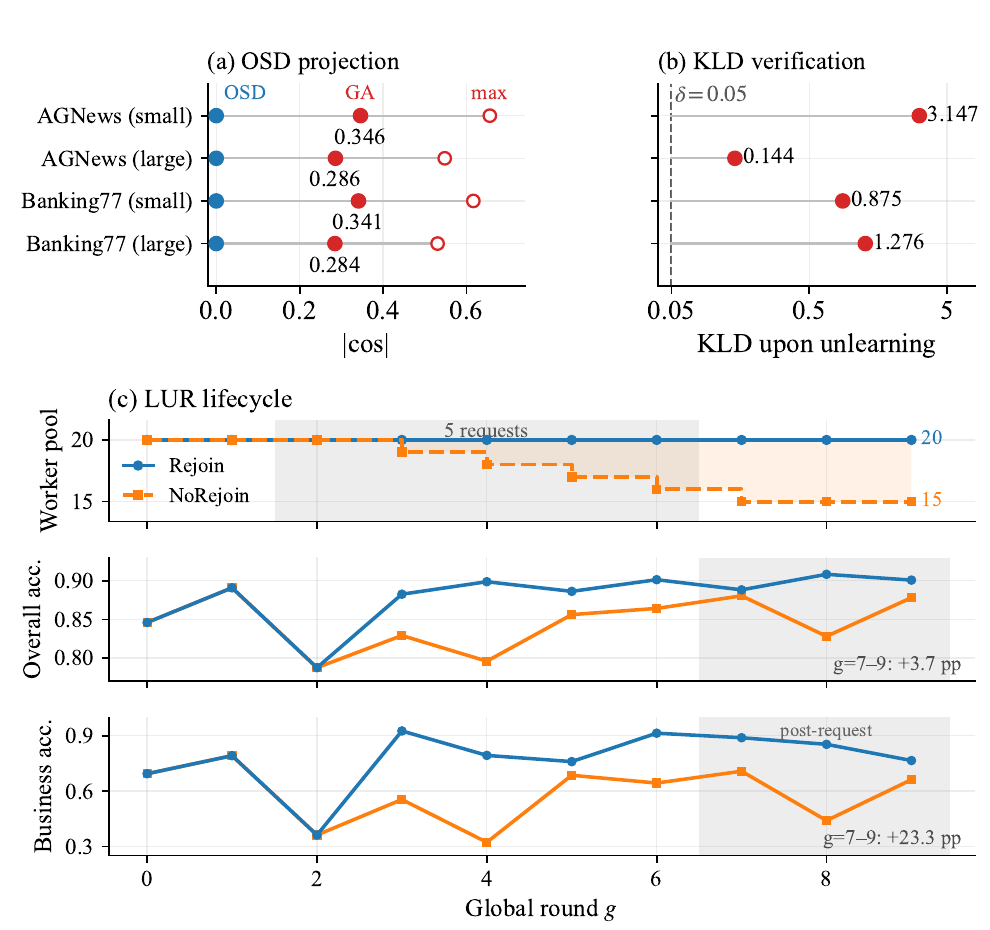}
		\vspace{-4mm}
		\caption{The HermesHFL unlearning mechanism: OSD projection, KLD verification, and the LUR lifecycle.}
		\label{fig:unlearning_mechanism}
		\vspace{-7mm}
	\end{figure}
	
	% Fig.~\ref{fig:unlearning_mechanism} illustrates why the mechanism takes this form. The GA direction of a departing worker can overlap with the retained workers' updates, so OSD projects it orthogonal to them to forget the target data without eroding retained knowledge. KLD certifies that the influence has been removed from the model's behavior before the worker is readmitted. Rejoin returns a verified worker's data to the pool with refreshed attributes, because that data is often scarce among the remaining workers and permanent removal degrades the model.
	% Fig.~\ref{fig:unlearning_mechanism} illustrates the effectiveness of the key components of our method. As shown in Fig.~\ref{fig:unlearning_mechanism}(a), OSD eliminates directional overlap between learning and unlearning updates (measured via $\cos(\cdot)$ similarity), mitigating interference between the two processes. Fig.~\ref{fig:unlearning_mechanism}(b) demonstrates that the KLD between the predictive distributions before and after unlearning exceeds the prescribed threshold $\delta$ across all considered settings. Finally, Fig.~\ref{fig:unlearning_mechanism}(c) highlights the importance of accommodating worker rejoining, which preserves the worker pool and improves both overall and class-specific model accuracy compared to permanently excluding workers after unlearning.

	Fig.~\ref{fig:unlearning_mechanism} depicts the effectiveness of the components of our method through experiments on two datasets (AGNews and Banking77) and two network sizes (detailed in Sec.~\ref{sec:evaluation}). Fig.~\ref{fig:unlearning_mechanism}(a) shows that OSD eliminates directional overlap between learning and unlearning updates (measured via $\cos(\cdot)$ similarity), mitigating interference between the two processes. Fig.~\ref{fig:unlearning_mechanism}(b) demonstrates that the KLD between the predictive distributions before and after unlearning exceeds the prescribed threshold $\delta$ across all four settings. Finally, Fig.~\ref{fig:unlearning_mechanism}(c) highlights the importance of accommodating worker rejoining, which preserves the worker pool and improves both overall classification accuracy and classification accuracy for the Business news category of AGNews, compared to permanently excluding workers after unlearning. The latter improvement highlights the particular importance of worker rejoining in preserving model performance on underrepresented data categories.

	% Upon completion of a global aggregation round, participating workers receive payments as stipulated in contracts, while some workers may subsequently request unlearning (e.g., due to market exit or privacy concerns). In such cases, they perform GA during subsequent training, whose direction is refined by an OSD projection \cite{pan2025federated} (Sec.~\ref{subsubsec:unlearning_process}), and compensate managers according to the agreed liquidated damages, which are incorporated into the managers' total budget for the $(g+1)$-th round. After several unlearning iterations, completion is verified via the Kullback--Leibler divergence (KLD) between the model’s predictive distributions on the unlearned worker’s data before and after the unlearning operation. At this time, some information of the worker (computing capacity, communication capability, privacy cost, and reputation score, e.g., estimated via a sliding time window reflecting the unlearning probability) is updated. Workers are allowed to rejoin the market after completing the unlearning process. This process repeats until reaching a predefined number of global rounds.
	
	In the following, we introduce basic modeling components of HermesHFL, including contract formulation, worker attributes, and budget modeling for the president and managers.
	
	\vspace{-4mm}
	\subsection{Basic Modeling}
	\label{subsec:basic_modeling}
	
	\subsubsection{Modeling of Contracts}
	\label{subsubsec:contract_model}
	
	We model the contract between workers and managers as:
	\begin{equation}
		\boldsymbol{C}_{i,j}^{(g)} = \left\{p_{i,j}^{(g)}, q_{i,j}^{(g)}, \theta_{i,j}^{(g)}\right\}
		\label{eq:contract_def}.
	\end{equation}
	
	\noindent
	Here, $p_{i,j}^{(g)}$ denotes the incentive payment provided by manager $m_j$ to worker $w_i$ for making contributions to model fine-tuning during the $g$-th round, $q_{i,j}^{(g)}$ represents the penalty imposed if $w_i$ requests withdrawal and unlearning after the completion of the $g$-th round, and $\theta_{i,j}^{(g)}$ specifies the target training quality or accuracy requirement assigned by manager \(m_j\) to worker \(w_i\).
	%In the considered scenarios, $p_{i,j}^{(g)}$ represents the incentive paid by a regional edge manager to a participating news organization or bank, while $q_{i,j}^{(g)}$ represents the penalty associated with the node's subsequent withdrawal and unlearning request. The quality target $\theta_{i,j}^{(g)}$ specifies the expected contribution quality for the corresponding news-topic or customer-service analysis service.

	\subsubsection{Modeling of Workers}
	\label{subsubsec:worker_model}
	
	Given the dynamic and uncertain nature of the clients' behaviors, we define the attributes of a worker $w_i \in \boldsymbol{W}$ participating in the $g$-th round as a six-tuple: {\small$\left\{ d_i^{\mathsf{distr},(g)}, \, d_i^{\mathsf{size},(g)}, \, f_i^{\mathsf{comp},(g)}, \, f_i^{\mathsf{comm},(g)}, \, \varepsilon_i^{(g)}, \, \rho_i^{\mathsf{fgt},(g)} \right\}$}. Here, $d_i^{\mathsf{distr},(g)}$ and $d_i^{\mathsf{size},(g)}$ denote the local data distribution and corresponding dataset size of the worker; $f_i^{\mathsf{comp},(g)}$ denotes the computation cost incurred by the worker per 1,000 data samples, and $f_i^{\mathsf{comm},(g)}$ denotes the communication cost incurred by the worker; both costs will be later used to model the worker's participation cost in \eqref{eq:worker_cost}. 
	% The computation term accounts for processing the corresponding local data, whereas the communication term accounts for exchanging the worker's LoRA update. In every global round, the worker uploads its LoRA factors in each of the edge aggregation rounds, whose number is fixed by the protocol; 
	In particular, $f_i^{\mathsf{comm},(g)}$ is the cost of communicating all LoRA parameters, which is a function of the number of transmitted bits, the transmit power, and the channel conditions, all of which we abstract into a single value\footnote{Due to worker dynamics (e.g., mobility), the transmit power and the channel conditions may vary across workers and rounds, e.g., depending on the distance to the associated manager; to capture the resulting randomness and uncertainty while maintaining analytical tractability, we aggregate these factors into $f_i^{\mathsf{comm},(g)}$.}. Moreover, $\varepsilon_i^{(g)}$ denotes the privacy cost and $\rho_i^{\mathsf{fgt},(g)}$ represents the worker's reputation score. The reputation score is updated over a sliding window of historical unlearning behavior, with more frequent withdrawal and unlearning requests resulting in higher  $\rho_i^{\mathsf{fgt},(g)}$ values (worse reputation), reflecting reduced reliability and stability during collaborative training (later detailed in \eqref{eq:unlearning_probability}).
	
	\subsubsection{Modeling of President and Managers}
	\label{subsubsec:manager_president_model}
	
	As the global coordinator of the HFUL market, the president maintains a per-round incentive budget pool  $\beta^{\mathsf{sum},(g)}$, representing the total amount of incentives available for allocation to managers during the \(g\)-th global aggregation round. The president distributes this budget across managers with the objective of improving the final global model utility by attracting high-quality workers. Also, each manager $m_j\in \boldsymbol{M}$ maintains a local budget pool:\begin{equation}
		\beta_{j}^{\mathsf{m},(g)} = \hat{\beta}_{j}^{\mathsf{m},(g-1)} + \beta_{j}^{\mathsf{new},(g)}.
		\label{eq:manager_budget}
	\end{equation}Here, $\hat{\beta}_{j}^{\mathsf{m},(g-1)}$ denotes the residual budget carried over from the previous round, which may include penalties collected from workers requesting unlearning, while $\beta_{j}^{\mathsf{new},(g)}$ is the newly allocated budget distributed by the president in the \(g\)-th round.

	\subsubsection{Core Decision Variables}
	\label{subsubsec:decision_variables}
	
	To operationalize HermesHFL in the \(g\)-th global aggregation round, we introduce the following decision variables:
	$x_i^{(g)} \in \{0,1\}$ indicating whether $w_i$ is selected for participation, $y_{i,j}^{(g)} \in \{0,1\}$ representing whether worker $w_i$ is associated with manager $m_j$ in the $g$-th round (i.e., its edge association), and $\boldsymbol{C}_{i,j}^{(g)}$ denoting the contract between $w_i$ and $m_j$, including variables $p_{i,j}^{(g)}, q_{i,j}^{(g)}$ and $\theta_{i,j}^{(g)}$ as given by~\eqref{eq:contract_def}. Determining appropriate values for these decision variables, along with $\beta_{j}^{\mathsf{new},(g)}$, constitutes our main goal.
	
	\vspace{-2mm}
	\section{Core Modeling and Problem Formulation of HermesHFL}
	\label{sec:core_modeling}
	
	\subsection{Procedure of the HFUL Market of Our Interest}
	\label{subsec:procedure}
	Henceforth, we detail the unified learning–unlearning procedure of HermesHFL. In a nutshell, in each global round, regular workers perform stochastic gradient descent (SGD)-based LoRA fine-tuning, while workers requesting unlearning apply GA-based updates refined by OSD. These updates are sequentially aggregated at manager and president levels, integrating learning and unlearning into one global model update. These operations are discussed below.

	\subsubsection{Local Training with LoRA}
	\label{subsubsec:local_training}
	
	Let $\mathbb{W}_{i}^{(g,e,l)}$ denote the local model parameter of worker $w_i$ in the $g$-th global aggregation, the $e$-th edge aggregation, and the $l$-th local training iteration. Let $E$ denote the total number of edge aggregation rounds involved in one global aggregation, and $L$ be the number of local training/fine-tuning iterations within one edge aggregation. Also, let the corresponding LoRA model update of worker $w_i$ be: $\Delta \mathbb{W}_{i}^{(g,e,l)} = \boldsymbol{B}_{i}^{(g,e,l)} \boldsymbol{A}_{i}^{(g,e,l)}$, where $\boldsymbol{A}_{i}^{(g,e,l)} \in \mathbb{R}^{r \times k}$ and $\boldsymbol{B}_{i}^{(g,e,l)} \in \mathbb{R}^{d \times r}$ are the LoRA parameters of $w_i$.
	
	In the $g$-th global aggregation, each worker first receives the global LoRA parameters $(\boldsymbol{A}^{(g)}, \boldsymbol{B}^{(g)})$ from the president, and synchronizes its local parameters as:
	\begin{align}
		\boldsymbol{A}_{i}^{(g,0,0)} \leftarrow \boldsymbol{A}^{(g)}, \; \boldsymbol{B}_{i}^{(g,0,0)} \leftarrow \boldsymbol{B}^{(g)}.
	\end{align}
	Then, in each edge aggregation round $e$, $w_i$ executes SGD to train LoRA parameters on the frozen LLM backbone $\mathbb{W}_0$ as:
	\begin{align}
		\boldsymbol{A}_{i}^{(g,e,l+1)} &\leftarrow \boldsymbol{A}_{i}^{(g,e,l)} - \eta \nabla_{A_i} \mathbb{F}_i(\mathbb{W}_0 + \boldsymbol{B}_{i}^{(g,e,l)} \boldsymbol{A}_{i}^{(g,e,l)}), \label{eq:700}\\
		\boldsymbol{B}_{i}^{(g,e,l+1)} &\leftarrow \boldsymbol{B}_{i}^{(g,e,l)} - \eta \nabla_{B_i} \mathbb{F}_i(\mathbb{W}_0 + \boldsymbol{B}_{i}^{(g,e,l)} \boldsymbol{A}_{i}^{(g,e,l)}),\label{eq:800}
	\end{align}
	where $\eta$ is the learning rate (LR) and the loss function is
	\begin{equation}
		\mathbb{F}_i(\mathbb{W}_{i}^{(g,e,l)}) = \frac{1}{d_i^{\mathsf{size},(g)}} \sum_{\kappa=1}^{d_i^{\mathsf{size},(g)}} {f}_i(\mathring{x}_\kappa, \mathring{y}_\kappa; \mathbb{W}_{i}^{(g,e,l)}).
		\label{eq:local_loss}
	\end{equation}
	In \eqref{eq:local_loss}, $\mathbb{W}_{i}^{(g,e,l)} = \mathbb{W}_0 + \boldsymbol{B}_{i}^{(g,e,l)} \boldsymbol{A}_{i}^{(g,e,l)}$; $f_i$ is the loss function; $\mathbb{F}_i$ represents the average loss; $\mathring{x}_\kappa$ and $\mathring{y}_\kappa$ denote the input and the label of the $\kappa$-th training sample, respectively. For notational simplicity, in 
	this paper, we use $\nabla$ to denote a stochastic 
	approximation of the corresponding gradient, computed over 
	a mini-batch of local training samples at each local 
	iteration. After completing local training, each worker uploads the final LoRA matrices 
	$\boldsymbol{A}_{i}^{(g,e,L)}$ and $\boldsymbol{B}_{i}^{(g,e,L)}$ to its affiliated manager.

	\subsubsection{Unlearning Process}
	\label{subsubsec:unlearning_process}
	During the $g$-th global round, workers with unlearning requests perform GA updates, whose direction is refined by an OSD projection (defined below), based on the latest global LoRA parameters, while regular workers continue standard SGD-based local training within the same global round. Both types of updates are jointly aggregated at the manager level and then at the president level, resulting in a single global model update. Let ${\widehat{\text{w}}}_j^{(g)}$ denote the set of workers selected for model training/fine-tuning and associated with manager $m_j$ during the $g$-th global round, i.e., ${\widehat{\text{w}}}_j^{(g)}= \{{w}_i | x_i^{(g)} y_{i,j}^{(g)} = 1, w_i \in \boldsymbol{W}\}$, and $\widetilde{{\text{w}}}_j^{(g)}$ denote the subset of workers associated with $m_j$ that request unlearning after the $(g-1)$-th global aggregation, where $\widetilde{{\text{w}}}_j^{(g)} \subseteq {\widehat{\text{w}}}_j^{(g)}$. After the $(g-1)$-th global round, the set of workers that submit unlearning requests is $\bigcup_{m_j \in \boldsymbol{M}} \widetilde{{\text{w}}}_j^{(g)} \subseteq \boldsymbol{W}$. Each unlearning worker $w_{i'} \in \widetilde{\text{w}}_j^{(g)}$ performs GA,\footnote{GA maximizes the loss associated with the worker’s local dataset, thereby driving the model away from the information contributed by that worker and facilitating its removal from the global model. Intuitively, GA moves the LoRA parameters along $+\widetilde{\eta}\nabla\mathbb{F}_{i'}$, i.e., opposite to the SGD direction $-\eta\nabla\mathbb{F}_{i'}$ that originally encoded this worker’s contribution during fine-tuning; since the manager- and president-level aggregations in \eqref{eq:manager_unlearning_A}--\eqref{eq:global_unlearning_B} are data-size-weighted convex combinations, the global model inherits this cancellation direction.} whose direction is refined by an OSD projection. This process is detailed below.

	\noindent
	$\bullet$ \textit{Initializing the Unlearning State:} 
	First, each worker in the set $\widetilde{{\text{w}}}_j^{(g)}$ receives the latest global LoRA parameters:
	\begin{align}
		\widetilde{\boldsymbol{A}}_{i'}^{(g,0,0)} \leftarrow \boldsymbol{A}^{(g)}, \quad
		\widetilde{\boldsymbol{B}}_{i'}^{(g,0,0)} \leftarrow \boldsymbol{B}^{(g)},
	\end{align}
	as the initial point of the unlearning process.
	
	\noindent
	$\bullet$ \textit{GA for Unlearning:} 
	During the local fine-tuning stage of the $g$-th round,
	regular workers perform SGD as in Sec.~\ref{subsubsec:local_training}, while workers with unlearning demands execute GA as:
	\begin{small}
		\begin{align}
			\widetilde{\boldsymbol{A}}_{i'}^{(g,e,l+1)}
			&\leftarrow
			\widetilde{\boldsymbol{A}}_{i'}^{(g,e,l)}
			+
			\widetilde{\eta}
			\nabla_{\widetilde{A}_{i'}}
			\mathbb{F}_{i'}
			\Big(
			\mathbb{W}_0
			+
			\widetilde{\boldsymbol{B}}_{i'}^{(g,e,l)}
			\widetilde{\boldsymbol{A}}_{i'}^{(g,e,l)}
			\Big),
			\\
			\widetilde{\boldsymbol{B}}_{i'}^{(g,e,l+1)}
			&\leftarrow
			\widetilde{\boldsymbol{B}}_{i'}^{(g,e,l)}
			+
			\widetilde{\eta}
			\nabla_{\widetilde{B}_{i'}}
			\mathbb{F}_{i'}
			\Big(
			\mathbb{W}_0
			+
			\widetilde{\boldsymbol{B}}_{i'}^{(g,e,l)}
			\widetilde{\boldsymbol{A}}_{i'}^{(g,e,l)}
			\Big),
			\label{eq:unlearning_gradient_ascent}
		\end{align}
	\end{small}
	
	\noindent
	from which the final LoRA factors  $\widetilde{\boldsymbol{A}}_{i'}^{(g,e,\widetilde{L})}$
	and
	$\widetilde{\boldsymbol{B}}_{i'}^{(g,e,\widetilde{L})}$ are obtained. Here,  $\widetilde{\eta} > 0$ denotes the unlearning rate (usually higher than the fine-tuning LR $\eta$), and $\widetilde{L}$ is the number of local unlearning steps within one edge aggregation round.

	\noindent
	$\bullet$ \textit{OSD Projection:}
	While the GA updates aim to remove the influence of the 
	workers requesting unlearning, their direct aggregation 
	with the SGD updates of the remaining workers may lead 
	to interference between the learning and unlearning 
	processes. Specifically, the GA update of an unlearning 
	worker may contain components that oppose the training 
	directions of the remaining workers, thereby hindering 
	their contributions to collaborative fine-tuning. To 
	mitigate this interference, inspired by 
	FedOSD~\cite{pan2025federated}, we employ an orthogonal 
	projection mechanism that refines the unlearning updates 
	prior to aggregation. 
	In doing so, to characterize the relationship between learning and 
	unlearning updates, we first represent the LoRA parameter 
	updates of all participating workers in a common vector 
	space. 
	Let $\mathrm{vec}(\cdot,\cdot)$ denote the vectorization 
	and concatenation of two matrices. For an unlearning 
	worker $w_{i'}$, we define its original GA-based 
	unlearning update during the $e$-th edge aggregation 
	round of the $g$-th global round as:
	\begin{equation}
		\hspace{-4mm}
		\begin{aligned}
			\boldsymbol{d}_{i'}^{\mathsf{ga},(g,e)}
			{=} \mathrm{vec}\Big(
			\widetilde{\boldsymbol{A}}_{i'}^{(g,e,\widetilde{L})}
			-\widetilde{\boldsymbol{A}}_{i'}^{(g,e,0)},
			\widetilde{\boldsymbol{B}}_{i'}^{(g,e,\widetilde{L})}
			-\widetilde{\boldsymbol{B}}_{i'}^{(g,e,0)}
			\Big).
		\end{aligned}
		\hspace{-4mm}
	\end{equation}
	Similarly, for each regular worker 
	$w_i\in\bigcup_{m_j\in\boldsymbol{M}}
	\big(\widehat{\text{w}}_j^{(g)}
	\setminus\widetilde{\text{w}}_j^{(g)}\big)$, 
	we define its SGD-based learning update as:
	\begin{equation}
		\hspace{-4mm}
		\begin{aligned}
			\boldsymbol{d}_{i}^{\mathsf{sgd},(g,e)}
			{=} \mathrm{vec}\Big(
			\boldsymbol{A}_{i}^{(g,e,L)}
			-\boldsymbol{A}_{i}^{(g,e,0)},\boldsymbol{B}_{i}^{(g,e,L)}
			-\boldsymbol{B}_{i}^{(g,e,0)}
			\Big).
		\end{aligned}
		\hspace{-4mm}
	\end{equation}
	
	Collectively, the SGD-based updates characterize the 
	subspace along which regular workers seek to adapt their 
	respective LoRA models during the current edge aggregation 
	round. Let $\mathcal{S}^{(g,e)}$ denote this subspace, 
	defined as:
	\begin{equation}
		\resizebox{0.85\columnwidth}{!}{$\mathcal{S}^{(g,e)}
			=
			\operatorname{span}
			\left\{
			\boldsymbol{d}_{i}^{\mathsf{sgd},(g,e)}
			\,\middle|\,
			w_i\in\bigcup_{m_j\in\boldsymbol{M}}
			\big(\widehat{\text{w}}_j^{(g)}
			\setminus\widetilde{\text{w}}_j^{(g)}\big)
			\right\}.$}
	\end{equation}
	
	Our objective is to remove the components of each 
	GA-based unlearning update that lie within this subspace, 
	thereby decoupling its direction from those of the 
	regular workers. To this end, let 
	$\{\boldsymbol{e}_k^{(g,e)}\}_{k=1}^{R^{(g,e)}}$ 
	denote an orthonormal basis of $\mathcal{S}^{(g,e)}$, 
	where $R^{(g,e)}=\dim(\mathcal{S}^{(g,e)})$. The 
	projection of $\boldsymbol{d}_{i'}^{\mathsf{ga},(g,e)}$ 
	onto the orthogonal complement of $\mathcal{S}^{(g,e)}$ 
	is obtained as:
	\begin{equation}
		\begin{aligned}
			\boldsymbol{d}_{i'}^{\mathsf{osd},(g,e)}
			=
			\boldsymbol{d}_{i'}^{\mathsf{ga},(g,e)}
			-
			\sum_{k=1}^{R^{(g,e)}}
			\left\langle
			\boldsymbol{d}_{i'}^{\mathsf{ga},(g,e)},
			\boldsymbol{e}_k^{(g,e)}
			\right\rangle
			\boldsymbol{e}_k^{(g,e)},
		\end{aligned}
		\label{eq:osd_projection}
	\end{equation}
	where $\langle\cdot,\cdot\rangle$ denotes the inner 
	product. Naturally, the projected unlearning 
	update satisfies
	$\langle
	\boldsymbol{d}_{i'}^{\mathsf{osd},(g,e)},
	\boldsymbol{d}_{i}^{\mathsf{sgd},(g,e)}
	\rangle=0$
	for every regular worker $w_i$, eliminating directional 
	overlap between the projected unlearning update of 
	worker $w_{i'}$ and the learning updates of the regular 
	workers in the LoRA parameter space.
	
	Since the projection may reduce the magnitude of the 
	original GA-based update, we rescale the projected 
	update as:
	\begin{equation*}
		\resizebox{0.83\columnwidth}{!}{$\widehat{\boldsymbol{d}}_{i'}^{\mathsf{osd},(g,e)}
			{=}
			\begin{cases}
				\displaystyle
				\frac{
					\|\boldsymbol{d}_{i'}^{\mathsf{ga},(g,e)}\|_2
				}{
					\|\boldsymbol{d}_{i'}^{\mathsf{osd},(g,e)}\|_2
				}
				\boldsymbol{d}_{i'}^{\mathsf{osd},(g,e)},
				& \text{if }
				\|\boldsymbol{d}_{i'}^{\mathsf{osd},(g,e)}\|_2>0,\\[6pt]
				\boldsymbol{d}_{i'}^{\mathsf{ga},(g,e)},
				& \text{otherwise}.
			\end{cases}
			$}
	\end{equation*}
	% When the projected update is nonzero, this rescaling 
	% preserves the magnitude of the original GA-based 
	% parameter displacement without altering the 
	% orthogonality of the projected update. If no nonzero 
	% learning update is available, the projection leaves 
	% the original GA-based update unchanged. If the 
	% projection yields a zero vector, we retain the original 
	% GA-based update to avoid completely suppressing the 
	% unlearning operation, albeit at the expense of the 
	% orthogonality guarantee in this particular case.
	
	Finally, the refined unlearning update is converted 
	back into the corresponding LoRA parameter updates 
	through:
	\begin{equation}
		\left(
		\Delta\widetilde{\boldsymbol{A}}_{i'}^{(g,e)},
		\Delta\widetilde{\boldsymbol{B}}_{i'}^{(g,e)}
		\right)
		=
		\mathrm{vec}^{-1}
		\left(
		\widehat{\boldsymbol{d}}_{i'}^{\mathsf{osd},(g,e)}
		\right),
	\end{equation}
	where $\mathrm{vec}^{-1}(\cdot)$ reverses the concatenation 
	and vectorization operations. The refined LoRA factors 
	are then obtained by applying these updates to the 
	local LoRA parameters at the beginning of the current 
	edge aggregation round as:
	\begin{align}
		\widetilde{\boldsymbol{A}}_{i'}^{(g,e,\widetilde{L})}
		&\leftarrow
		\widetilde{\boldsymbol{A}}_{i'}^{(g,e,0)}
		+\Delta\widetilde{\boldsymbol{A}}_{i'}^{(g,e)},
		\label{eq:199}\\
		\widetilde{\boldsymbol{B}}_{i'}^{(g,e,\widetilde{L})}
		&\leftarrow
		\widetilde{\boldsymbol{B}}_{i'}^{(g,e,0)}
		+\Delta\widetilde{\boldsymbol{B}}_{i'}^{(g,e)}.
		\label{eq:200}
	\end{align}
	These refined LoRA factors replace those obtained through 
	the original GA operations and are subsequently used in 
	the hierarchical aggregation procedures described below.

	\noindent
	$\bullet$ \textit{Uploading Unlearning Updates to Managers:}
	Each unlearning worker uploads its final LoRA factors
	$\widetilde{\boldsymbol{A}}_{i'}^{(g,e,\widetilde{L})}$
	and
	$\widetilde{\boldsymbol{B}}_{i'}^{(g,e,\widetilde{L})}$ (i.e.,~\eqref{eq:199} and~\eqref{eq:200}), while each regular worker uploads its final LoRA factors ${\boldsymbol{A}}_{i}^{(g,e,{L})}$
	and
	${\boldsymbol{B}}_{i}^{(g,e,{L})}$ (i.e.,~\eqref{eq:700} and~\eqref{eq:800}) to its affiliated manager.
	
	\noindent
	$\bullet$ \textit{Managers Aggregate Learning and Unlearning Results:} 
	Manager $m_j$ performs \textit{FedAvg-style} aggregation on ${\widehat{\text{w}}}_j^{(g)}$ as (${\widehat{\text{w}}}_j^{(g)} \setminus \widetilde{{\text{w}}}_j^{(g)}$ gives the set that excludes $\widetilde{{\text{w}}}_j^{(g)}$ from the set ${\widehat{\text{w}}}_j^{(g)}$):
	\begin{equation} \resizebox{0.88\columnwidth}{!}{$
			\hspace{-7.5mm}\begin{aligned}
				\widetilde{\boldsymbol{A}}_{j}^{(g,e+1)}
				{=}
				\Bigg(
				\sum_{w_{i'} \in \widetilde{{\text{w}}}_j^{(g)}}\hspace{-3mm}
				\varsigma^{\mathsf{ul}} d_{i'}^{\mathsf{size},(g)}
				\widetilde{\boldsymbol{A}}_{i'}^{(g,e,\widetilde{L})}
				+ \hspace{-6mm}
				\sum_{w_i \in {\widehat{\text{w}}}_j^{(g)}
					\setminus
					\widetilde{{\text{w}}}_j^{(g)}}\hspace{-6mm}
				d_i^{\mathsf{size},(g)}
				\boldsymbol{A}_i^{(g,e,{L})}
				\Bigg)
				\Big/
				\widetilde{D}_j^{(g)} - (1 - \varsigma^{\mathsf{ul}}) \hspace{-3mm} \sum_{w_{i'} \in \widetilde{{\text{w}}}_j^{(g)}} d_{i'}^{\mathsf{size},(g)},
			\end{aligned}
			$}\hspace{-4mm}
		\label{eq:manager_unlearning_A}
	\end{equation}
	\begin{equation}
		\resizebox{0.88\columnwidth}{!}{$
			\hspace{-7.5mm}
			\begin{aligned}
				\widetilde{\boldsymbol{B}}_{j}^{(g,e+1)}
				{=}
				\Bigg(
				\sum_{w_{i'} \in \widetilde{{\text{w}}}_j^{(g)}}\hspace{-2mm}
				\varsigma^{\mathsf{ul}} d_{i'}^{\mathsf{size},(g)}
				\widetilde{\boldsymbol{B}}_{i'}^{(g,e,\widetilde{L})}
				+\hspace{-5mm}
				\sum_{w_i \in {\widehat{\text{w}}}_j^{(g)}
					\setminus
					\widetilde{{\text{w}}}_j^{(g)}}\hspace{-6mm}
				d_i^{\mathsf{size},(g)}
				\boldsymbol{B}_i^{(g,e,{L})}
				\Bigg)
				\Big/
				\widetilde{D}_j^{(g)} - (1 - \varsigma^{\mathsf{ul}})\hspace{-3mm} \sum_{w_{i'} \in \widetilde{{\text{w}}}_j^{(g)}} d_{i'}^{\mathsf{size},(g)},
			\end{aligned}
			$}\hspace{-4mm}
		\label{eq:manager_unlearning_B}
	\end{equation}
	where
	$
	\widetilde{D}_j^{(g)}
	=
	\sum_{w_i \in {\widehat{\text{w}}}_j^{(g)}}
	d_i^{\mathsf{size},(g)}$,
	and $\varsigma^{\mathsf{ul}}$ denotes the unlearning weight scale at the manager-level aggregation (introduced to prevent potentially aggressive GA-based unlearning updates from dominating the aggregated model update).
	
	The aggregated LoRA factors 
	$\left(\widetilde{\boldsymbol{A}}_{j}^{(g,e+1)}, \widetilde{\boldsymbol{B}}_{j}^{(g,e+1)}\right)$
	are transmitted back to affiliated workers to initialize local fine-tuning for the next edge aggregation round $e+1$. Once the final edge-level aggregation round is completed, i.e., when $e+1=E$, the manager forwards the final aggregated LoRA factors 
	$\left(\widetilde{\boldsymbol{A}}_{j}^{(g,E)}, \widetilde{\boldsymbol{B}}_{j}^{(g,E)}\right)$
	to the president for global aggregation and subsequent coordination of the HFUL process.

	\noindent
	$\bullet$ \textit{ President Aggregates All Managers' Updates: } The president performs aggregation of learning and unlearning updates of managers, following the \textit{FedAvg} principle \cite{pmlr-v54-mcmahan17a} as follows:
	\begin{equation}
		\resizebox{0.88\columnwidth}{!}{$
			\widetilde{\boldsymbol{A}}^{(g+1)}
			=
			\frac{
				\sum_{j=1}^{|\boldsymbol{M}|}
				\widetilde{D}_j^{(g)}
				\widetilde{\boldsymbol{A}}_j^{(g,E)}
			}{
				\sum_{j=1}^{|\boldsymbol{M}|}
				\widetilde{D}_j^{(g)}
			},~~
			\widetilde{\boldsymbol{B}}^{(g+1)}
			=
			\frac{
				\sum_{j=1}^{|\boldsymbol{M}|}
				\widetilde{D}_j^{(g)}
				\widetilde{\boldsymbol{B}}_j^{(g,E)}
			}{
				\sum_{j=1}^{|\boldsymbol{M}|}
				\widetilde{D}_j^{(g)}
			},
			$}
		\label{eq:global_unlearning_B}
	\end{equation}
	where
	$\left(
	\widetilde{\boldsymbol{A}}^{(g+1)},
	\widetilde{\boldsymbol{B}}^{(g+1)}
	\right)$
	denote the final global LoRA parameters. The resulting global LoRA parameters are then used as 
	$
	\boldsymbol{A}^{(g+1)} \leftarrow 
	\widetilde{\boldsymbol{A}}^{(g+1)}, 
	\boldsymbol{B}^{(g+1)} \leftarrow 
	\widetilde{\boldsymbol{B}}^{(g+1)}.
	$
	Through the above procedure, 
	the president integrates both learning and refined 
	unlearning updates into the global LoRA parameters, 
	which are  broadcast to the managers to 
	initialize the next global round.

	\noindent
	$\bullet$ \textit{Worker Rejoining after Unlearning:}
	Following the hierarchical aggregation of learning and 
	unlearning updates, worker $w_{i'}$ assesses the 
	effectiveness of the unlearning operation before becoming 
	eligible to rejoin the system. To this end, we examine 
	the extent to which unlearning has altered the global 
	model's predictions on the worker's dataset. Let 
	$P_{i'}^{\mathsf{old}}(z)$ and 
	$P_{i'}^{\mathsf{new}}(z)$ denote the predictive 
	probability distributions of the global LoRA-adapted 
	model for sample $z$ before and after the unlearning 
	operation, respectively. We quantify the change between 
	these predictive distributions using the average KLD 
	over the dataset of worker $w_{i'}$ as:
	\begin{equation}
		\hspace{-3mm}
		\resizebox{0.9\columnwidth}{!}{$
			\mathsf{KL}(P_{i'}^{\mathsf{old}}\|P_{i'}^{\mathsf{new}})
			{=}
			\frac{1}{d_{i'}^{\mathsf{size},(g)}}
			\hspace{-.5mm}
			\sum_{n=1}^{d_{i'}^{\mathsf{size},(g)}}
			\hspace{-1mm}D_{KL}\Big(
			P_{i'}^{\mathsf{old}}(z_n)
			\|
			P_{i'}^{\mathsf{new}}(z_n)
			\Big).
			$}
		\label{eq:unlearning_kl}
		\hspace{-3mm}
	\end{equation}
	
	\noindent
	Once the KLD exceeds a prescribed threshold $\delta$, 
	the unlearning criterion is deemed satisfied, and 
	worker $w_{i'}$ becomes eligible to rejoin the system. 
	However, rejoining does not entail restoring the 
	worker's previous participation state, as its local 
	model parameters, data characteristics, and operational 
	conditions may have changed since its departure. 
	Instead, HermesHFL discards the worker's stale local 
	model parameters and synchronizes its local LoRA factors 
	with the latest global LoRA parameters.
	
	Further, to account for potential changes in the 
	worker's attributes during its absence, HermesHFL 
	refreshes its profile upon rejoining: 
	assuming that worker $w_{i'}$ rejoins at the beginning 
	of global round $g'+1$, its updated attributes are 
	given by
	{\small $d_{i'}^{\mathsf{distr},(g'+1)},
		\allowbreak d_{i'}^{\mathsf{size},(g'+1)},
		\allowbreak f_{i'}^{\mathsf{comp},(g'+1)},
		\allowbreak f_{i'}^{\mathsf{comm},(g'+1)},
		\allowbreak \varepsilon_{i'}^{(g'+1)},
		\allowbreak \rho_{i'}^{\mathsf{fgt},(g'+1)}$},
	following the worker characterization in Sec.~\ref{subsec:basic_modeling}.
	Accordingly, the returning worker is treated as a 
	newly available participant whose selection, manager 
	association, and contractual terms are jointly 
	re-optimized based on its updated attributes and the 
	current system conditions. If selected, the worker 
	resumes collaborative fine-tuning from the latest 
	global model using its currently authorized local data. 
	% Through this mechanism, HermesHFL accommodates dynamic 
	% worker rejoining without restoring outdated model 
	% states or previously established contractual agreements.
	
	% \noindent
	% $\bullet$ \textit{Recovery Projection against Re-learning:}
	% After unlearning, let $\boldsymbol{d}_{\mathsf{f}} = \boldsymbol{d}_{\mathsf{osd}} / \|\boldsymbol{d}_{\mathsf{osd}}\|$ denote the unit forgetting direction actually applied (the raw GA direction normalized, if the projection degenerated). In a subsequent round, a candidate update $\boldsymbol{s}$ (relative to that round's global LoRA parameters $(\boldsymbol{A}^{(g)}, \boldsymbol{B}^{(g)})$, vectorized) may re-introduce the forgotten information when it moves back along $-\boldsymbol{d}_{\mathsf{f}}$. The recovery projection filters only this re-learning component:
	% \begin{equation}
		% 	\boldsymbol{s}^{\mathsf{rec}} = \boldsymbol{s} - \min\!\big( \langle \boldsymbol{s}, \boldsymbol{d}_{\mathsf{f}} \rangle, \, 0 \big)\, \boldsymbol{d}_{\mathsf{f}}.
		% 	\label{eq:recovery_projection}
		% \end{equation}
	
	% \noindent
	% When $\langle \boldsymbol{s}, \boldsymbol{d}_{\mathsf{f}} \rangle \geq 0$, the update is left untouched; only the negative alignment along the forgetting direction is removed. This prevents later training rounds from re-injecting the knowledge that has just been unlearned, and is adopted by the FedOSD strategy in the experiments.
	
	\subsection{Utility of Various Players}
	\label{subsec:utility}
	\subsubsection{Utility of Workers}
	\label{subsubsec:worker_utility}
	We design the utility of each worker to capture three aspects: \textit{(i)} the net revenue obtained from participating in the fine-tuning process, \textit{(ii)} the potential privacy-related benefit of unlearning, and \textit{(iii)} the penalty incurred when submitting an unlearning request. Accordingly, the utility of worker $w_i$ during the $g$-th global round is given by:
	\begin{equation}
		\hspace{-4mm}\begin{aligned}
			u_{i}^{\mathsf{work},(g)} =
			& \sum_{m_j \in \boldsymbol{M}} x_i^{(g)} y_{i,j}^{(g)}
			\big( p_{i,j}^{(g)} - c_{i,j}^{\mathsf{work},(g)} \big) \\
			& \hspace{-0.8em}+ x_i^{(g)} \rho_{i}^{\mathsf{fgt},(g)} \Delta \xi_i^{(g)}
			- \sum_{m_j \in \boldsymbol{M}} x_i^{(g)} y_{i,j}^{(g)}
			\rho_{i}^{\mathsf{fgt},(g)} q_{i,j}^{(g)},
		\end{aligned}\hspace{-4mm}
		\label{eq:worker_utility}
	\end{equation}
	where $c_{i,j}^{\mathsf{work},(g)}$ denotes the cost incurred by worker $w_i$ when participating in the model fine-tuning under manager $m_j$; it combines the computation and  communication costs along with the privacy cost, introduced in Sec.~\ref{subsubsec:worker_model}, and modeled as:
	\begin{equation}
		c_{i,j}^{\mathsf{work},(g)} = f_i^{\mathsf{comp},(g)} \cdot d_i^{\mathsf{size},(g)}/1000 + f_i^{\mathsf{comm},(g)} + \varepsilon_i^{(g)}.
		\label{eq:worker_cost}
	\end{equation}
	% This cost model is adopted uniformly by HermesHFL's contract design and by the fixed-price benchmark strategies in Sec.~\ref{sec:evaluation}. 
	% Also, $\Delta \xi_i^{(g)}$ denotes the privacy gain achieved when the contribution of worker $w_i$ is successfully removed from the trained model through unlearning. 
	Also, $\Delta \xi_i^{(g)}$ denotes the privacy benefit that worker $w_i$ anticipates obtaining through the removal of its historical contributions to the trained model via unlearning. Accordingly, $\rho_{i}^{\mathsf{fgt},(g)} \Delta \xi_i^{(g)}$ captures the expected privacy-related benefit associated with unlearning. 
	To characterize the uncertainty and historical tendency of unlearning behavior, we introduce the notion of an \emph{unlearning window}, connected to the reputation score $\rho_i^{\mathsf{fgt},(g)}$, which estimates the probability that worker $w_i$ requests unlearning after the $g$-th global aggregation round. Specifically, we consider an unlearning history spanning the latest $H$ global rounds and let $\tilde{x}_i^{(g-1)},..., \tilde{x}_i^{(g-H)}$ denote the historical unlearning indicators of worker $w_i$, where $\tilde{x}_i^{(g-h)}$ is a binary variable indicating whether the worker submitted an unlearning request in round $g-h$, for $h = 1, \ldots, H$ ($\tilde{x}_i^{(g-h)}=1$) or not ($\tilde{x}_i^{(g-h)}=0$). To place greater emphasis on recent behavior, we adopt an exponentially decaying weighting mechanism and define the reputation score as follows:
	\begin{equation}
		\rho_{i}^{\mathsf{fgt},(g)} = \frac{\sum_{h=1}^{H} \tau (1-\tau)^{h-1} \tilde{x}_i^{(g-h)}}{\sum_{h=1}^{H} \tau (1-\tau)^{h-1}}
		\label{eq:unlearning_probability},
	\end{equation}
	
	\noindent
	where $0 < \tau < 1$ controls the \textit{memory decay rate} (larger values of $\tau$ place more emphasis on recent unlearning behavior). Moreover, in \eqref{eq:worker_utility}, $q_{i,j}^{(g)}$ denotes the penalty of the unlearning request of worker $w_i$ under manager $m_j$. Intuitively, workers with higher probabilities of future unlearning requests may incur larger penalties due to the increased instability and retraining overhead imposed on the system. Accordingly, the overall utility of the workers during the $g$-th global round is given by:
	\begin{equation}
		U^{\mathsf{Work},(g)} = \sum_{w_i \in \boldsymbol{W}} u_i^{\mathsf{work},(g)}.
	\end{equation}
	
	\subsubsection{Utility of Managers and President}
	\label{subsubsec:manager_president_utility}
	Each manager $m_j$ aims to recruit high-quality workers for model fine-tuning and achieve satisfactory model accuracy under its budget constraint. The utility of manager $m_j$ captures the tradeoff among the model performance contributed by its assigned workers, the incentives paid to workers, and the penalties received from workers who request unlearning, defined as:
	\begin{equation}
		\begin{aligned}
			u_{j}^{\mathsf{mana},(g)} =
			& \lambda^{\mathsf{mana}} \sum_{w_i \in \boldsymbol{W}}
			x_i^{(g)} y_{i,j}^{(g)} \theta_{i,j}^{(g)} \\
			&-  \sum_{w_i \in \boldsymbol{W}} x_i^{(g)} y_{i,j}^{(g)}
			\left(p_{i,j}^{(g)} - q_{i,j}^{(g)}\rho_{i}^{\mathsf{fgt},(g)} \right),
		\end{aligned}
		\label{eq:manager_utility}
	\end{equation}
	where $\lambda^{\mathsf{mana}}$ is a normalization coefficient that balances model performance and monetary cost, and $\theta_{i,j}^{(g)}$ denotes the target training accuracy achieved by worker $w_i$ under manager $m_j$. Subsequently, the overall utility of the managers is:\begin{equation}
		U^{\mathsf{Mana},(g)} = \sum_{m_j \in \boldsymbol{M}} u_j^{\mathsf{mana},(g)}.
	\end{equation}

	\noindent
	The president, as the global aggregator, aims to maximize the system performance while controlling the budget allocated across managers and workers, with its utility defined as:
	\begin{equation}
		U^{\mathsf{Pres},(g)} = \sum_{m_j \in \boldsymbol{M}} \sum_{w_i \in \boldsymbol{W}} x_i^{(g)} y_{i,j}^{(g)} \theta_{i,j}^{(g)} - \lambda^{\mathsf{pres}} \beta^{\mathsf{sum},(g)},
		\label{eq:president_utility}
	\end{equation}
	where $\lambda^{\mathsf{pres}}$ is a normalization coefficient that balances global model performance and budget expenditure, and $\beta^{\mathsf{sum},(g)}$ denotes the total budget available for allocation in the $g$-th global round, which dictates $\beta^{\mathsf{new},(g)}$ (as detailed in \eqref{eq:budget_allocation_proportion}).
	
	\subsection{Problem Formulation and Transformation}
	
	\textit{Problem Formulation.} We consider the coordination of our HFUL market of interest as a  cross-layer optimization. In its hierarchical setting, the president allocates budgets across managers to optimize a system-level objective $U^{\mathsf{Pres},(g)}$. Under allocated budgets, each manager selects suitable workers and signs the corresponding incentive contracts to maximize the aggregate manager utility $U^{\mathsf{Mana},(g)}$. Meanwhile, workers decide whether to participate in the current global round by negotiating contracts with their managers, to maximize the aggregate worker utility $U^{\mathsf{Work},(g)}$. These coupled decisions form a multi-objective optimization problem during each global round $g$, which we formulate as $\boldsymbol{\mathcal{P}}_0^{(g)}$: 
	\begin{equation} \label{eq:three_layer_obj}
		\boldsymbol{\mathcal{P}}_0^{(g)}:
		\left\{
		\begin{aligned}
			& \mathop{\arg\max}_{\{p_{i,j}^{(g)}, q_{i,j}^{(g)}\}} U^{\mathsf{Work},(g)} \\
			& \mathop{\arg\max}_{\{x_i^{(g)}, y_{i,j}^{(g)}, q_{i,j}^{(g)}, p_{i,j}^{(g)}\}} U^{\mathsf{Mana},(g)} \\
			& \mathop{\arg\max}_{\{\beta_j^{\mathsf{new},(g)}\}} U^{\mathsf{Pres},(g)}
		\end{aligned}
		\right.
	\end{equation}
	\begin{align*}
		&\quad\text{\textbf{s.t.}} \\
		\text{(C1)}: \quad & x_i^{(g)} \in \{0,1\}, \; y_{i,j}^{(g)} \in \{0,1\}, \quad \forall w_i \in \boldsymbol{W}, \; m_j \in \boldsymbol{M} \\
		\text{(C2)}: \quad & \sum_{m_j \in \boldsymbol{M}} y_{i,j}^{(g)} = x_i^{(g)}, \quad \forall w_i \in \boldsymbol{W} \\
		\text{(C3)}: \quad & p_{i,j}^{(g)} \ge 0, \; q_{i,j}^{(g)} \ge 0, \quad \forall w_i \in \boldsymbol{W}, \; m_j \in \boldsymbol{M} \\
		\text{(C4)}: \quad & \theta_{i,j}^{(g)} \ge \theta_{i,j}^{\min,(g)} \text{ if } x_i^{(g)} y_{i,j}^{(g)} = 1, \quad \forall w_i \in \boldsymbol{W}, \; m_j \in \boldsymbol{M} \\
		\text{(C5)}: \quad & \sum_{w_i \in \boldsymbol{W}} x_i^{(g)} y_{i,j}^{(g)} p_{i,j}^{(g)} \le \beta_j^{\mathsf{m},(g)}, \quad \forall m_j \in \boldsymbol{M} \\
		\text{(C6)}: \quad & \sum_{m_j \in \boldsymbol{M}} \beta_j^{\mathsf{new},(g)} \le \beta^{\mathsf{sum},(g)} \\
		\text{(C7)}: \quad & u_i^{\mathsf{work},(g)} \ge 0, \quad \forall w_i \in \boldsymbol{W} \\
		\text{(C8)}: \quad & \sum_{m_j \in \boldsymbol{M}} \sum_{w_i \in \boldsymbol{W}} x_i^{(g)} y_{i,j}^{(g)} \theta_{i,j}^{(g)} > 0
	\end{align*}
	
	\noindent
	In $\boldsymbol{\mathcal{P}}_0^{(g)}$, constraint (C1) defines the worker selection indicator, where $x_i^{(g)} \in \{0,1\}$ indicates whether worker $w_i$ is selected by managers. It also specifies the manager association indicator $y_{i,j}^{(g)} \in \{0,1\}$. Constraint (C2) ensures that if a worker is selected, it is assigned to a manager. Constraint (C3) enforces the non-negativity of incentive payments and unlearning penalties. Constraint (C4) guarantees the training quality requirement, i.e., the model performance of each selected worker should satisfy the minimum threshold $\theta_{i,j}^{(g)} \ge \theta_{i,j}^{\min,(g)}$ when $x_i^{(g)} y_{i,j}^{(g)} = 1$. Constraint (C5) ensures that the total incentive payments do not exceed the allocated budget $\beta_j^{\mathsf{m},(g)}$ given by~\eqref{eq:manager_budget}. Constraint (C6) ensures that the total budget allocated across all managers remains within the system-level budget. Constraint (C7) guarantees non-negative worker utilities, ensuring individual rationality. Finally, constraint (C8) ensures that the aggregate target training quality across selected worker--manager pairs is strictly positive, thereby ensuring that the budget allocation subsequently defined in \eqref{eq:budget_allocation_proportion} is well-defined. Notably, we allow managers to attain negative utilities, as they are assumed to remain committed to collaborative training even when their costs exceed their economic benefits.

	\textit{Problem Transformation. }The original problem $\boldsymbol{\mathcal{P}}_0^{(g)}$ involves a tightly coupled three-layer optimization across the president, managers, and workers. Its computational difficulty stems from the interdependence of discrete worker selection, continuous budget allocation, and nonlinear incentive decisions. Specifically, upper-layer budget decisions shape lower-layer coalition formation and incentive allocation, which in turn determine global system utility. Such a bidirectional coupling yields a non-convex mixed-integer nonlinear program with a combinatorial search space. Crucially, despite their hierarchical roles, the president and managers share aligned objectives of maximizing training utility under budget constraints, differing primarily in their decision scopes rather than strategic interests. This structural alignment motivates consolidating their optimization decisions, thereby reducing hierarchical complexity while preserving essential cross-layer dependencies.
	Accordingly, we reformulate the original three-layer problem into a two-level optimization by adopting a contribution-proportional budget allocation scheme (see Appx. \ref{sec:appendix_examples} for practical examples). Specifically, in each global round, the president allocates the system budget among managers according to the aggregate training contributions of their associated workers. As such, the budget allocated to manager $m_j$ is determined by the relative contribution quality of its participating workers:
	\begin{align}
		\beta_{j}^{\mathsf{new},(g)} &= \beta^{\mathsf{sum},(g)} \cdot 
		\frac{\sum_{w_i \in \boldsymbol{W}} x_i^{(g)} y_{i,j}^{(g)} \theta_{i,j}^{(g)}}
		{\sum_{m_j \in \boldsymbol{M}} \sum_{w_i \in \boldsymbol{W}} x_i^{(g)} y_{i,j}^{(g)} \theta_{i,j}^{(g)}}.
		\label{eq:budget_allocation_proportion}
	\end{align}

	Under this contribution-based assignment, the president’s budget decision is determined analytically, reducing the original three-layer optimization to a two-level problem involving managers and workers. The resulting formulation jointly optimizes worker selection, manager association, incentive contracts, and participation decisions as follows:
	\begin{equation} \label{eq:two_layer_obj}
		\boldsymbol{\mathcal{P}}_1^{(g)}:
		\left\{
		\begin{aligned}
			& \mathop{\arg\max}_{\{p_{i,j}^{(g)}, q_{i,j}^{(g)}\}} U^{\mathsf{Work},(g)} \\[4pt]
			& \mathop{\arg\max}_{\{x_i^{(g)}, y_{i,j}^{(g)}, p_{i,j}^{(g)}, q_{i,j}^{(g)}\}} U^{\mathsf{Mana},(g)} 
		\end{aligned}
		\right.
	\end{equation}
	\begin{align*}
		\text{\textbf{s.t.}} \quad
		& \text{(C1)--(C4),(C7)--(C8)} \nonumber \\
		& \text{(C9)}: \sum_{w_i \in \boldsymbol{W}} x_i^{(g)} y_{i,j}^{(g)} p_{i,j}^{(g)}
		\le \hat{\beta}_{j}^{\mathsf{m},(g-1)} \nonumber \\
		& \quad + \beta^{\mathsf{sum},(g)} \cdot
		\frac{\displaystyle\sum_{w_i \in \boldsymbol{W}} x_i^{(g)} y_{i,j}^{(g)} \theta_{i,j}^{(g)}}
		{\displaystyle\sum_{m_j \in \boldsymbol{M}} \sum_{w_i \in \boldsymbol{W}} x_i^{(g)} y_{i,j}^{(g)} \theta_{i,j}^{(g)}},
		\quad \forall m_j \in \boldsymbol{M}
	\end{align*}
	\noindent
	Constraint (C9) replaces (C5) and (C6) by ensuring that the total payments issued by each manager do not exceed its effective available budget during the $g$-th global round. This effective budget consists of two components: \textit{(i)} the remaining unused budget from the previous round, denoted by $\hat{\beta}_{j}^{\mathsf{m},(g-1)}$, and \textit{(ii)} the newly assigned system budget determined according to the contribution quality of the manager’s associated workers.

	\textit{Analysis of $\boldsymbol{\mathcal{P}}_1^{(g)}$. } Problem $\boldsymbol{\mathcal{P}}_1^{(g)}$ remains challenging to solve due to the interplay between discrete and continuous  variables and the coupling among managers. Specifically, the continuous incentive and unlearning penalty variables $\{p_{i,j}^{(g)}, q_{i,j}^{(g)}\}$ are coupled with the binary worker selection and association variables $\{x_i^{(g)}, y_{i,j}^{(g)}\}$. Furthermore, the fractional budget constraint (C9) couples the feasible decisions of each manager with the worker selection and association decisions of all other managers through their aggregate target training quality. These interdependencies give rise to a non-convex, non-separable mixed-integer optimization problem with a combinatorial search space. Consequently, conventional gradient-based optimization and Karush--Kuhn--Tucker (KKT)-based solution approaches cannot be  applied to solve $\boldsymbol{\mathcal{P}}_1^{(g)}$, motivating the development of a customized solution methodology.
	
	% %\noindent
	% Conventional optimization approaches, including Karush-Kuhn-Tucker (KKT)-based methods, are difficult to apply to $\boldsymbol{\mathcal{P}}_1^{(g)}$ due to the coupled bilevel structure and discrete association decisions. Specifically, lower-level worker selection and association $\{x_i^{(g)}, y_{i,j}^{(g)}\}$ respond to upper-level incentive decisions $\{p_{i,j}^{(g)}, q_{i,j}^{(g)}\}$, while the resulting worker contributions influence budget availability and subsequent incentive allocation. This bidirectional dependency complicates analytical reformulation and precludes straightforward decomposition into independently solvable subproblems.
	
	\section{Solution Methodology}
	\label{subsec:solution}
	
	To address $\boldsymbol{\mathcal{P}}_1^{(g)}$, we propose Neogen, decomposing the optimization process into two interconnected levels. The upper level determines $\mathbf{u}^{(g)} = \{p_{i,j}^{(g)}, q_{i,j}^{(g)}\}$, which corresponds to the incentive payments and unlearning penalties associated with worker–manager pairs, while the lower level determines $\mathbf{v}^{(g)} = \{x_i^{(g)}, y_{i,j}^{(g)}\}$, which represents worker selection and worker–manager association decisions under the upper-level incentive strategy. Neogen integrates two complementary evolutionary mechanisms along with a tailored neural network (NN)-based guidance module: \textit{(i)} CMA-ES at the upper level, and \textit{(ii)} a CHC evolutionary algorithm at the lower level. 
	
	%\noindent
	At the upper level, we employ CMA-ES to optimize the continuous incentive variables. The rationale behind adopting CMA-ES is that the upper-level objective depends on the optimal solution to the lower-level combinatorial optimization problem, making explicit gradient information unavailable. To formalize this dependency, we introduce the lower-level optimal response mapping $\Phi(\cdot)$. Specifically, for a given upper-level incentive decision $\mathbf{u}^{(g)}$, the managers optimize the lower-level worker selection and association decision vector $\mathbf{v}$ to maximize their aggregate utility $U^{\mathsf{Mana},(g)}$, subject to constraints (C1)--(C4) and (C7)--(C9). The resulting optimal response, denoted by $\mathbf{v}^{*}=\Phi(\mathbf{u}^{(g)})$, is characterized as follows:
	\begin{equation}
		\begin{aligned}
			\mathbf{v}^{*}=\Phi(\mathbf{u}^{(g)})
			\in
			\mathop{\arg\max}_{\mathbf{v}}
			U^{\mathsf{Mana},(g)}\!\left(\mathbf{v},\mathbf{u}^{(g)}\right),\\
			\qquad \text{s.t. (C1)--(C4), (C7)--(C9)}.
		\end{aligned}
		\label{eq:lower_response_mapping}
	\end{equation}

	For each candidate upper-level decision $\mathbf{u}^{(g)}$, the corresponding lower-level optimal response $\mathbf{v}^{*}=\Phi(\mathbf{u}^{(g)})$ determines the worker utility as follows:
	\begin{equation}
		U^{\mathsf{Work},(g)}
		\!\left(\mathbf{u}^{(g)},\mathbf{v}^{*}\right)
		=
		U^{\mathsf{Work},(g)}
		\!\left(\mathbf{u}^{(g)},\Phi(\mathbf{u}^{(g)})\right).
		\label{eq:upper_response_evaluation}
	\end{equation}
	Accordingly, the upper-level optimization seeks to maximize $U^{\mathsf{Work},(g)}(\mathbf{u}^{(g)},\Phi(\mathbf{u}^{(g)}))$ with respect to $\mathbf{u}^{(g)}$. Since changes in the continuous incentive variables may induce abrupt changes in the optimal worker selection and manager association decisions, the resulting upper-level objective is potentially non-smooth and discontinuous. This motivates the adoption of CMA-ES, which explores the continuous decision space without requiring gradient information.
	Nevertheless, the adoption of CMA-ES introduces a computational challenge: evaluating the fitness of each candidate $\mathbf{u}^{(g)}$ requires solving the lower-level combinatorial optimization problem to obtain the corresponding response $\Phi(\mathbf{u}^{(g)})$. To tackle this problem, we employ CHC, whose Hamming distance-based mating control and cataclysmic mutation mechanisms promote search diversity and mitigate premature convergence in the discrete worker selection and manager association space. However, repeatedly executing CHC for every candidate generated by CMA-ES constitutes a computational bottleneck in the bilevel optimization process. To address this bottleneck, Neogen incorporates an NN that learns the mapping between upper-level incentive decisions and their corresponding lower-level solutions. By exploiting the solutions obtained in previous optimization iterations, the NN predicts promising lower-level solutions for newly generated upper-level candidates, allowing CHC to concentrate its search around these predictions. We next discuss these processes in detail (see Alg.~\ref{alg:neogen} in Appx.~\ref{sec:appendix_algorithm} for a summary of the overall optimization procedure).

	\subsection{Upper-Level Sampling and CMA-ES Update}
	\label{subsec:cma_es}
	
	During the $g$-th global aggregation round, Neogen performs multiple CMA-ES generations; we use $a$ to index these successive CMA-ES generations within each global round. In each generation, Neogen samples ${pop}^u$ candidate upper-level solutions for evolutionary exploration, and the resulting population is denoted as $\{\mathbf{u}_k^{(a)}\}_{k=1}^{pop^u}$. Each upper-level candidate corresponds to a set of incentive and unlearning penalty variables, $\mathbf{u}_k^{(a)}$ ($\{p_{i,j}^{(g)}, q_{i,j}^{(g)}\}$).
	Each $\mathbf{u}_k^{(a)}$ is sampled from a multivariate Gaussian distribution, following the standard CMA-ES sampling and update rules~\cite{hansen2006cma,hansen2001completely}:
	\begin{equation}\label{eq:34}
		\mathbf{u}_k^{(a)} \sim \mathcal{N}\!\left(\mathbf{m}_u^{(a)}, \bigl(\sigma_u^{(a)}\bigr)^2\mathbf{C}_u^{(a)}\right),
	\end{equation}
	where $\mathbf{m}_u^{(a)}$ denotes the mean vector of the sampling distribution, $\mathbf{C}_u^{(a)}$ is the covariance matrix governing the shape and orientation of the search distribution, and $\sigma_u^{(a)}$ is the global step size controlling its overall spread. Here, $\mathbf{C}_u^{(a)}$ is defined
	in the upper-level decision space, whose coordinates correspond to the
	incentive and unlearning-penalty variables
	$\mathbf{u}^{(g)}=\{p_{i,j}^{(g)},q_{i,j}^{(g)}\}$, and is subsequently
	learned from the historical movement of high-fitness candidate solutions. Its diagonal entries describe the search variance of
	individual decision variables, whereas its off-diagonal entries describe
	how pairs of decision variables vary jointly. Each sampled upper-level candidate $\mathbf{u}_k^{(a)}$ is evaluated
	by the objective $U^{\mathsf{Work},(g)}
	\left(\mathbf{u}_k^{(a)},\mathbf{v}_k^*\right)$ as in
	\eqref{eq:upper_response_evaluation}, where $\mathbf{v}_k^*$ is
	obtained through lower-level optimization.

	Among the $pop^u$ upper-level candidate solutions sampled according to \eqref{eq:34}, we select the top-$K$ candidates, where $K=\lfloor pop^u/2\rfloor$, based on their fitness values in descending order. Let $\mathbf{u}^{\mathsf{topK},(a)}_k$ denote the $k$-th highest-ranked candidate. These candidates are then used to update the three parameters of the Gaussian sampling distribution in \eqref{eq:34}, namely, the mean vector $\mathbf{m}_u^{(a)}$, the global step size $\sigma_u^{(a)}$, and the covariance matrix $\mathbf{C}_u^{(a)}$, to guide subsequent sampling toward regions of higher upper-level utility. These updates are done as follows.
	
	\textit{(i) Mean Vector Update:}
	To shift the center of the sampling distribution toward 
	regions associated with higher fitness values, we update 
	the mean vector as an average of the candidates:
	\begin{equation}\label{eq:35}
		\mathbf{m}_u^{(a+1)} =
		\sum_{k=1}^{K} \gamma'_k \,
		\mathbf{u}^{\mathsf{topK},(a)}_k,
	\end{equation}
	where $\gamma'_k>0$ denotes the recombination weight 
	associated with the $k$-th selected candidate, satisfying 
	$\sum_{k=1}^{K}\gamma'_k=1$.
	
	\textit{(ii) Global Step-Size Update:}
	While the mean vector update determines the center of the 
	sampling distribution, the global step size controls the 
	overall extent of exploration around this center. To adapt 
	the step size, CMA-ES maintains an evolution path that 
	accumulates successive normalized displacements of the 
	mean vector, updated as:
	\begin{equation}
		\begin{aligned}
			\mathbf{p}_\sigma^{(a+1)}
			&= (1-c_\sigma)\mathbf{p}_\sigma^{(a)} \\
			&\quad + \sqrt{c_\sigma(2-c_\sigma)\mu_{\text{eff}}}
			\left(\mathbf{C}_u^{(a)}\right)^{-\frac{1}{2}}
			\frac{\mathbf{m}_u^{(a+1)}-\mathbf{m}_u^{(a)}}
			{\sigma_u^{(a)}} ,
		\end{aligned}
	\end{equation}
	where $c_\sigma$ is the learning rate governing the 
	adaptation of the evolution path, 
	$\mu_{\text{eff}}=(\sum_{k=1}^{K}(\gamma'_k)^2)^{-1}$
	is the effective selection mass, and 
	$\left(\mathbf{C}_u^{(a)}\right)^{-\frac{1}{2}}$
	normalizes the mean displacement with respect to the 
	current covariance matrix. 
	The length of the updated evolution path is then compared 
	with its expected length under isotropic Gaussian sampling 
	to adapt the global step size as:
	\begin{equation}
		\sigma_u^{(a+1)} = \sigma_u^{(a)}
		\exp\Bigg(
		\frac{c_\sigma}{d_\sigma}
		\Big(
		\frac{\|\mathbf{p}_\sigma^{(a+1)}\|}
		{\mathbb{E}[\|\boldsymbol{X}\|]}-1
		\Big)
		\Bigg),
	\end{equation}
	where $d_\sigma$ is the damping factor and 
	$\mathbb{E}[\|\boldsymbol{X}\|]$ denotes the expected 
	Euclidean norm of a standard Gaussian random vector 
	$\boldsymbol{X}\sim\mathcal{N}(\boldsymbol{0},\boldsymbol{I})$.
	Intuitively, consistently aligned search directions lead 
	to a longer evolution path and consequently a larger step 
	size, whereas shorter evolution paths lead to a smaller 
	step size, encouraging more localized exploration.
	
	\textit{(iii) Covariance Matrix Update:}
	While step-size adaptation controls the overall spread of 
	the sampling distribution, covariance matrix adaptation 
	adjusts its shape to favor promising search directions. 
	To this end, we maintain a separate evolution path for 
	covariance adaptation, updated as:
	\begin{equation}
		\begin{aligned}
			\mathbf{p}_c^{(a+1)}
			{=} (1-c_c)\mathbf{p}_c^{(a)}  {+} \sqrt{c_c(2-c_c)\mu_{\text{eff}}}
			\frac{\mathbf{m}_u^{(a+1)}-\mathbf{m}_u^{(a)}}
			{\sigma_u^{(a)}},
		\end{aligned}
	\end{equation}
	where $c_c$ governs the adaptation rate of the covariance 
	evolution path. In addition to the accumulated search 
	direction captured by this path, we exploit the individual 
	search directions of the selected candidates, defined as:
	\begin{equation}\label{eq:y_def}
		\mathbf{y}_k^{(a+1)}
		=
		{\left(\mathbf{u}^{\mathsf{topK},(a)}_k
			-\mathbf{m}_u^{(a)}\right)}\big/{\sigma_u^{(a)}}.
	\end{equation}
	
	Combining these two sources of information, the covariance 
	matrix is adapted through the conventional rank-one and 
	rank-$\mu$ updates of CMA-ES~\cite{hansen2001completely}, where $\mu=K$ denotes the
	number of selected candidates:
	\begin{equation}\label{eq:cma_c_update}
		\begin{aligned}
			\mathbf{C}_u^{(a+1)}
			&= (1-c_1-c_\mu)\mathbf{C}_u^{(a)}
			+ c_1\mathbf{p}_c^{(a+1)}
			{\mathbf{p}_c^{(a+1)}}^\top \\
			&\quad + c_\mu\sum_{k=1}^{K}\gamma'_k
			\mathbf{y}_k^{(a+1)}
			{\mathbf{y}_k^{(a+1)}}^\top.
		\end{aligned}
	\end{equation}
	Here, $c_1$ and $c_\mu$ denote the learning rates associated 
	with the rank-one and rank-$\mu$ updates, respectively.  In essence, the rank-one update captures the accumulated direction of 
	mean displacement across successive generations, whereas 
	the rank-$\mu$ update captures the weighted second moment 
	of the selected candidates' search directions. Together, 
	these updates enable CMA-ES to adapt its sampling 
	distribution according to both historical search directions 
	and the distribution of high-fitness candidates in the 
	current generation.
	
	The updated parameters $\mathbf{m}_u^{(a+1)}$, $\sigma_u^{(a+1)}$, and $\mathbf{C}_u^{(a+1)}$ are used to sample the next generation of upper-level candidates according to \eqref{eq:34}. Through these iterative updates, CMA-ES refines its search over the continuous incentive and penalty variables. We next discuss how the corresponding lower-level solutions are obtained for each sampled upper-level candidate.
	
	\subsection{Lower-Level Optimization with CHC and NN Guidance}
	\label{subsec:chc_nn}
	
	% For each upper-level solution $\mathbf{u}_k^{(a)}$ sampled according to \eqref{eq:34} with the CMA-ES parameters tuned as in Sec.~\ref{subsec:cma_es}, the lower-level solution is obtained by solving: 
	% \begin{equation}\label{eq:41}
		% 	\begin{aligned}
			% 		\mathbf{v}_k^* &= \arg\max_{\mathbf{v}} \; U^{\mathsf{Mana},(g)}(\mathbf{v}, \mathbf{u}_k^{(a)}) \\ 
			% 		&\text{s.t. (C1)--(C4), (C7)--(C9)}.
			% 	\end{aligned}
		% \end{equation}
	% Let $len^\mathsf{w}$ denote the length of the binary lower-level solution vector, which is  $|\boldsymbol{W}|+|\boldsymbol{W}||\boldsymbol{M}|$. Since the lower-level optimization involves binary decision variables and combinatorial constraints, Neogen adopts the CHC evolutionary algorithm to efficiently explore the feasible discrete search space. Specifically, during the $g$-th global  round, the CHC population associated with upper-level candidate $\mathbf{u}_k^{(a)}$ is initialized as:
	% \begin{equation}
		% 	\mathbf{V}_k^{(0)} = \{\mathbf{v}_{k,i}^{(0)}\}_{i=1}^{pop^l}, \quad 
		% 	\mathbf{v}_{k,i}^{(0)} \in \{0,1\}^{len^\mathsf{w}},
		% \end{equation}
	
	% \noindent
	% where ${pop}^l$ denotes the lower-level population size. To ensure feasible worker–manager associations, the initialized solutions have to satisfy:
	% \begin{equation}\label{eq:43}
		% 	x_i^{(g)} \ge y_{i,j}^{(g)}, \quad \forall i,j.
		% \end{equation}
	
	For each upper-level candidate $\mathbf{u}_k^{(a)}$ sampled according to \eqref{eq:34} using the CMA-ES parameters updated as in Sec.~\ref{subsec:cma_es}, the corresponding lower-level solution is obtained by solving:
	\begin{equation}\label{eq:41}
		\begin{aligned}
			\mathbf{v}_k^* &\in \arg\max_{\mathbf{v}} \; 
			U^{\mathsf{Mana},(g)}(\mathbf{v},\mathbf{u}_k^{(a)}) \\
			&\text{s.t. (C1)--(C4), (C7)--(C9)}.
		\end{aligned}
	\end{equation}
	Here, to facilitate the evolutionary search, we encode the lower-level decision variables into a binary vector as:
	\begin{equation}
		\mathbf{v}=
		\left[\mathbf{x}^{\top},
		\operatorname{vec}(\mathbf{Y})^{\top}\right]^{\top}
		\in \{0,1\}^{len^{\mathsf{w}}},
	\end{equation}
	where $\mathbf{x}=[x_i^{(g)}]_{i=1}^{|\boldsymbol{W}|}$ denotes the worker selection vector, $\mathbf{Y}=[y_{i,j}^{(g)}]_{i,j}$ denotes the worker--manager association matrix, and $\operatorname{vec}(\cdot)$ denotes matrix vectorization. Accordingly, the length of the binary solution vector is given by $len^{\mathsf{w}}=|\boldsymbol{W}|+|\boldsymbol{W}||\boldsymbol{M}|$.
	
	Given the combinatorial nature of \eqref{eq:41}, Neogen adopts the CHC algorithm~\cite{eshelman1991chc} to explore the resulting binary decision space. Specifically, during the $g$-th global round, the CHC population associated with the upper-level candidate $\mathbf{u}_k^{(a)}$ is initialized as:
	\begin{equation}
		\mathbf{V}_k^{(0)}
		=\{\mathbf{v}_{k,s}^{(0)}\}_{s=1}^{pop^l},
		\quad
		\mathbf{v}_{k,s}^{(0)}\in\{0,1\}^{len^{\mathsf{w}}},
	\end{equation}
	where $pop^l$ denotes the lower-level population size and $\mathbf{v}_{k,s}^{(0)}$ represents the $s$-th initial lower-level candidate associated with $\mathbf{u}_k^{(a)}$. 
	To prevent the association of unselected workers with managers, 
	the worker selection and association variables encoded in each 
	initialized candidate should satisfy:
	\begin{equation}\label{eq:43}
		x_i^{(g)}\geq y_{i,j}^{(g)},\quad \forall i,j.
	\end{equation}
	
	Starting from this initial population, CHC progressively 
	explores the binary decision space by generating new candidates 
	through crossover. To prevent premature convergence, CHC 
	restricts crossover to sufficiently dissimilar parent solutions, 
	thereby preserving population diversity. The dissimilarity 
	between two binary solutions is measured by their Hamming 
	distance, i.e., the number of bit positions at which they differ. 
	Accordingly, we initialize the minimum Hamming distance 
	threshold for crossover as:
	\begin{equation}
		d^{\mathsf{ham}}=d_{\text{init}}^{\mathsf{ham}}
		=\lfloor len^\mathsf{w}/4\rfloor.
	\end{equation}
	
	Let $t$ denote the inner generation (IG) index of the 
	lower-level optimization. During the $t$-th IG, pairs of 
	candidate solutions $(\mathbf{v}_{k,i}^{(t)}, 
	\mathbf{v}_{k,j}^{(t)})$ are randomly selected from the 
	current population for potential crossover. A selected pair 
	is permitted to undergo crossover if it satisfies the 
	following incest prevention criterion:
	\begin{equation}
		\operatorname{Ham}(\mathbf{v}_{k,i}^{(t)}, 
		\mathbf{v}_{k,j}^{(t)})>d^{\mathsf{ham}},
	\end{equation}
	where $\operatorname{Ham}(\cdot,\cdot)$ denotes the Hamming 
	distance. For each eligible parent pair, Half Uniform 
	Crossover (HUX)~\cite{eshelman1991chc} generates two offspring as:
	\begin{equation}
		(\mathbf{v}_{k,\text{child},1}, 
		\mathbf{v}_{k,\text{child},2})
		=\text{HUX}(\mathbf{v}_{k,i}^{(t)}, 
		\mathbf{v}_{k,j}^{(t)}),
	\end{equation}
	by exchanging half (rounded down) of the differing bits 
	between the two parents. This operation generates new 
	combinations of worker selection and manager association 
	decisions. The generated offspring satisfying \eqref{eq:43} 
	collectively form the offspring population 
	$\mathbf{V}_{k,\text{offspring}}^{(t)}$.
	To retain high-quality solutions while incorporating newly 
	generated candidates, CHC employs elitist survival selection, where the next-generation population is:
	\begin{equation}\label{eq:chc_selection}
		\mathbf{V}_k^{(t+1)}
		=\text{Top}_{pop^l}
		\big(\mathbf{V}_k^{(t)}
		\cup\mathbf{V}_{k,\text{offspring}}^{(t)}\big).
	\end{equation}
	Here, $\text{Top}_{pop^l}(\cdot)$ selects the $pop^l$ 
	highest-ranked solutions according to the lower-level objective 
	$U^{\mathsf{Mana},(g)}$.
	
	As the population evolves, its diversity may decrease, 
	making it increasingly difficult to identify eligible parent 
	pairs. To sustain the evolutionary search, whenever no valid 
	offspring can be generated, we reduce the Hamming distance 
	threshold $d^{\mathsf{ham}}$ by one, provided it is positive, 
	thereby gradually relaxing the crossover restriction. If the 
	threshold reaches zero and no improvement in the best objective 
	value is observed for $T_{\text{stag}}$ consecutive generations, 
	a cataclysmic mutation is triggered to restore population diversity. 
	Specifically, we retain the best solution obtained thus far and 
	regenerate the remaining population through randomized bit flips:
	\begin{equation}
		\mathbf{v}_{k,i}^{(t+1)}=
		\begin{cases}
			\mathbf{v}_{k,\text{best}}, & i=1,\\
			\mathbf{v}_{k,\text{best}}\oplus
			\boldsymbol{\delta}_{k,i}, & i=2,\dots,pop^l,
		\end{cases}
	\end{equation}
	where $\mathbf{v}_{k,\text{best}}$ denotes the best lower-level 
	solution found thus far for upper-level candidate 
	$\mathbf{u}_k^{(a)}$, and $\boldsymbol{\delta}_{k,i}$ is a 
	binary mutation vector whose entries are independently sampled 
	from a Bernoulli distribution with parameter $p_{\text{mut}}$. 
	Here, $\oplus$ denotes the bitwise XOR operation, which flips 
	each bit of $\mathbf{v}_{k,\text{best}}$ for which the 
	corresponding entry of $\boldsymbol{\delta}_{k,i}$ equals one. 
	The regenerated candidates are filtered to satisfy \eqref{eq:43}, 
	and the Hamming distance threshold is reset to 
	$d_{\text{init}}^{\mathsf{ham}}$ before resuming the evolutionary search.

	\textit{NN Guidance Design:} The CHC procedure described above is executed for each 
	candidate generated by the upper-level CMA-ES, terminating 
	when the maximum number of IGs, denoted by $T$, is reached. 
	However, repeatedly initializing CHC from scratch overlooks 
	the information acquired from previously solved lower-level 
	optimization problems. To exploit this information, we 
	progressively construct a training dataset 
	$\mathcal{D}=\{(\mathbf{u}_k^{(a)},\mathbf{v}_k^*)\}$ 
	by collecting upper-level candidates and their corresponding 
	CHC solutions throughout the optimization process. Using 
	this dataset, we train an NN to approximate the lower-level 
	response mapping $\Phi(\cdot)$ as:
	\begin{equation}
		\mathbf{v}^*=\Phi(\mathbf{u})\approx NN(\mathbf{u};\theta),
	\end{equation}
	where $\theta$ denotes the trainable parameters of the NN. 
	The training objective is to minimize the discrepancy between 
	the NN predictions and the corresponding CHC solutions 
	in $\mathcal{D}$.
	
	Once trained, the NN provides a promising initial lower-level 
	solution for each newly sampled upper-level candidate 
	$\mathbf{u}_k^{(a)}$ as:
	\begin{equation}
		\hat{\mathbf{v}}_k=NN(\mathbf{u}_k^{(a)};\theta).
	\end{equation}
	Rather than randomly initializing the CHC population across 
	the entire binary decision space, we leverage this prediction 
	to initialize the population within a Hamming neighborhood 
	centered at $\hat{\mathbf{v}}_k$, defined as:
	\begin{equation}
		\mathcal{N}(\hat{\mathbf{v}}_k)
		=\{\mathbf{v}\in\{0,1\}^{len^\mathsf{w}}
		\mid \operatorname{Ham}(\mathbf{v},\hat{\mathbf{v}}_k)
		\leq r^{\mathsf{ham}}\},
	\end{equation}
	where $r^{\mathsf{ham}}$ controls the extent of exploration 
	around the predicted solution. The initialized population 
	subsequently undergoes the CHC refinement procedure described 
	above. The resulting solution is used both to evaluate the 
	fitness of the corresponding upper-level candidate and to 
	enrich $\mathcal{D}$ for subsequent NN training. Through this 
	feedback mechanism, Neogen exploits previously acquired 
	optimization experience to guide subsequent lower-level 
	searches, thereby reducing the computational overhead of 
	repeated CHC executions.
	% The CHC process terminates when the maximum number of IGs, denoted by $T$, is reached. After termination, the obtained solution pairs are stored in the dataset $\mathcal{D} = \{(\mathbf{u}_k^{(a)}, \mathbf{v}_k^*)\}$, which is used to train an NN that approximates the implicit mapping from upper-level decisions to lower-level optimal responses:
	% \begin{equation}
		% 	\mathbf{v}^* \approx NN(\mathbf{u}; \theta).
		% \end{equation}
	% The NN is trained to minimize the discrepancy between predicted lower-level solutions and the corresponding optimized CHC solutions stored in $\mathcal{D}$. For a newly sampled upper-level candidate $\mathbf{u}_k^{(a)}$, the trained NN produces an initial lower-level prediction:
	% \begin{equation}
		% 	\hat{\mathbf{v}}_k = NN(\mathbf{u}_k^{(a)}; \theta).
		% \end{equation}
	
	% \noindent
	% Instead of performing a full combinatorial search, a localized CHC refinement is conducted within a restricted Hamming neighborhood:
	% \begin{equation}
		% 	\mathcal{N}(\hat{\mathbf{v}}_k) = \{\mathbf{v} \mid \operatorname{Ham}(\mathbf{v}, \hat{\mathbf{v}}_k) \le r^{\mathsf{ham}}\},
		% \end{equation}
	% where the initial population is generated within $\mathcal{N}(\hat{\mathbf{v}}_k)$ to refine the predicted solution. This NN-guided initialization narrows the search space and accelerates convergence of the lower-level optimization.

	% \subsection{Convergence Analysis of Neogen}
	\subsection{Convergence Analysis of Neogen}
	
	In the following, we discuss the convergence properties of Neogen for solving 
	$\boldsymbol{\mathcal{P}}_1^{(g)}$ by examining its three 
	interconnected components  (see Appx.~\ref{sec:appendix_convergence} for further 
	details).

	\textit{(i) Lower-Level Optimization:}
	For a given upper-level candidate $\mathbf{u}_k^{(a)}$, 
	the lower-level problem in \eqref{eq:41} has a finite 
	search space, since both worker selection and manager 
	association decisions are binary. Through elitist survival 
	selection, CHC preserves the best solution obtained thus 
	far, ensuring that the best objective value is monotonically 
	non-decreasing across successive IGs. Moreover, since the 
	objective function takes a finite set of values over the 
	feasible decision space, the sequence of best objective 
	values is upper-bounded and therefore guaranteed to converge. 
	To mitigate premature convergence, the cataclysmic mutation 
	mechanism restores population diversity by generating new 
	candidates while preserving the incumbent best solution. 
	Together, elitist survival selection and cataclysmic mutation 
	enable CHC to balance solution refinement with continued 
	exploration of the feasible decision space, while ensuring 
	the convergence of the best objective value.

	\textit{(ii) NN-Based Guidance:}
	While CHC progressively refines lower-level solutions, 
	its repeated execution for different upper-level candidates 
	provides an opportunity to exploit previously acquired 
	optimization experience. To this end, we leverage the 
	universal approximation property of multilayer feedforward 
	networks~\cite{HORNIK1989359} to learn the underlying 
	relationship between upper-level decisions and their 
	corresponding lower-level responses. Specifically, as 
	additional solution pairs are incorporated into $\mathcal{D}$, 
	the NN is trained on an increasingly rich set of previously 
	explored decisions (see Appx.~\ref{sec:appendix_convergence} 
	for details). By exploiting previously obtained solutions, the trained NN 
	guides the initialization of CHC toward promising regions of 
	the binary decision space for newly sampled upper-level 
	candidates, thereby facilitating faster convergence while 
	preserving the monotonic best-fitness property of CHC.

	\textit{(iii) Upper-Level Optimization:}
	Convergence properties of CMA-ES have been established 
	for a broad range of objective 
	functions~\cite{toure2023global}. In our bilevel optimization 
	framework, the effectiveness of the CMA-ES search further 
	depends on the quality of the lower-level solutions used 
	to evaluate the fitness of upper-level candidates. By 
	combining NN-based initialization with CHC refinement, 
	Neogen aims to obtain high-quality lower-level solutions 
	within a limited computational budget, thereby facilitating 
	reliable fitness ranking at the upper level. Based on 
	these rankings, CMA-ES progressively adapts the mean vector, 
	global step size, and covariance matrix of its Gaussian 
	sampling distribution, concentrating the search around 
	regions associated with higher upper-level utility. 
	This iterative interplay between lower-level solution 
	refinement and upper-level distribution adaptation promotes 
	the stabilization of the bilevel optimization process 
	toward high-quality solutions.

	\section{Evaluation}
	\label{sec:evaluation}
	
	\subsection{Experimental Setup}
	We conduct our experiments in Python 3.12 and PyTorch 2.8, on an Intel Xeon Platinum 8358P CPU with 120 GB RAM and an NVIDIA GeForce RTX 4090 (24 GB) GPU under CUDA 12.8. We conduct evaluations on AGNews\footnote{\url{https://huggingface.co/datasets/fancyzhx/ag_news}} (4-class news-topic classification) and Banking77\footnote{\url{https://huggingface.co/datasets/PolyAI/banking77}} (banking customer-service intent classification, with the 77 fine-grained intents grouped into 4 service categories), using the GPT-2 (124M) model with LoRA-based PEFT ($r=8$, $\alpha=16$, dropout $0.1$, target modules $[c_{\text{attn}}, c_{\text{proj}}]$)\footnote{The target modules $c_{\text{attn}}$ (the combined query-key-value (QKV) projection at the attention input) and $c_{\text{proj}}$ (the output projection after multi-head attention) are selected following standard practice for GPT-2 LoRA fine-tuning.}. AGNews and Banking77 represent \textit{Scenario 1 (regional news analysis)} and \textit{Scenario 2 (banking customer service)}, respectively.  We consider two sets of system configurations: \textit{Set\#1} with 10 workers and 3 managers, $\beta^{\mathsf{sum},(g)}$ = 30; \textit{Set\#2} with 20 workers and 4 managers, $\beta^{\mathsf{sum},(g)}$ = 60. 
	For both datasets, we construct heterogeneous local datasets across clients by partitioning the training data according to a \textit{Dirichlet distribution}~\cite{11413826} with a concentration parameter of $0.3$. In the simulator, the realized communication cost $f_i^{\mathsf{comm},(g)}$ ranges over $[0,1)$.
	% \footnote{For instance, on Banking77, one 1,000-sample worker in \textit{Set\#1} holds 7, 970, 1, and 22 samples of the four classes (severe skew), whereas one 400-sample worker in \textit{Set\#2} holds 109, 189, 1, and 101 samples (moderate skew).} 
	We set the number of global rounds to $G=6$, with $E=5$ edge-level aggregation rounds per global round. For CMA-ES, we set the population size to $pop^u=20$ and the initial step size to $\sigma_u=0.3$, and we run at most 30 CMA-ES generations per global round, terminating earlier when the best fitness fails to improve by more than $1\%$ over five consecutive generations. For computational efficiency, we restrict the covariance matrix $\mathbf{C}_u^{(a)}$ to be diagonal. For CHC, we set the population size to $pop^l=20$ and the maximum number of generations to $T=20$, employing HUX crossover, incest prevention, and cataclysmic mutation with mutation probability $p_{\mathsf{mut}}=0.35$.
	The NN surrogate is a 3-layer multilayer perceptron (MLP) with 128 hidden units, trained with mean squared error (MSE) loss and a learning rate of $10^{-3}$. 
	
	To capture performance, we adopt the following evaluation metrics: \textit{(i)} the test accuracy of the globally aggregated LoRA model ({Acc}), reflecting the performance of the trained model on unseen test data; \textit{(ii)} the manager utility $U^{\mathsf{Mana},(g)}$ ({MgU}) and worker utility $U^{\mathsf{Work},(g)}$ ({WkU}); and \textit{(iii)} the average number of fitness evaluations per round ({Feval/r}) and total wall-clock time (s), demonstrating the time efficiency. In the comparison tables, {Acc}, {MgU}, and {WkU} are averaged over the last three global rounds, i.e., the steady state after the unlearning round, whereas {Time(s)} is the total runtime over all six global rounds. The key hyperparameters are summarized in Table \ref{tab:config} in Appx.~\ref{sec:appendix_algorithm}, where the tuning ranges of the tunable hyperparameters are given in brackets. The hyperparameters were selected via grid search over the specified ranges, with the best-performing configuration adopted for all experiments. For example, the learning rate for normal training is set to $\eta = 5\times10^{-4}$, while that for GA-based unlearning is $\widetilde{\eta} = 1\times10^{-3}$. We also adopted a KLD threshold $\delta = 0.05$. 
	
	%.

	\vspace{-3mm}
	
	\subsection{Evaluations of Unlearning Strategy}
	\label{subsec:unlearning_strategy}
	To evaluate the impact of LUR lifecyle, we compare HermesHFL with four strategies: \textit{(i)} \textit{NoRejoin}, where the unlearned worker is permanently removed from the system; \textit{(ii)} \textit{Retrain}, where the global model is fully reset and retrained from scratch without the data contribution of the unlearned worker; \textit{(iii)} \textit{Rejoin (GA)}, an internal variant of HermesHFL that replaces the OSD unlearning projection with plain GA-based unlearning; and \textit{(iv)} \textit{FedOSD} \cite{pan2025federated}, which extends \textit{NoRejoin} by projecting subsequent model updates to remove components that oppose the previously applied unlearning direction, thereby mitigating the reversal of prior unlearning operations.
	All strategies employ Neogen, with unlearning trigger round $g = 2$.

\begin{figure}[htbp]
		\centering
		% 一行三张图并排
		\includegraphics[width=0.32\linewidth]{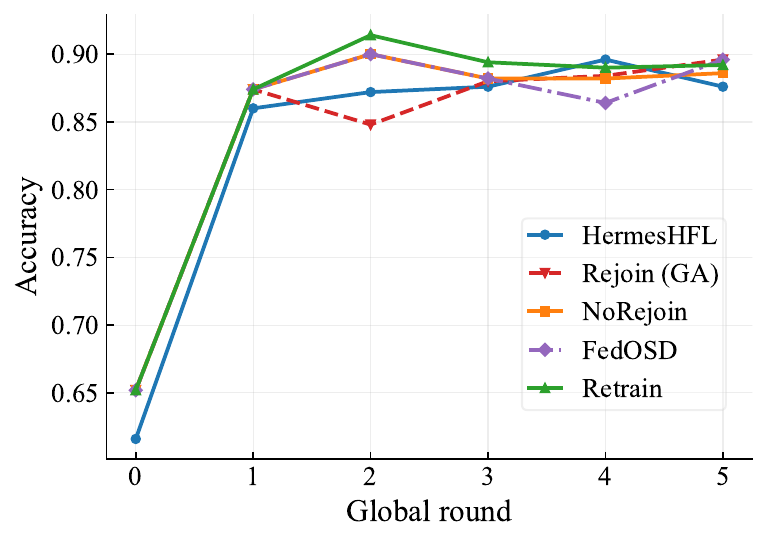}\hfill
		\includegraphics[width=0.32\linewidth]{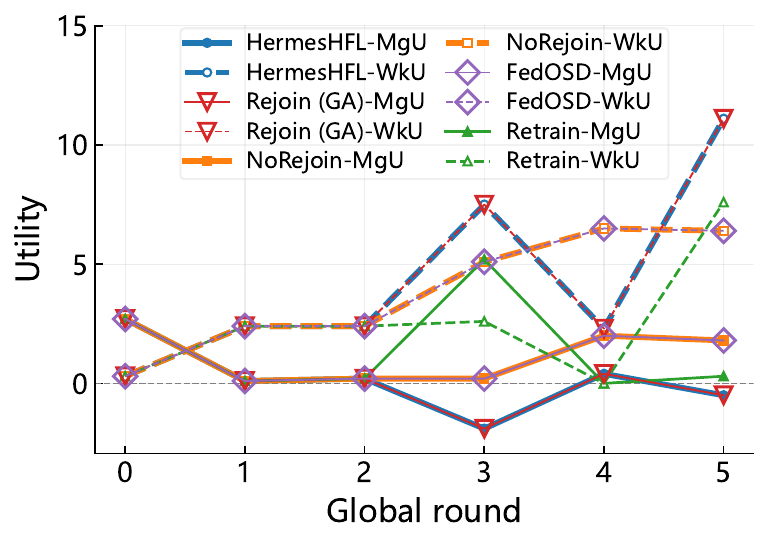}\hfill
		\includegraphics[width=0.32\linewidth]{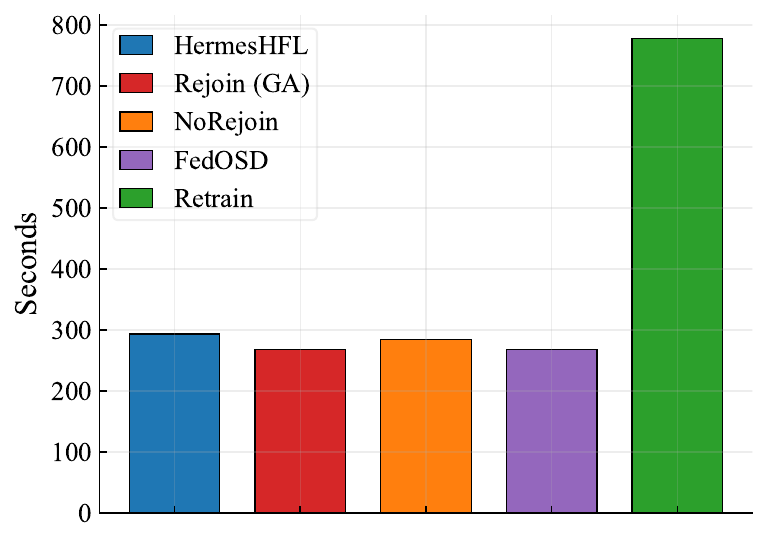}
		% 下方子图标注
		\par\footnotesize
		% 三段分别对齐三张图的下方，仅给(c)加负空格左拉
		\makebox[0.32\linewidth]{(a) Accuracy}
		\hfill
		\makebox[0.32\linewidth]{(b) Utility}
		\hfill
		\makebox[0.32\linewidth]{\hspace{-1em}(c) Time cost}
		\vspace{-2mm}
		\caption{Performance on AGNews regarding unlearning (\textit{Set\#1}) (See detailed values in Table \ref{tab:ablation_unlearn_10w}, Appx. \ref{sec:table}.)}
		\label{fig:eval4}
		\vspace{-3mm}
	\end{figure}

\begin{figure}[htbp]
		\centering
		% 一行三张图并排
		\includegraphics[width=0.32\linewidth]{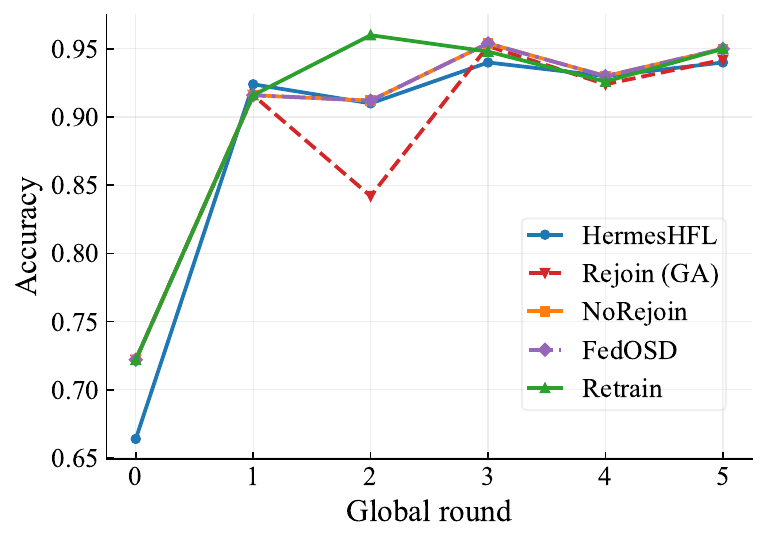}\hfill
		\includegraphics[width=0.32\linewidth]{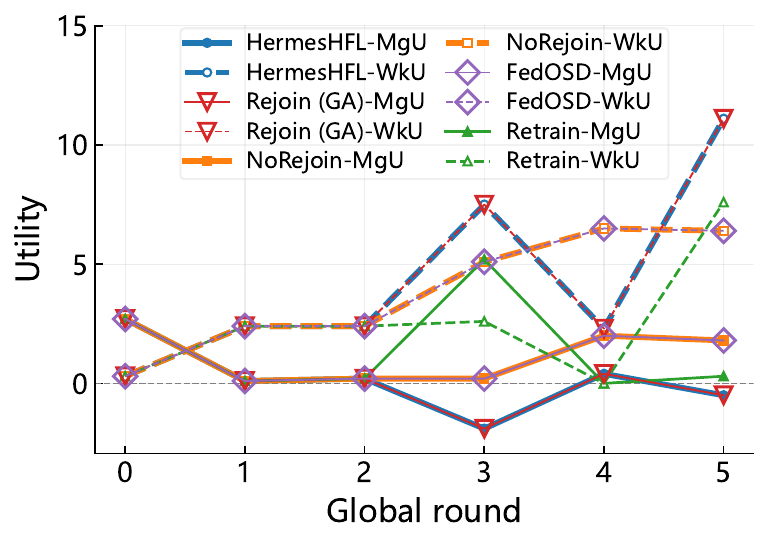}\hfill
		\includegraphics[width=0.32\linewidth]{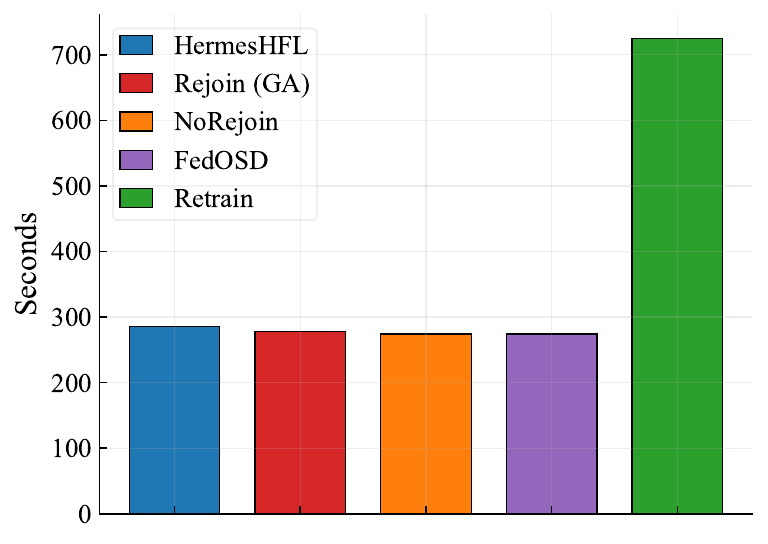}
		% 下方子图标注
		\par\footnotesize
		% 三段分别对齐三张图的下方，仅给(c)加负空格左拉
		\makebox[0.32\linewidth]{(a) Accuracy}
		\hfill
		\makebox[0.32\linewidth]{(b) Utility}
		\hfill
		\makebox[0.32\linewidth]{\hspace{-1em}(c) Time cost}
		\vspace{-2mm}
		\caption{Performance on Banking77 regarding unlearning (\textit{Set\#1}) (See detailed values in Table \ref{tab:ablation_unlearn_10w}, Appx. \ref{sec:table}.)}
		\label{fig:eval3}
		\vspace{-3mm}
	\end{figure}

\begin{figure}[htbp]
		\vspace{-2mm}
		\centering
		% 一行三张图并排
		\includegraphics[width=0.32\linewidth]{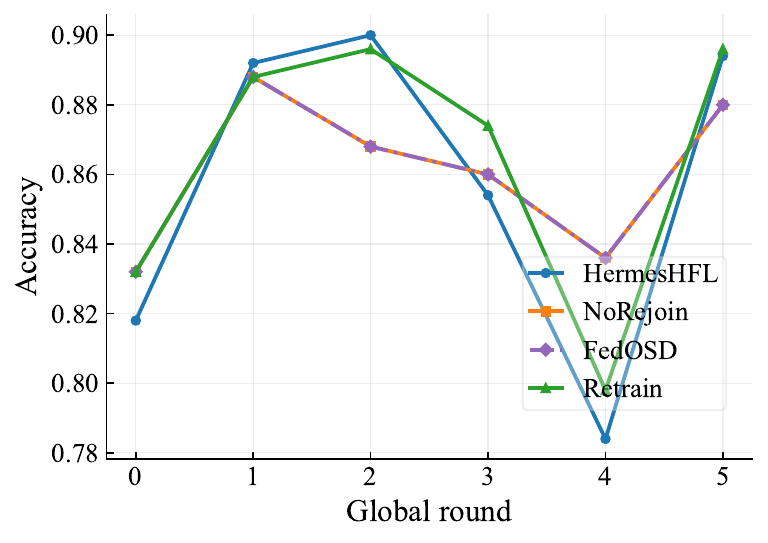}\hfill
		\includegraphics[width=0.32\linewidth]{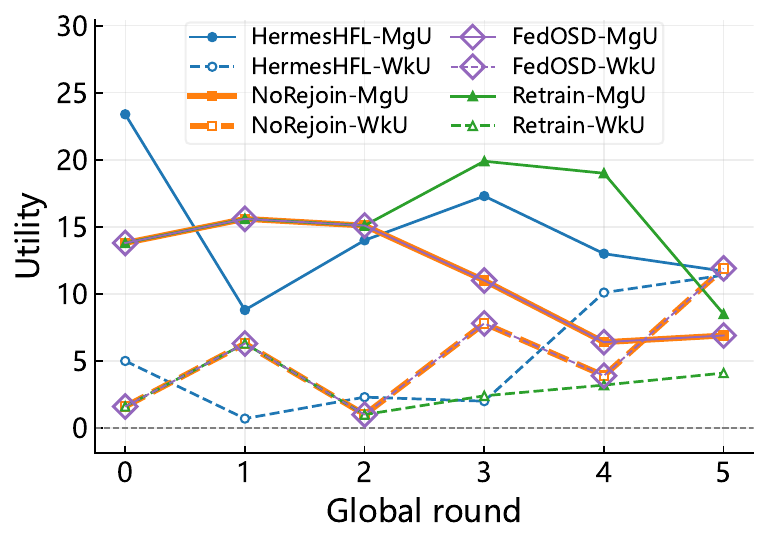}\hfill
		\includegraphics[width=0.32\linewidth]{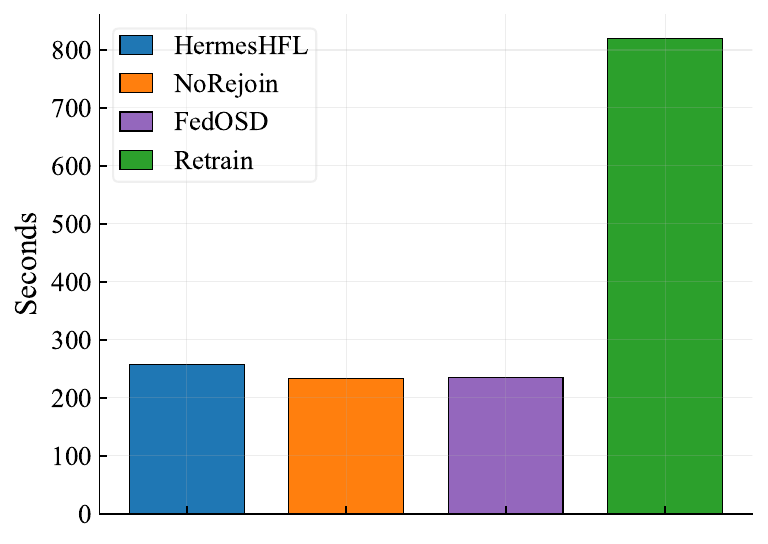}
		% 下方子图标注
		\par\footnotesize
		% 三段分别对齐三张图的下方，仅给(c)加负空格左拉
		\makebox[0.32\linewidth]{(a) Accuracy}
		\hfill
		\makebox[0.32\linewidth]{(b) Utility}
		\hfill
		\makebox[0.32\linewidth]{\hspace{-1em}(c) Time cost}
		\vspace{-2mm}
		\caption{Performance on AGNews regarding unlearning (\textit{Set\#2}) (See detailed values in Table \ref{tab:ablation_unlearn_20w}, Appx. \ref{sec:table}.)}
		\label{fig:eval7}
		\vspace{-3mm}
	\end{figure}

\begin{figure}[htbp]
		\centering
		% 一行三张图并排
		\includegraphics[width=0.32\linewidth]{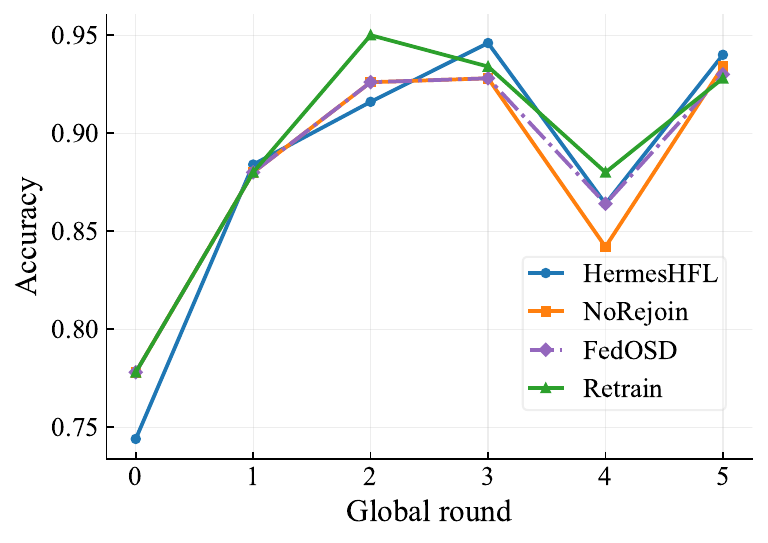}\hfill
		\includegraphics[width=0.32\linewidth]{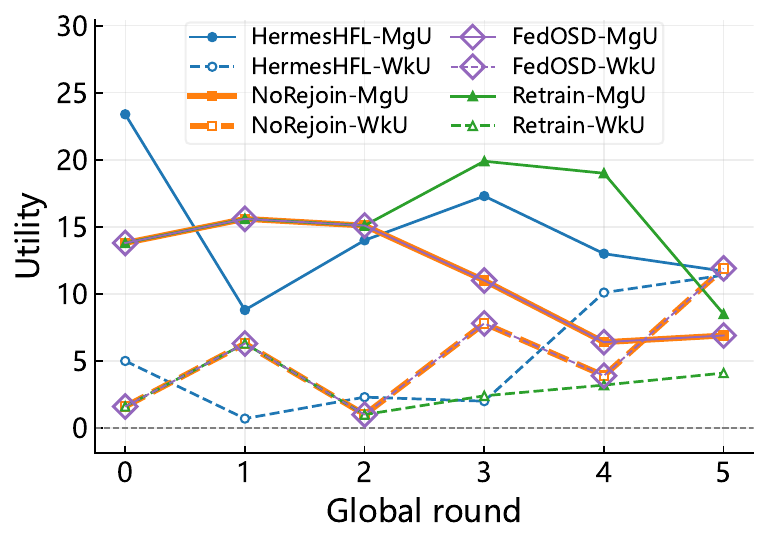}\hfill
		\includegraphics[width=0.32\linewidth]{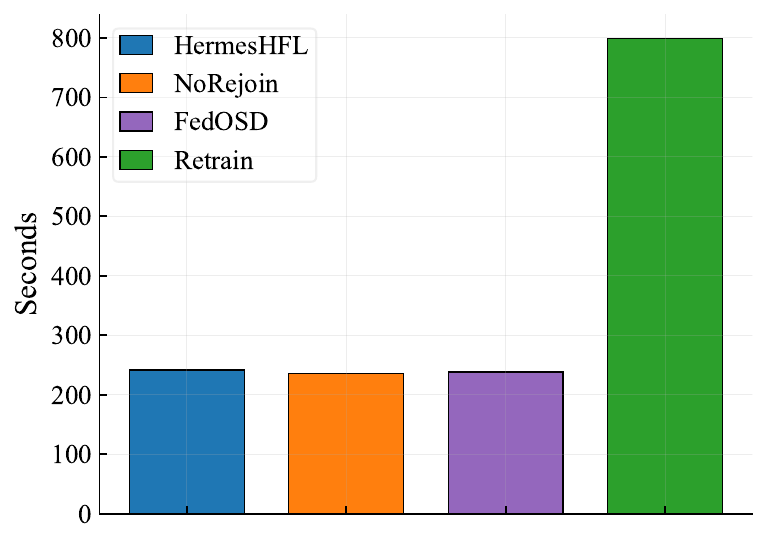}
		% 下方子图标注
		\par\footnotesize
		% 三段分别对齐三张图的下方，仅给(c)加负空格左拉
		\makebox[0.32\linewidth]{(a) Accuracy}
		\hfill
		\makebox[0.32\linewidth]{(b) Utility}
		\hfill
		\makebox[0.32\linewidth]{\hspace{-1em}(c) Time cost}
		\vspace{-2mm}
		\caption{Performance on Banking77 regarding unlearning (\textit{Set\#2}) (See detailed values in Table \ref{tab:ablation_unlearn_20w}, Appx. \ref{sec:table}.)}
		\label{fig:eval5}
		\vspace{-2mm}
	\end{figure}

	\textit{Accuracy Preservation:}
	Inspecting subplot (a) of Figs. \ref{fig:eval4}-\ref{fig:eval5}, the compared strategies achieve comparable average accuracy under all configurations: the accuracy gap between HermesHFL and Retrain is marginal at \textit{Set\#1} (0.883 vs.\ 0.892 on AGNews, 0.937 vs.\ 0.941 on Banking77); at \textit{Set\#2} it remains marginal on AGNews (0.844 vs.\ 0.856) and reverses in favor of HermesHFL on Banking77 (0.917 vs.\ 0.914). This validates that HermesHFL's OSD-based unlearning can achieve retrain-level performance fidelity at a drastically lower computational cost compared to retraining the model. Moreover, post-unlearning recovery is fast: on Banking77 \textit{Set\#1}, HermesHFL rebounds from 0.910 ($g = 2$) to 0.940 ($g = 3$) within a single round. 
	A transient accuracy dip is observed at $g=4$ in 
	Figs.~\ref{fig:eval7}-\ref{fig:eval5} across all  
	strategies, despite their different approaches to unlearning 
	and worker rejoining. This is caused by the variations in worker selection across 
	global rounds that alters the collective data distribution. Specifically, under \textit{Set\#2}, 
	all strategies select the same four workers at $g=4$, 
	whose collective training data contain only 2.5\% and 2.4\% 
	of samples belonging to the Sports class (AGNews) and the 
	card-payment class (Banking77), respectively, compared to 
	25\% in the test sets. Consequently, the 
	aggregated model update is dominated by contributions from 
	other classes, degrading the  
	performance on the underrepresented classes. In the 
	subsequent round ($g=5$), changes in worker selection 
	increase the proportion of samples belonging to the 
	previously underrepresented classes to 26.7--28.5\%, 
	accompanied by accuracy recovery. 
	% Taken together, these observations suggest 
	% that the transient accuracy degradation and subsequent 
	% recovery are primarily attributable to variations in the 
	% collective data distribution of the selected workers rather 
	% than to the proposed rejoining mechanism.
	
	\textit{Incentive Efficiency:}
	Inspecting subplot (b) of Figs. \ref{fig:eval4}-\ref{fig:eval5}, where per-round trajectories are shown, we focus on the utility values averaged over the last three global rounds ($g=3$--$5$). In terms of total system utility (MgU $+$ WkU), HermesHFL remains above Retrain at both scales (6.3 vs.\ 5.2 at \textit{Set\#1}; 21.8 vs.\ 19.0 at \textit{Set\#2} on both datasets) and achieves the highest total utility at the 20-worker scale, exceeding Retrain by 14.7\% and NoRejoin/FedOSD (15.9) by 37.1\%. This benefit comes from the economic mechanism in HermesHFL: unlearning penalties $q_{i,j}$ from departing workers are recycled into the residual budget $\hat{\beta}_{j}^{\mathsf{m},(g-1)}$, turning exit costs into future recruitment capacity, and rejoined workers further contribute refreshed data after compensating the system for the earlier disruption. In terms of worker utility, HermesHFL attains the highest or tied-highest WkU at both scales (7.0 vs.\ 6.0 for NoRejoin at \textit{Set\#1}; 7.8, tied with NoRejoin and FedOSD, at \textit{Set\#2} on both datasets), as it relies on Neogen-optimized contracts that satisfy the individual rationality constraint (C7) for the rejoined workers. In contrast, Retrain results in much lower worker utility (3.4 at \textit{Set\#1}, 3.2 at \textit{Set\#2}), since it reinitializes the model and restarts the contract optimization from scratch.
	
	\textit{Computational Cost:}
	Subplot (c) of Figs. \ref{fig:eval4}-\ref{fig:eval5} presents the average runtime per global round. At \textit{Set\#1}, Retrain averages 777.6s (AGNews) and 724.5s (Banking77) per round, versus 293.3s and 285.7s for HermesHFL, about 2.7$\times$ and 2.5$\times$ higher; at \textit{Set\#2}, it averages 819.2s and 798.5s per round, versus 257.2s and 241.7s, about 3.2$\times$ and 3.3$\times$ higher. In total (i.e., over the six global rounds), Retrain consumes 4,665s (AGNews) and 4,347s (Banking77) at \textit{Set\#1}, and 4,915s and 4,791s at \textit{Set\#2}, compared with 1,760s and 1,714s at \textit{Set\#1} and 1,543s and 1,451s at \textit{Set\#2} for HermesHFL. Such overhead would amplify with frequent unlearning requests over long training, making HermesHFL highly favorable for large-scale, long-term HFUL.
	
	\textit{KLD-Based Unlearning Verification: }
	% At the unlearning round ($g = 2$), the KLD exceeds the threshold $\delta=0.05$ in all scenarios: 3.147 and 0.144 on AGNews \textit{Set\#1} and \textit{Set\#2}, and 0.875 and 1.276 on Banking77 \textit{Set\#1} and \textit{Set\#2}, validating effective knowledge removal. Compared with the plain-GA variant, whose KLD reaches 4.466 on Banking77 \textit{Set\#1}, the OSD projection yields a smaller but still clearly detectable output-distribution shift on the target worker's data. The same-target directional analysis provides direct evidence about interference with the retained knowledge: the OSD unlearning direction is exactly orthogonal to the retained workers' update directions ($|\cos|=0.000$), whereas the GA direction exhibits a mean alignment of 0.28--0.36 with them, indicating substantially less interference with the retained updates. In all cases, accuracy recovers within 1--3 rounds, confirming that OSD-based unlearning effectively eliminates target worker knowledge without permanent damage to model capability.
	We examine the effectiveness of unlearning by measuring the change in the model's predictive distribution on the data of the worker requesting unlearning. At the unlearning round ($g=2$), the KLD exceeds the prescribed threshold of $\delta=0.05$ in all considered settings for HermesHFL: the KLD reaches 3.147 and 0.144 for AGNews and 0.875 and 1.276 for Banking77 under \textit{Set\#1} and \textit{Set\#2}, respectively. Thus, HermesHFL consistently satisfies the prescribed KLD-based unlearning criterion.
	Interestingly, under Banking77 \textit{Set\#1}, plain GA-based unlearning induces a larger change in the model's predictive distribution than HermesHFL's OSD-based unlearning (KLD of 4.466 versus 0.875). However, our objective is not to maximize the KLD but to induce sufficient changes in the model's predictions on the targeted data while minimizing interference with the training of retained workers. To assess this trade-off, we measure the alignment between the unlearning direction and the retained workers' update directions for the same target worker. The OSD-based direction exhibits zero absolute cosine similarity with the retained workers' updates, whereas the GA-based direction exhibits a mean absolute cosine similarity of 0.28--0.36. These results highlight the advantage of OSD: it satisfies the unlearning criterion while eliminating directional interference with the retained workers' updates. Further, model accuracy recovers within one to three global rounds across all considered settings, indicating that the unlearning operation does not cause lasting degradation in classification performance.
	
\begin{figure}[htbp]
		\centering
		\includegraphics[width=0.62\linewidth]{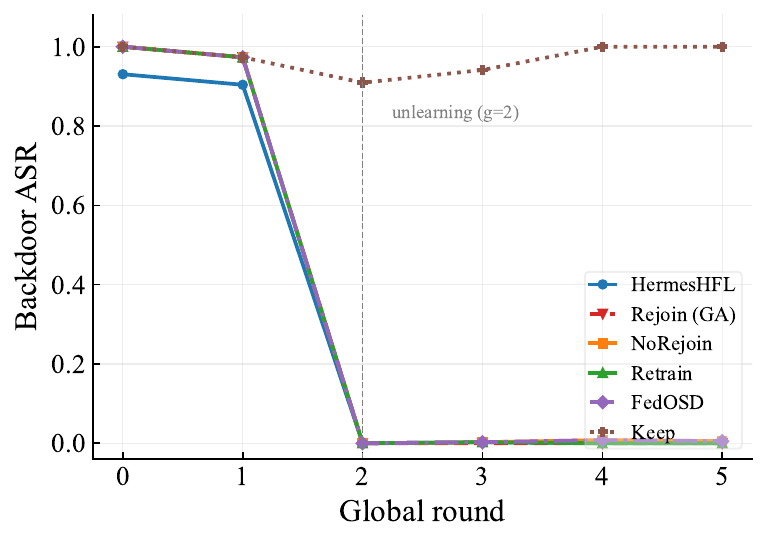}
		\caption{Backdoor ASR per global round on Banking77 (\textit{Set\#1}). Unlearning at $g = 2$ reduces the ASR to $\approx$0 for all strategies, while the \textit{Keep} control (no unlearning) rises to 1.00.}
		\label{fig:backdoor}
	\end{figure}

	%  \textit{Deletion Verification via Backdoor Canary: }
	% As a complementary, directly observable verification of deletion, we run a backdoor canary on Banking77 \textit{Set\#1} with the setup detailed in Appx. \ref{sec:table}: the unlearning request targets a poisoned worker at $g=2$. The poisoned model reaches an attack success rate (ASR) of 0.904--0.973 at $g=1$; from $g=2$ onward the ASR drops to $\approx$0 for all unlearning strategies, while the \textit{Keep} control, which ignores the request, rises from 0.909 to 1.00. The per-round ASR trajectories are reported in Table \ref{tab:backdoor_asr} and Fig. \ref{fig:backdoor}.
	\textit{Deletion Verification via Backdoor Canary: }
	To complement the KLD-based evaluation, we investigate whether HermesHFL can remove a specific behavior intentionally introduced into the global model by a worker. To this end, under Banking77 \textit{Set\#1}, we poison the local dataset of one worker by inserting a predefined trigger into its training samples and assigning them a common target label (see Appx.~\ref{sec:table} for details). The effectiveness of unlearning is then assessed by the attack success rate (ASR), defined as the fraction of triggered test samples classified into the target class. Before initiating unlearning, the ASR reaches 0.904--0.973 at $g=1$, indicating that the global model has learned the backdoor behavior. Upon initiating unlearning for the poisoned worker at $g=2$, the ASR drops to nearly zero across all unlearning strategies and remains negligible in subsequent rounds. In contrast, under the \textit{Keep} baseline, where the unlearning request is ignored, the ASR increases from 0.909 to 1.00. As illustrated in Fig.~\ref{fig:backdoor} and quantified in Table~\ref{tab:backdoor_asr} in Appx.~\ref{sec:table}, these results demonstrate that HermesHFL successfully removes the targeted backdoor behavior without requiring full model retraining.

	\subsection{Evaluations on Optimization}
	To assess the effectiveness of Neogen, we benchmark it against four representative alternatives, enabling evaluation against different solution strategies: 
	\textit{(i) EAOnly}, a surrogate-free variant of Neogen that combines CMA-ES and CHC without neural network guidance to isolate the benefit of the proposed surrogate;
	\textit{(ii) GenSA}, a general-purpose simulated annealing (SA) algorithm with 200 iterations and a geometric cooling schedule;
	\textit{(iii) GdyRatio}, a greedy heuristic that sequentially associates workers with managers in descending order of their worker-to-cost ratios; and
	\textit{(iv) RandAssign}, a non-optimized baseline that processes workers in a randomly shuffled order and assigns them to randomly selected eligible managers under a fixed pricing policy.
	The rationale behind the selection of these baselines is provided in Appx.~\ref{sec:appendix_optimization}.

\begin{figure}[htbp]
		\centering
		% 第一行 3张子图 a b c
		\includegraphics[width=0.32\linewidth]{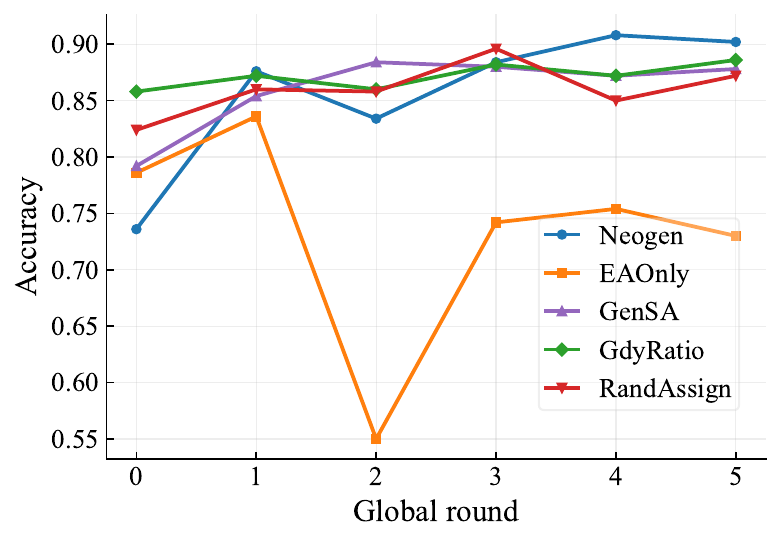}\hfill
		\includegraphics[width=0.32\linewidth]{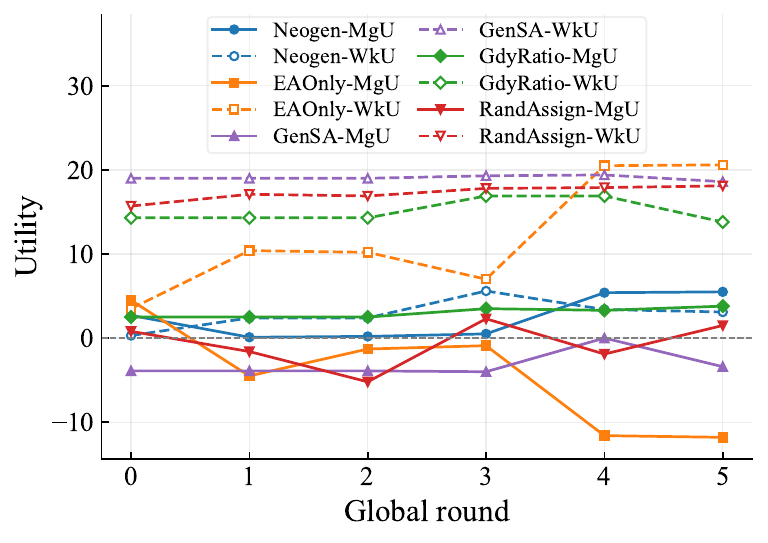}\hfill
		\includegraphics[width=0.32\linewidth]{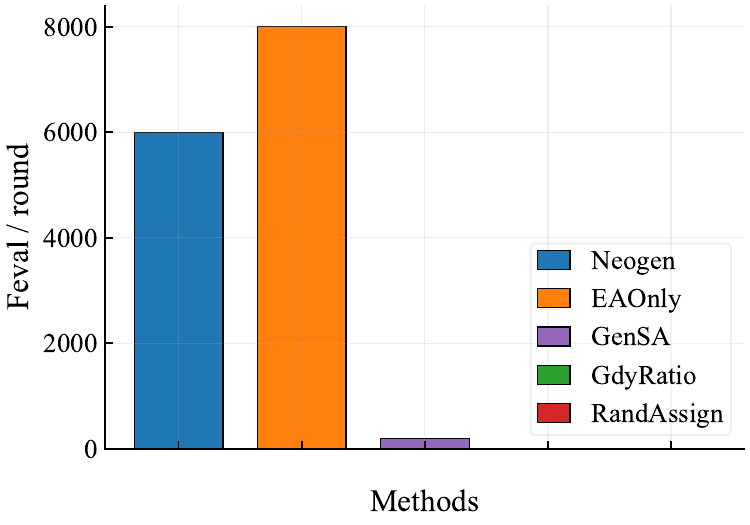}
		\par\footnotesize
		% 三段分别对齐三张图的下方，仅给(c)加负空格左拉
		\makebox[0.32\linewidth]{(a) Accuracy}
		\hfill
		\makebox[0.32\linewidth]{(b) Utility}
		\hfill
		\makebox[0.32\linewidth]{\hspace{-1em}(c) Feval}
		
		\vspace{1em}
		% 第二行 3张子图 d e f
		\includegraphics[width=0.32\linewidth]{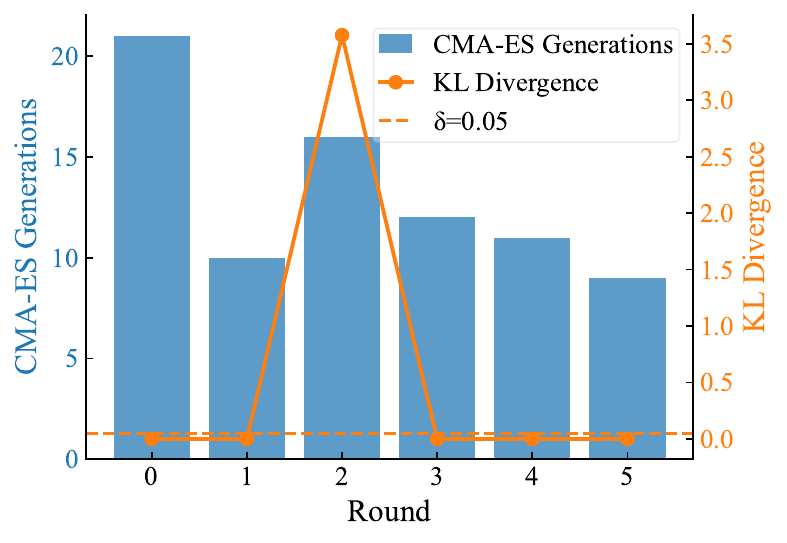}\hfill
		\includegraphics[width=0.32\linewidth]{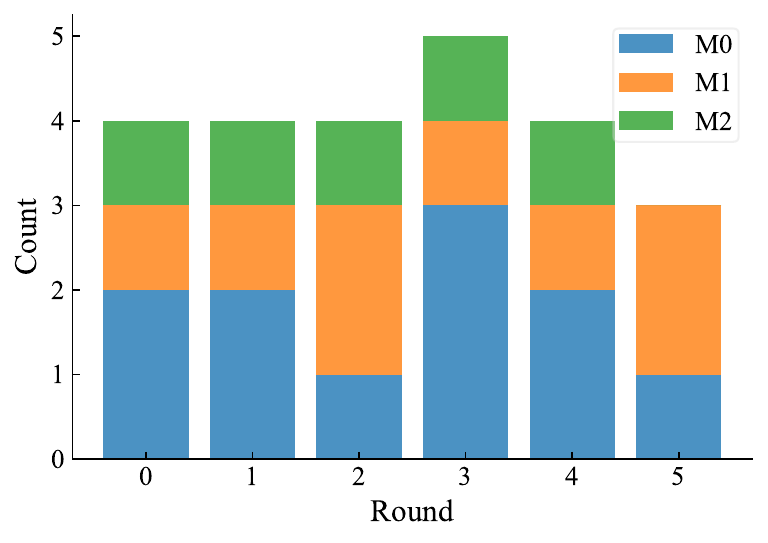}\hfill
		\includegraphics[width=0.32\linewidth]{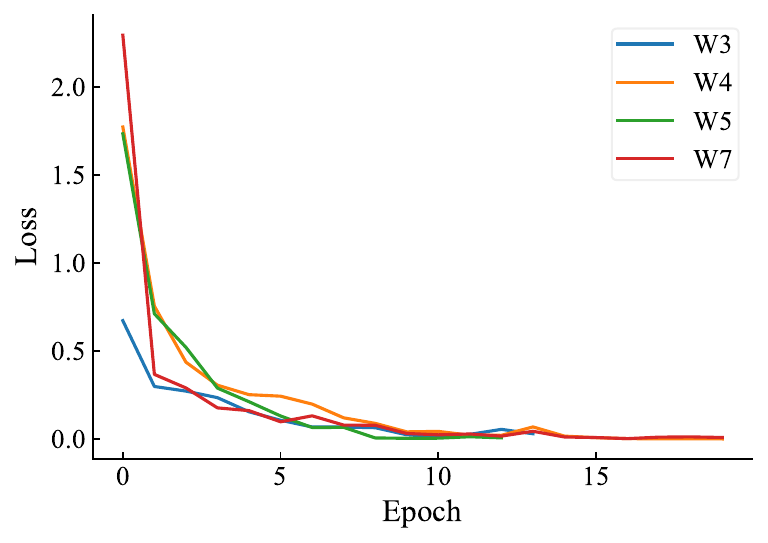}
		\par\footnotesize(d) Neogen \& KLD \hfill (e) Selected workers \hfill (f) Training loss
		\vspace{-2mm}
		\caption{Performance on AGNews regarding optimization (\textit{Set\#1}) (See detailed values in Table \ref{tab:ablation_mechanism}, Appx. \ref{sec:table}.)}
		\label{fig:eval2}
		\vspace{-5.5mm}
	\end{figure}

\begin{figure}[htbp]
		\centering
		% 第一行 3张子图 a b c
		\includegraphics[width=0.32\linewidth]{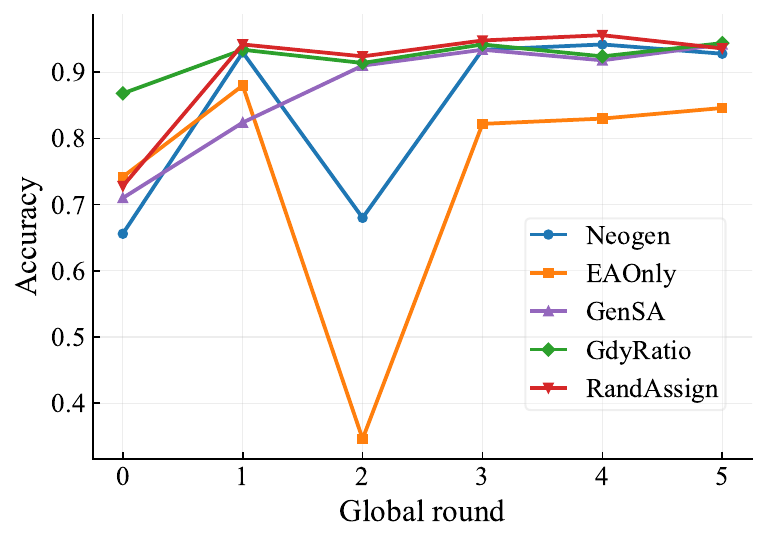}\hfill
		\includegraphics[width=0.32\linewidth]{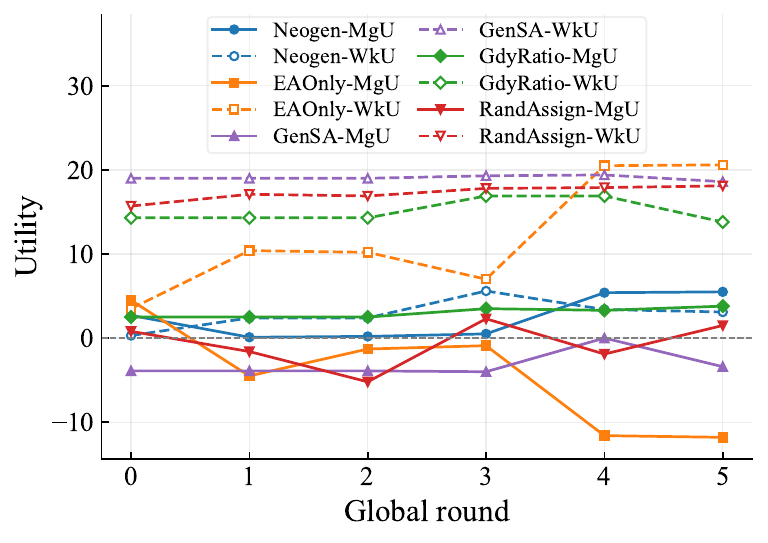}\hfill
		\includegraphics[width=0.32\linewidth]{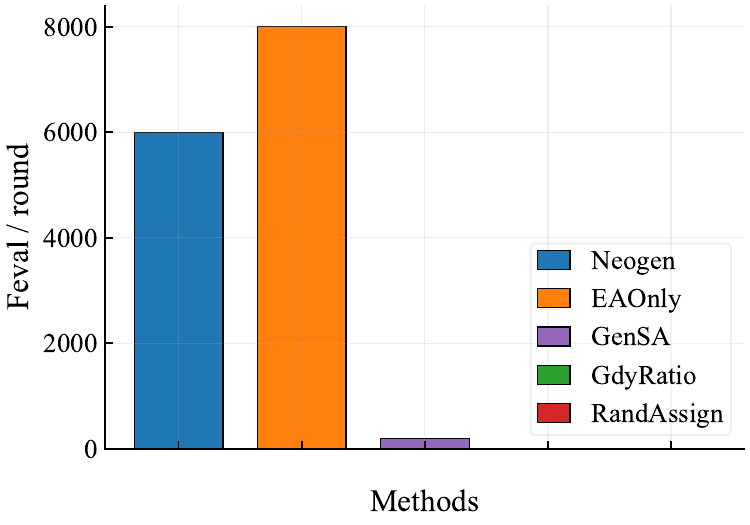}
		\par\footnotesize
		% 三段分别对齐三张图的下方，仅给(c)加负空格左拉
		\makebox[0.32\linewidth]{(a) Accuracy}
		\hfill
		\makebox[0.32\linewidth]{(b) Utility}
		\hfill
		\makebox[0.32\linewidth]{\hspace{-1em}(c) Feval}
		
		\vspace{1em}
		% 第二行 3张子图 d e f
		\includegraphics[width=0.32\linewidth]{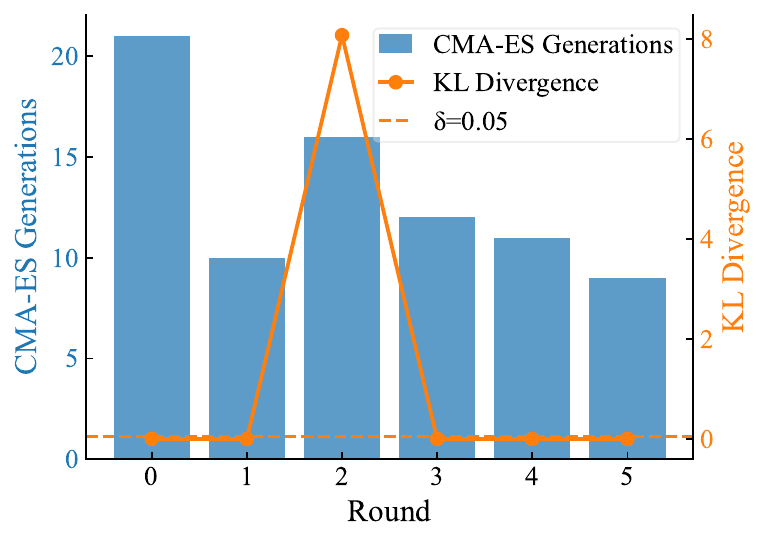}\hfill
		\includegraphics[width=0.32\linewidth]{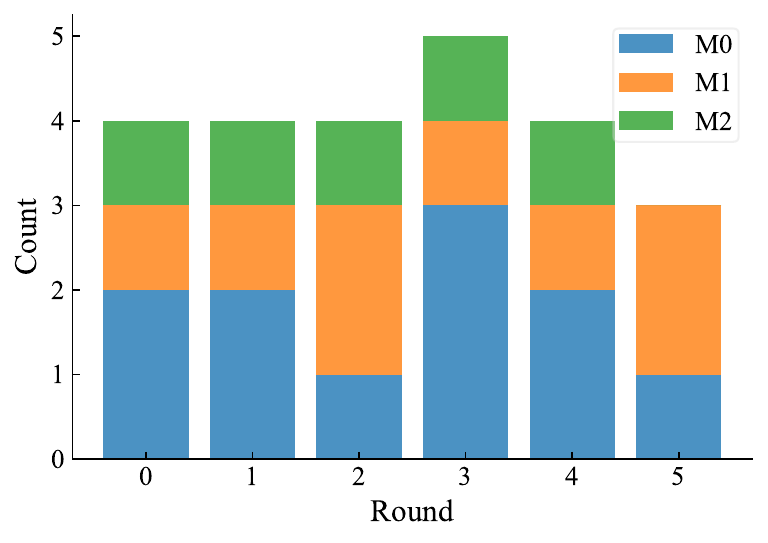}\hfill
		\includegraphics[width=0.32\linewidth]{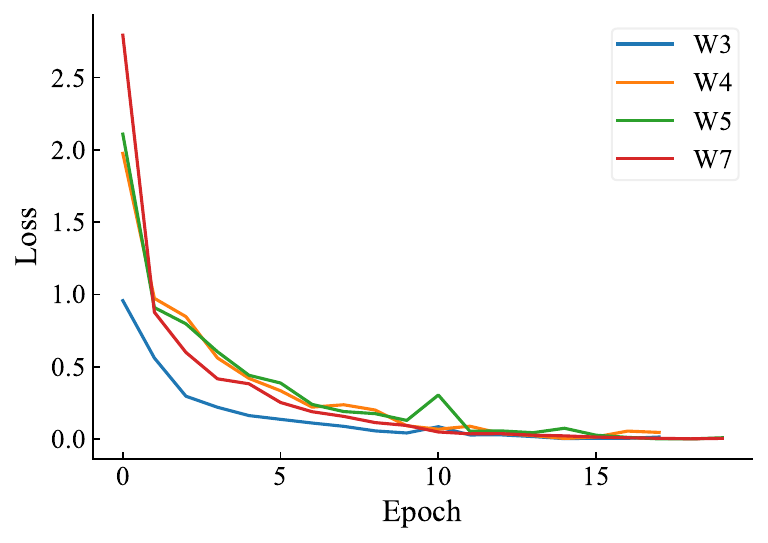}
		\par\footnotesize(d) Neogen \& KLD \hfill (e) Selected workers \hfill (f) Training loss
		\vspace{-2mm}		
		\caption{Performance on Banking77 regarding optimization (\textit{Set\#1}) (See detailed values in Table \ref{tab:ablation_mechanism}, Appx. \ref{sec:table}.)}
		\label{fig:eval1}
		\vspace{-5.5mm}
	\end{figure}

\begin{figure}[htbp]
		\centering
		% 第一行 3张子图 a b c
		\includegraphics[width=0.32\linewidth]{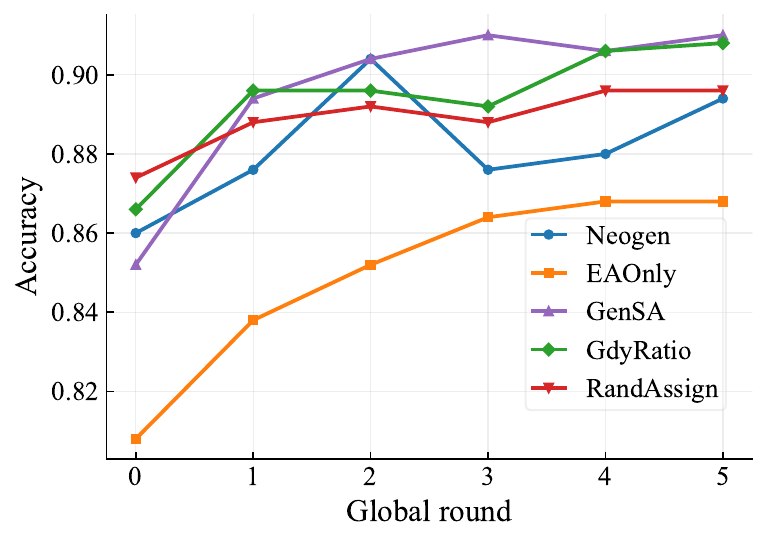}\hfill
		\includegraphics[width=0.32\linewidth]{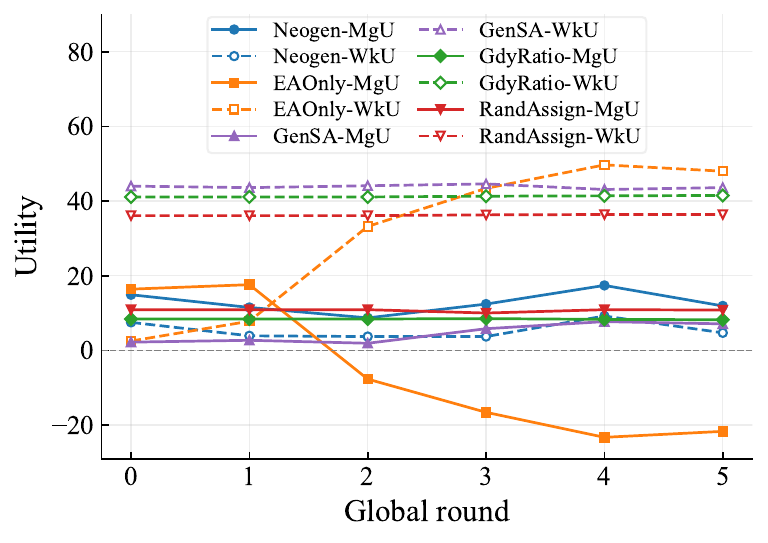}\hfill
		\includegraphics[width=0.32\linewidth]{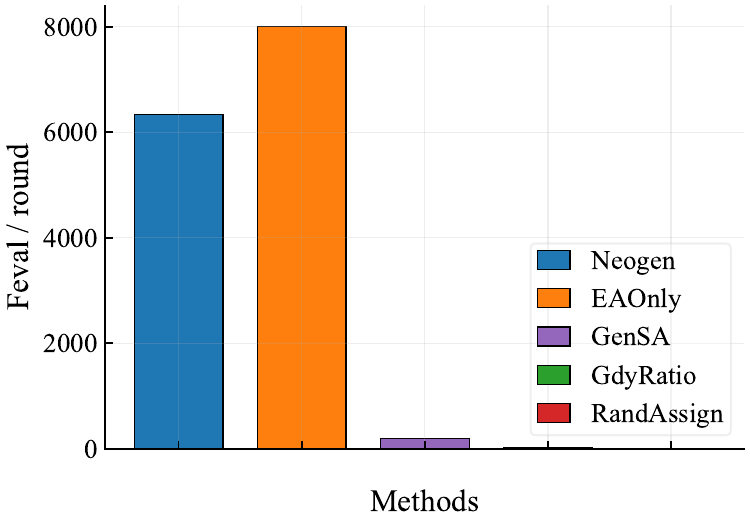}
		\par\footnotesize
		% 三段分别对齐三张图的下方，仅给(c)加负空格左拉
		\makebox[0.32\linewidth]{(a) Accuracy}
		\hfill
		\makebox[0.32\linewidth]{(b) Utility}
		\hfill
		\makebox[0.32\linewidth]{\hspace{-1em}(c) Feval}
		
		\vspace{1em}
		% 第二行 3张子图 d e f
		\includegraphics[width=0.32\linewidth]{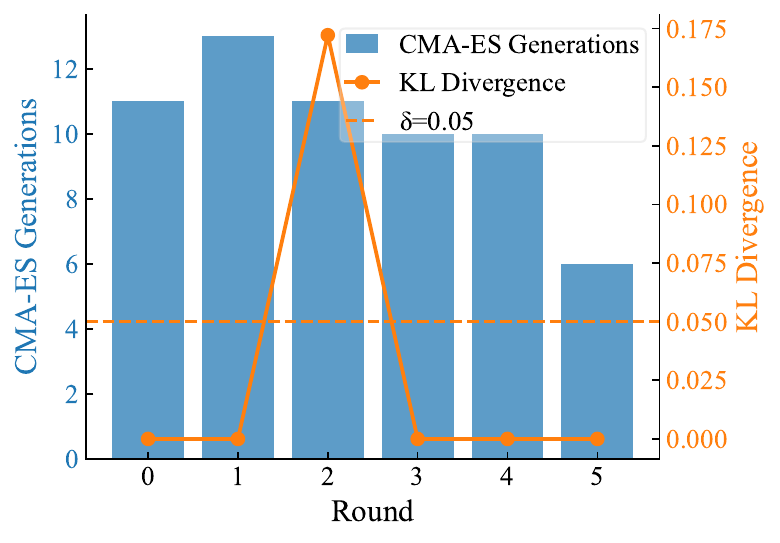}\hfill
		\includegraphics[width=0.32\linewidth]{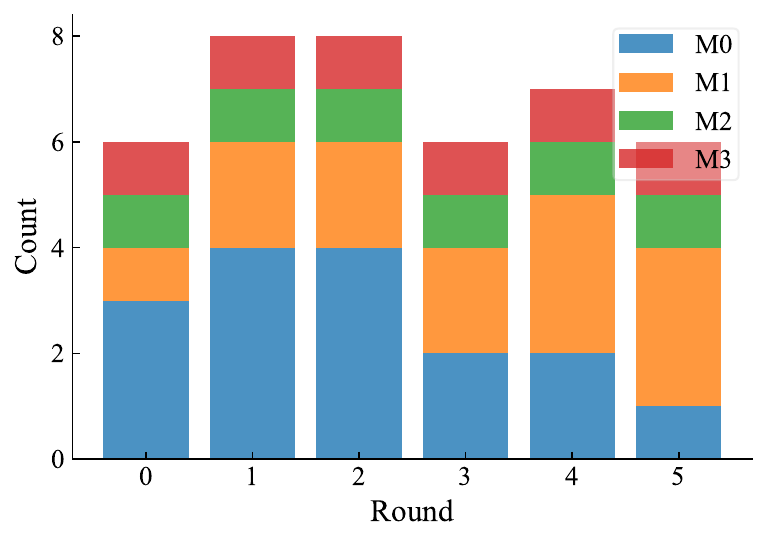}\hfill
		\includegraphics[width=0.32\linewidth]{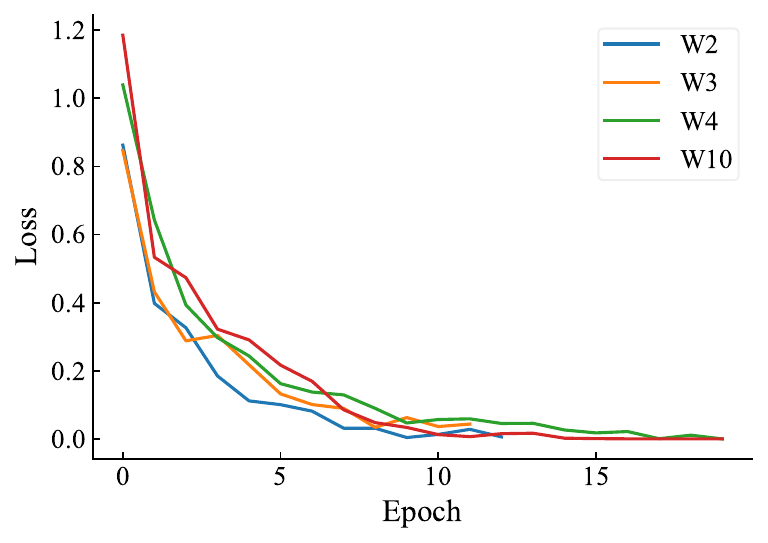}
		\par\footnotesize(d) Neogen \& KLD \hfill (e) Selected workers \hfill (f) Training loss
		\vspace{-2mm}
		\caption{Performance on AGNews regarding optimization (\textit{Set\#2}) (See detailed values in Table \ref{tab:ablation_20w4m}, Appx. \ref{sec:table}.)}
		\label{fig:eval8}
		\vspace{-6.5mm}
	\end{figure}

\begin{figure}[htbp]
		\centering
		% 第一行 3张子图 a b c
		\includegraphics[width=0.32\linewidth]{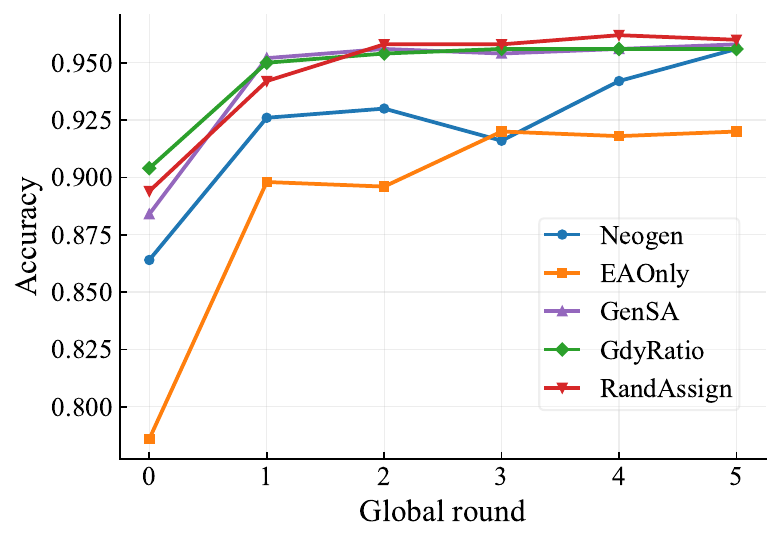}\hfill
		\includegraphics[width=0.32\linewidth]{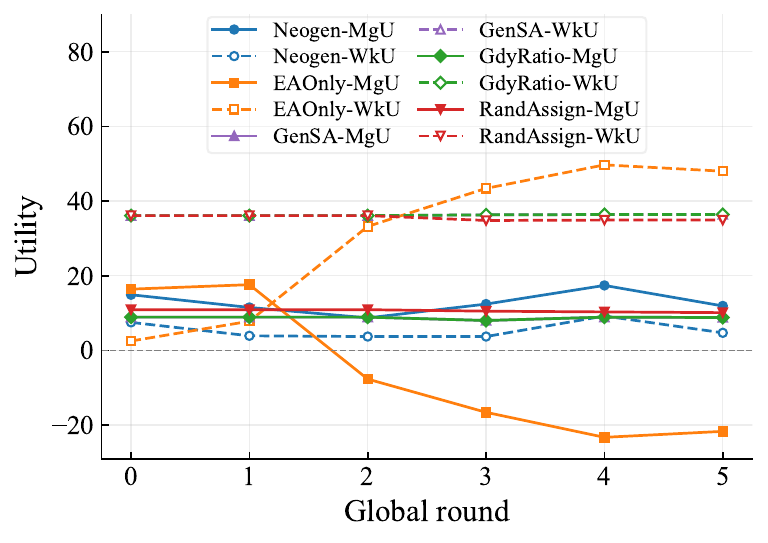}\hfill
		\includegraphics[width=0.32\linewidth]{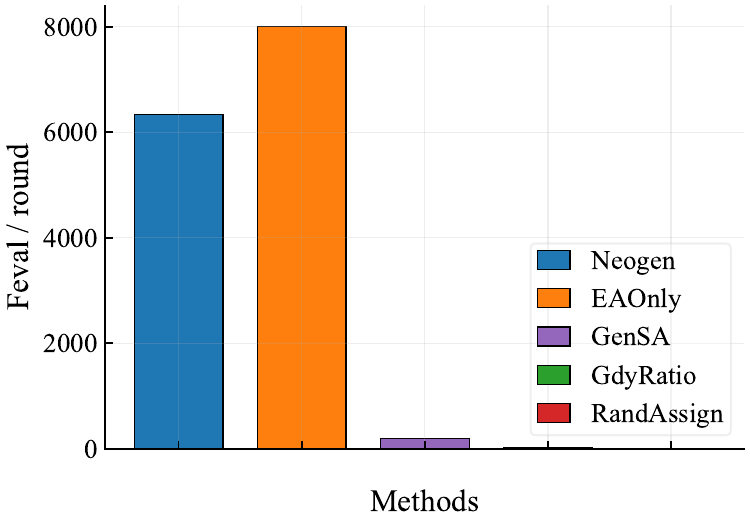}
		\par\footnotesize
		% 三段分别对齐三张图的下方，仅给(c)加负空格左拉
		\makebox[0.32\linewidth]{(a) Accuracy}
		\hfill
		\makebox[0.32\linewidth]{(b) Utility}
		\hfill
		\makebox[0.32\linewidth]{\hspace{-1em}(c) Feval}
		
		\vspace{1em}
		% 第二行 3张子图 d e f
		\includegraphics[width=0.32\linewidth]{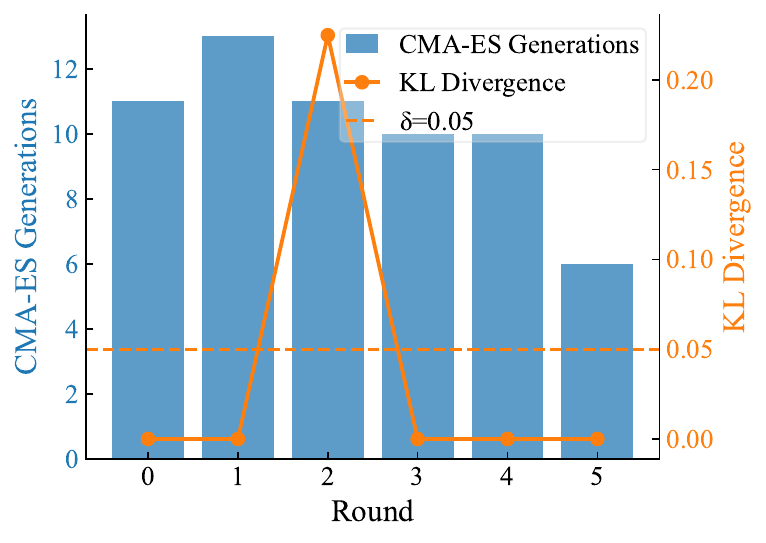}\hfill
		\includegraphics[width=0.32\linewidth]{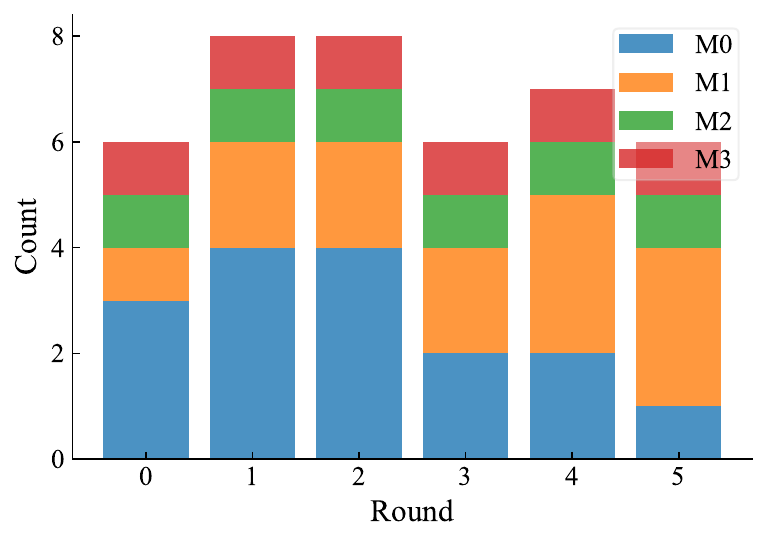}\hfill
		\includegraphics[width=0.32\linewidth]{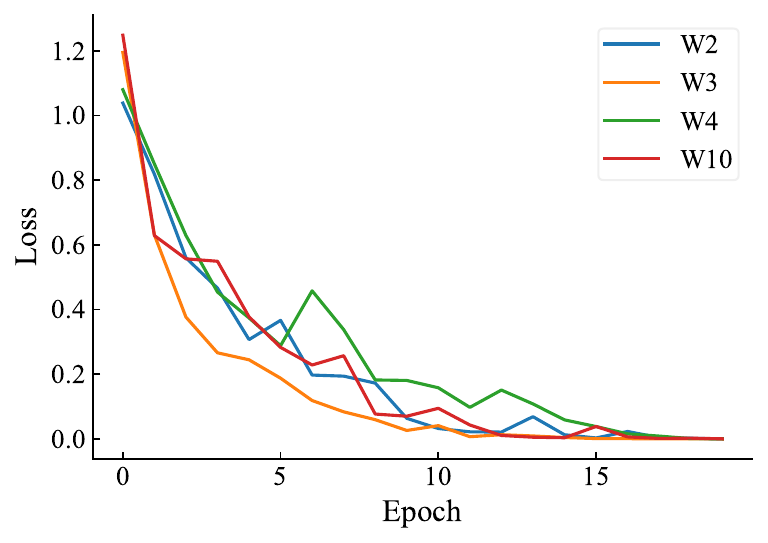}
		\par\footnotesize(d) Neogen \& KLD \hfill (e) Selected workers \hfill (f) Training loss
		\vspace{-2mm}		
		\caption{Performance on Banking77 regarding optimization (\textit{Set\#2}) (See detailed values in Table \ref{tab:ablation_20w4m}, Appx. \ref{sec:table}.)}
		\label{fig:eval6}
		\vspace{-3mm}
	\end{figure}

	To ensure a consistent comparison, we adopt a common cost-based pricing policy for the benchmarks that do not explicitly optimize worker--manager contracts. Specifically, the worker cost is computed by the unified cost model \eqref{eq:worker_cost} with the same attribute values used for HermesHFL,
	and the payment offered to each worker is set to $p_{i,j}^{(g)} = 4\,c_{i,j}^{\mathsf{work},(g)}$. The multiplier of 4 is selected empirically based on the payments obtained by Neogen, which range from approximately one to seven times the corresponding worker costs. By selecting the midpoint of this range, we establish a representative fixed pricing policy that avoids favoring either extreme of the empirically observed payment range. The corresponding results are presented in Figs.~\ref{fig:eval2}--\ref{fig:eval6}, where M0, M1, $\ldots$ in the legend of subfigure (e) denote the individual managers.

	\textit{Accuracy Degradation without NN Guidance: }
	% Inspecting subplot (a) of Figs. \ref{fig:eval2}-\ref{fig:eval6}, our Neogen achieves average accuracies of 0.898 on AGNews and 0.935 on Banking77 under \textit{Set\#1}, outperforming EAOnly (0.742 and 0.833) by 21.0\% and 12.2\%, respectively. Under more challenging \textit{Set\#2} with $\beta^{\mathsf{sum},(g)}=60$, Neogen still maintains stable accuracies of 0.883 on AGNews and 0.938 on Banking77, slightly above EAOnly (0.867 and 0.919). Without NN guidance, EAOnly suffers a severe accuracy drop at the unlearning round: on Banking77 \textit{Set\#1}, its accuracy falls from 0.880 ($g = 1$) to 0.346 ($g = 2$) and only recovers to 0.846 by $g = 5$. The lack of a surrogate model to constrain CHC search leads to noisy binary solutions in the high-dimensional solution space. As a result, CMA-ES accumulates such errors in its fitness ranking and gradually drifts toward low-quality contract regions. In contrast, Neogen stabilizes between 0.88 and 0.91 on AGNews in the subsequent rounds and recovers to 0.934 at $g = 3$ on Banking77 after the unlearning round.
	As shown in subplot (a) of Figs.~\ref{fig:eval2}--\ref{fig:eval6}, incorporating the NN surrogate into the evolutionary optimization framework consistently improves model accuracy compared with its surrogate-free counterpart, EAOnly. Under \textit{Set\#1}, Neogen achieves average accuracies of 0.898 and 0.935 on AGNews and Banking77, respectively, exceeding those of EAOnly by 15.6 and 10.2 percentage points. Under the larger-scale \textit{Set\#2}, the corresponding improvements are 1.6 and 1.9 percentage points. The benefit of NN guidance is particularly pronounced in the model's ability to recover from unlearning-induced perturbations. For instance, on Banking77 under \textit{Set\#1}, the accuracy of EAOnly drops from 0.880 to 0.346 at the unlearning round ($g=2$) and reaches only 0.846 by $g=5$. In contrast, Neogen recovers to an accuracy of 0.934 immediately after the unlearning round ($g=3$). These observations suggest that initializing the lower-level CHC search with NN-predicted worker selection and association decisions helps identify promising solutions within a limited computational budget, thereby providing more reliable fitness evaluations for the upper-level CMA-ES search.
	
	\textit{Incentive Efficiency: } 
	Considering subplot (b) of Figs.~\ref{fig:eval2}--\ref{fig:eval6}, which shows the per-round trajectories, we focus on the utility values averaged over the last three global rounds ($g=3$--$5$): Neogen consistently maintains positive manager and worker utilities across both datasets, achieving average MgU and WkU values of 3.8 and 4.0 under \textit{Set\#1}, and 13.9 and 5.9 under \textit{Set\#2}, respectively. In contrast, removing the NN guidance (EAOnly) results in negative average manager utilities under both settings ($-8.1$ and $-20.5$, respectively), accompanied by substantially higher worker utilities (16.0 and 47.0 on average, reaching 20.6 and 49.7 per round); its per-round manager utility falls to $-11.8$ and $-23.3$ in the later rounds, well below the reported averages, which accounts for the low EAOnly curves in subplot (b). These results highlight the importance of NN guidance in identifying contract terms and worker selection and association decisions that balance the economic interests of managers and workers.
	The limitations of fixed pricing become particularly evident when comparing Neogen with GenSA, GdyRatio, and RandAssign. Under \textit{Set\#1}, these benchmarks yield worker utilities ranging from 15.9 to 19.1, while manager utilities remain between $-2.5$ and 3.5. This disparity becomes  pronounced under \textit{Set\#2}, where worker utilities increase to 34.9--43.8, while manager utilities range from 6.9 to 10.5. In contrast, Neogen achieves the highest manager utility across all considered settings while maintaining positive worker utilities. These results highlight the benefit of jointly optimizing contract terms and worker selection and edge association decisions in sustaining the economic interests of managers without violating workers' individual rationality. The ranges above are averages over the same $g=3$--$5$ window.

	\textit{Computational Cost: }Subplot (c) of Figs.~\ref{fig:eval2}--\ref{fig:eval6} reports the average number of fitness evaluations per round. Neogen requires 6,000 evaluations under \textit{Set\#1}, 25\% fewer than EAOnly (8,000).
	For the larger-scale \textit{Set\#2}, it adaptively early-stops at 6,333 evaluations per round on average, 21\% fewer than EAOnly (8,000).
	After the first 1--2 rounds of training, the NN surrogate initializes CHC search within a Hamming neighborhood of radius $r^{\mathsf{ham}}=3$ around its prediction, reducing CHC generations from 20 to 15 without performance degradation.
	GenSA requires only 200 fitness evaluations per round across all settings, but incurs heavy overhead from full constraint validation in each evaluation.
	GdyRatio and RandAssign are computationally lightweight with 1--20 fitness evaluations per round, yet this efficiency is offset by severe utility imbalance, particularly under large-scale \textit{Set\#2} as discussed above.
	
	\textit{Unlearning Robustness: }Subplots (a) and (d) of Figs. \ref{fig:eval2}-\ref{fig:eval6} show the accuracy and KLD trajectories. At the unlearning round ($g = 2$), the KLD exceeds the threshold $\delta=0.05$ on both datasets, confirming effective knowledge removal by the OSD-refined GA unlearning used in this ablation (the KLD is evaluated only at the round in which an unlearning request is processed and is set to zero in the remaining rounds, and thus the zero values after $g=2$ reflect this measurement protocol rather than a physical restoration of the pre-unlearning predictive distribution). Neogen absorbs the perturbation and returns to its pre-unlearning accuracy already at $g = 3$ ($0.680\rightarrow 0.934$ on Banking77) under \textit{Set\#1}, and retains strong recovery ability and positive utility even in the large-scale \textit{Set\#2} scenario, where the perturbation is milder because the unlearned worker contributes only a small share of the round-2 aggregate. This shows that the bilevel optimizer can smoothly adapt to the unlearning-induced structural perturbation, and maintain strong robustness across different task scales. Moreover, the bars in subplot (d) record the number of CMA-ES generations executed by Neogen in each round under its early-stopping criterion. The count drops sharply after the first round as the optimizer warm-starts from the previous round's solution (from 21 to 10 on \textit{Set\#1}), rises temporarily at the unlearning round, when the perturbed fitness landscape requires more generations (16 at $g=2$), and then declines again as convergence resumes (to 9 at $g=5$); under \textit{Set\#2} the count varies between 6 and 13 across rounds.

	\textit{Dynamic Worker Selection and Local Training: }Subplots (e) and (f) of Figs.~\ref{fig:eval2}--\ref{fig:eval6} record, respectively, the number of workers selected and associated with each manager per round and the epoch-wise local training losses of the selected workers. The selection and association are re-optimized every round rather than fixed: on \textit{Set\#1}, the per-manager counts shift from 2/1/1 to 1/2/1 (M0/M1/M2) at the unlearning round ($g=2$), and the selection expands to five workers at $g=3$, consistent with the penalty-recycled residual budget; on \textit{Set\#2}, workload is progressively transferred from the busiest manager to another after the unlearning round (4/2 to 1/3 for M0/M1), while the total selection stays between six and eight workers. These adjustments reflect the optimizer's response to the rejoining worker's refreshed attributes and the evolving budgets. In subplot (f), the local training losses of the selected workers decrease rapidly and stabilize at small values within a few epochs, confirming that the workers chosen by Neogen perform stable local fine-tuning without divergence, which underlies the accuracy recovery in subplot (a).

	\vspace{-3mm}
	
	\section{Conclusion and Future Work}
	In this work, we proposed HermesHFL, an HFUL framework that integrates the client LUR lifecycle with economic incentive design for collaborative LLM fine-tuning. To solve the resulting cross-layer optimization problem, we developed Neogen, a NN-guided bilevel evolutionary optimizer. Experimental results demonstrated that HermesHFL achieves effective unlearning and competitive model accuracy compared with existing unlearning strategies, while substantially reducing the computational cost associated with full retraining. Furthermore, compared with conventional metaheuristics, heuristic benchmarks, and its surrogate-free counterpart, Neogen achieves the highest manager utility while maintaining positive worker utilities and competitive model accuracy. 
	Our experiments focused on scenarios with limited data distribution shifts among rejoining workers and two representative classification tasks. These considerations motivate two immediate directions for future research: \textit{(i)} investigating the impact of substantial data distribution shifts upon worker rejoining on incentive allocation, worker selection, and model performance; and \textit{(ii)} studying multilingual, domain-specific, and proprietary industrial datasets. Beyond these directions, promising research avenues include extending HermesHFL to multimodal foundation models and asynchronous hierarchical aggregation, as well as developing unlearning mechanisms with formal guarantees on information removal and privacy protection.
	\vspace{-3mm}
	
	\bibliographystyle{IEEEtran}
	\bibliography{reference}
	\newpage
	\appendices
	\clearpage
	\section{Practical Scenarios of the LUR Lifecycle}
	\label{sec:appendix_drift_scenarios}
	The LUR lifecycle considered in HermesHFL is governed by two coupled dynamics observed in practical collaborative services: \emph{participation dynamics}, under which an entity temporarily suspends or resumes model-training participation due to operational, regulatory, or resource conditions, and \emph{authorization dynamics}, under which data providers grant, pause, or restore consent for their data to be used in fine-tuning. A key observation is that, in many real-world cases, both dynamics alter only the participation status of the entity, while its data remain essentially unchanged: the data sources, collection pipelines, content categories, and task objectives persist throughout the suspension period. Consequently, the data distribution upon rejoining stays close to the distribution at the time of departure, i.e., the data change is small. We illustrate this observation with three representative scenarios from different application domains.
	
	\noindent
	$\bullet$ \textit{Example 1 (News organizations):} A news organization participating in a federated news-analysis service temporarily suspends its involvement while the authorization of an editorial data source is under review. Once the review concludes and authorization is restored, the organization rejoins the collaboration. During the suspension, its article streams, topic taxonomy, and classification task remain unchanged, so its local data distribution upon rejoining differs from the departure-time distribution only through routine news turnover. The data change is therefore small.
	
	\noindent
	$\bullet$ \textit{Example 2 (Banking institutions):} A bank contributing to a federated customer-service intelligence service pauses model updates while undergoing a regulatory audit or while customer authorizations for service dialogues are being renewed. After compliance approval is restored, the bank rejoins the collaboration. Its customer inquiries still follow the same service intents, product categories, and dialogue formats, so the underlying data distribution drifts only mildly over the suspension period. The data change remains small.
	
	\noindent
	$\bullet$ \textit{Example 3 (Industrial Internet of Things (IoT) devices):} Sensors and controllers on a production line of a federated predictive-maintenance service drop out during equipment maintenance or communication failures, and rejoin once the devices resume normal operation. The sensing targets, sampling frequencies, and production conditions remain within the same category of industrial process data, so the data change is small and limited to the downtime interval itself.
	
	In the above scenarios, the departure originates from participation dynamics or authorization dynamics rather than from a fundamental change in the nature of the data. The worker's historical contribution therefore remains a reliable estimate of its post-rejoin contribution, and completely discarding the accumulated knowledge through full retraining is disproportionate. This is precisely the regime for which the LUR mechanism is designed: the unlearning phase removes the historical influence at a controlled cost, and the rejoining phase readmits the worker with a refreshed profile, restoring a valuable participant while preserving model continuity. By contrast, substantial changes in the returning workers’ data distributions may alter their contributions to collaborative fine-tuning, requiring further investigation of the effectiveness of the proposed rejoining mechanism.
	% , and the associated limitations are analyzed in Sec.~\ref{subsec:threats_validity}.
	
	\section{Practical Motivation for the Two-Layer Reformulation}
	\label{sec:appendix_examples}
	The transformation from the original three-layer optimization problem to its two-layer counterpart is motivated by the observation that hierarchical decision-making with aligned objectives is ubiquitous in real-world systems. In such systems, upper-level entities determine the allocation of global resources, whereas lower-level entities optimize local decisions subject to the allocated resources. Although the decision variables differ across hierarchical levels, both optimize the same system-level objective.
	
	\noindent
	$\bullet$ \textit{Example 1 (Healthcare resource management).} In healthcare systems, hospital administrators allocate medical resources across departments to maximize system-wide healthcare quality, whereas each department optimizes treatment scheduling and resource utilization under the allocated budget. Despite the hierarchical decision process, both levels share the common objective of maximizing overall healthcare outcomes.
	
	\noindent
	$\bullet$ \textit{Example 2 (Financial resource allocation).} In financial institutions, headquarters allocate investment budgets across regional branches to maximize the overall return on limited capital, while each branch distributes its assigned budget among local projects. Although operating at different decision levels, both headquarters and branches pursue the common objective of maximizing the institution's overall investment return.
	
	Motivated by these practical hierarchical systems, we absorb the president-level optimization into the manager-level decision process through the contribution-aware budget allocation mechanism defined in \eqref{eq:budget_allocation_proportion}. Specifically, the allocated budget is explicitly coupled with each manager's contribution to the global objective, allowing the upper-level optimization objective to be faithfully preserved within the manager-level formulation. Consequently, the original three-layer optimization problem can be reformulated as a two-layer framework, substantially reducing computational complexity.
	
	\section{Benchmark Design Principle for Neogen}
	\label{sec:appendix_optimization}
	Existing optimization approaches for problems of this nature can be broadly grouped into four categories: evolutionary computation, classical metaheuristics, greedy/random heuristics, and learning-based methods such as RL \cite{LI2026129953}. To ensure a comprehensive and representative evaluation, our benchmark suite is designed to cover the first three categories, which collectively represent fundamentally different optimization philosophies and search mechanisms.
	
	Specifically, the evolutionary computation paradigm is represented by both EAOnly and Neogen, where EAOnly serves as a surrogate-free ablation to isolate the contribution of the proposed neural surrogate, while our Neogen further incorporates surrogate-assisted evolutionary optimization. The classical metaheuristic paradigm is represented by GenSA, a widely adopted SA algorithm. Finally, GdyRatio and RandAssign represent deterministic greedy optimization and randomized heuristic search, respectively, providing efficient heuristic baselines and a computational lower-bound reference.
	
	\noindent
	\textbf{\textit{Why RL Is Not Included?}} Although RL constitutes another important optimization method, it is not adopted as a baseline because its underlying assumptions are fundamentally mismatched with the considered bilevel optimization problem.
	
	\noindent
	\textit{High interaction cost.} Each fitness evaluation requires computationally expensive LLM training and hierarchical aggregation. Consequently, the interaction cost is prohibitively high for RL algorithms, which typically require thousands to millions of environment interactions to learn an effective policy.
	
	\noindent
	\textit{Dynamic mixed-integer decision space.} The optimization variables are mixed-integer, and the number of participating workers changes over time due to departure, unlearning, and rejoining events. Such dynamically varying state and action spaces violate the fixed-dimensional assumptions adopted by most existing RL algorithms.
	
	\noindent
	\textit{Delayed objective feedback.} The bilevel optimization structure provides only terminal objective evaluations: the lower-level combinatorial optimization should be completed before the upper-level objective can be evaluated. Thus, the problem does not provide the sequential reward signals required for effective temporal credit assignment in RL.
	
	The proposed benchmark suite provides representative coverage of the major optimization paradigms applicable to the considered problem, while excluding approaches whose underlying assumptions are incompatible with the problem formulation. By contrast, CMA-ES directly optimizes black-box objective evaluations without relying on intermediate rewards or differentiable structures, making it inherently suited to the considered bilevel optimization problem. Exploring RL-based optimization under this setting remains an interesting direction for future work.

	\section{Overall Procedure of Neogen and Key Hyperparameters}
	\label{sec:appendix_algorithm}
	
	For completeness, the overall procedure of Neogen described in Sec.~\ref{subsec:solution} is summarized in Alg.~\ref{alg:neogen}. Further, the key hyperparameters are summarized in Table \ref{tab:config}, with the search ranges for tunable hyperparameters indicated in brackets. We performed a grid search over these ranges and adopted the best-performing configuration for all experiments.
	
	\begin{table}[htbp]
		\vspace{-3mm}
		\centering
		\scriptsize
		\caption{Simulation Parameters}
		\vspace{-2mm}
		\label{tab:config}
		\begin{tabular}{ll}
			\hline
			Component & Value \\
			\hline
			Base model & GPT-2 (124M) \\
			Datasets & AGNews / Banking77 \\
			Workers $\times$ Managers & 10$\times$3 / 20$\times$4 \\
			Global rounds $G$ & 6 \\
			Edge rounds $E$ & 5 \\
			Budget $\beta^{\mathsf{sum},(g)}$ & 30 (10W) / 60 (20W) \\
			\hline
			LoRA rank $r$ / $\alpha$ / dropout & 8 / 16 / 0.1 \\
			Target modules & [$c\_attn$, $c\_proj$] \\
			Normal LR $\eta$ &  $[10^{-4}, 10^{-3}]$ \\
			Unlearning LR $\widetilde{\eta}$ & $[5\times10^{-4}, 5\times10^{-3}]$ \\
			Unlearning weight scale $\varsigma^{\mathsf{ul}}$ & 0.1 \\
			Max GA epochs & 5 (stop if stale $\ge 2$) \\
			KLD threshold $\delta$ &  $[0.01, 0.1]$ \\
			Memory $H$ / decay $\tau$ & 5 / 0.6 \\
			$\lambda^{\mathsf{mana}}$ / $\lambda^{\mathsf{pres}}$ & $[4, 16]$ / $[0.1, 1]$ \\
			\hline
			CMA-ES pop.\ $pop^u$ / $\sigma_u$ & 20 / 0.3  \\
			CHC pop.\ $pop^l$ / $T$ / $T_{\mathsf{stag}}$ & 20 / 20 / 10 \\
			CHC mutation prob.\ $p_{\mathsf{mut}}$ & 0.35 \\
			Hamming radius $r^{\mathsf{ham}}$ & 3 \\
			NN hidden dim.\ / LR & 128 / (LR range $[10^{-4}, 10^{-2}]$) \\
			NN min samples / acc.\ thresh. & 10 / 80\% \\
			\hline
			Budget constraint (C9) & Proportional \eqref{eq:budget_allocation_proportion} \\
			Hardware & Xeon 8358P + RTX 4090 \\
			Framework & PyTorch 2.8 + CUDA 12.8 \\
			\hline
		\end{tabular}
		\vspace{-1mm}
	\end{table}
	
	\begin{algorithm}[t]
		\scriptsize
		\caption{Proposed Neogen}
		\label{alg:neogen}
		\begin{algorithmic}[1]
			\Require{Upper-level population size $pop^u$, lower-level population size $pop^l$, number of global rounds $G$; CMA-ES parameters $(\mathbf{m}_u^{(0)}, \mathbf{C}_u^{(0)}, \sigma_u^{(0)})$; Hamming radius $r^{\mathsf{ham}}$ for local CHC search; NN training thresholds $D_{\min}=10$ and $\alpha_{\min}=80\%$}
			\Ensure{Best upper-level and corresponding refined lower-level solution $(\mathbf{u}^*, \mathbf{v}^*)$}
			
			\State Initialize dataset $\mathcal{D} = \emptyset$ and NN $NN(\cdot;\theta)$
			\State Initialize $(\mathbf{u}^*, \mathbf{v}^*)$ and best fitness $F^*=-\infty$
			
			\For{$g = 0$ \textbf{to} $G-1$}
			\Repeat
			
			\State Sample upper-level population:
			\[
			\mathbf{u}_k^{(a)} \sim \mathcal{N}\!\left(\mathbf{m}_u^{(a)}, \bigl(\sigma_u^{(a)}\bigr)^2\mathbf{C}_u^{(a)}\right), \quad k=1,\dots,pop^u
			\]
			
			\For{$k = 1$ \textbf{to} $pop^u$}
			
			\If{NN not well-trained (i.e., $|\mathcal{D}|<D_{\min}$ or accuracy $<\alpha_{\min}$)}
			\State Solve lower-level problem via CHC:
			\[
			\begin{aligned}
				\mathbf{v}_k^* &= \arg\max_{\mathbf{v}} 
				U^{\mathsf{Mana},(g)}(\mathbf{v}, \mathbf{u}_k^{(a)}) \\
				&\text{s.t. (C1)--(C4), (C7)--(C9)}
			\end{aligned}
			\]
			\State Perform CHC with constraint $x_i^{(g)} \ge y_{i,j}^{(g)}$, HUX, incest prevention, and cataclysmic mutation upon stagnation
			\State Add $(\mathbf{u}_k^{(a)}, \mathbf{v}_k^*)$ to dataset $\mathcal{D}$
			\Else
			\State Predict lower-level solution:
			\[
			\hat{\mathbf{v}}_k = NN(\mathbf{u}_k^{(a)}; \theta)
			\]
			\State Initialize CHC population around $\hat{\mathbf{v}}_k$ within Hamming radius $r^{\mathsf{ham}}$
			\State Obtain $\mathbf{v}_k^*$ via localized CHC search around $\hat{\mathbf{v}}_k$ with constraints $x_i^{(g)} \ge y_{i,j}^{(g)}$
			\State Add $(\mathbf{u}_k^{(a)}, \mathbf{v}_k^*)$ to dataset $\mathcal{D}$
			\EndIf
			
			\State Evaluate upper-level fitness:
			\[
			F_k^{(a)}=U^{\mathsf{Work},(g)}(\mathbf{u}_k^{(a)}, \mathbf{v}_k^*)
			\]
			\If{$F_k^{(a)} > F^*$}
			\State Update $(\mathbf{u}^*, \mathbf{v}^*) \gets (\mathbf{u}_k^{(a)}, \mathbf{v}_k^*)$ and $F^* \gets F_k^{(a)}$
			\EndIf
			\EndFor
			
			\State Train NN using dataset $\mathcal{D}$:
			\[
			\min_{\theta} \frac{1}{|\mathcal{D}|} \sum_{(\mathbf{u},\mathbf{v}^*) \in \mathcal{D}} 
			\|\mathbf{v}^* - NN(\mathbf{u};\theta)\|^2
			\]
			\State Select top $K$ individuals based on fitness
			\State Update CMA-ES parameters $(\mathbf{m}_u^{(a+1)}, \mathbf{C}_u^{(a+1)}, \sigma_u^{(a+1)})$
			\Until{CMA-ES early-stopping criterion is met}
			
			\EndFor
			
			\State \Return best upper-level and corresponding refined lower-level solution $(\mathbf{u}^*, \mathbf{v}^*)$
		\end{algorithmic}
	\end{algorithm}
	
	\section{Convergence Analysis Details of Neogen}
	\label{sec:appendix_convergence}
	
	This appendix provides the detailed convergence analysis of Neogen for solving $\boldsymbol{\mathcal{P}}_1^{(g)}$. Recall that the upper-level decision variables are $\mathbf{u}^{(g)}=\{p_{i,j}^{(g)},q_{i,j}^{(g)}\}$, while the lower-level decision variables are $\mathbf{v}^{(g)}=\{x_i^{(g)},y_{i,j}^{(g)}\}$. For each upper-level candidate solution $\mathbf{u}_k^{(a)}$, the corresponding lower-level solution $\mathbf{v}_k^*$ can be obtained by solving \eqref{eq:41} through the CHC evolutionary optimization procedure (Sec.~\ref{subsec:chc_nn}). The corresponding upper-level fitness evaluation is therefore given by \eqref{eq:upper_response_evaluation}.
	
	We first show lower-level convergence. Note that under an upper-level solution $\mathbf{u}_k^{(a)}$, the lower-level optimization problem contains only binary decision variables, and given finite worker and manager populations, the resulting combinatorial search space is finite. Further, during each IG, the population is updated by elitist survival selection as in \eqref{eq:chc_selection}. Let
	\begin{equation}
		B_k^{(t)}
		=
		\max_{\mathbf{v} \in \mathbf{V}_k^{(t)}}
		U^{\mathsf{Mana},(g)}
		\left(
		\mathbf{v},
		\mathbf{u}_k^{(a)}
		\right)
		\label{eq:best_fitness}
	\end{equation}
	
	\noindent
	denote the best fitness value within the population at generation $t$. Since the next-generation population is selected as the top-ranked solutions from the merged parent–offspring population, the best solution discovered at generation $t$ is always retained in generation $t+1$. This yields a monotonically non-decreasing fitness sequence:
	\begin{equation}
		B_k^{(t+1)}
		\ge
		B_k^{(t)}.
	\end{equation}
	
	\noindent
	The sequence $\{B_k^{(t)}\}$ is therefore also upper-bounded. Moreover, the cataclysmic mutation mechanism helps prevent the search process from being trapped in inferior stationary populations by periodically restoring population diversity while retaining the incumbent best solution. Thus, the lower-level CHC procedure can stabilize at a locally optimal or stationary solution $\mathbf{v}_k^*$.
	
	Then, we analyze the behavior of the designed NN guidance, where the NN is trained on the dataset $\mathcal{D}$ (defined in Sec.~\ref{subsec:chc_nn}), which continuously accumulates optimized upper-level and lower-level solution pairs during the evolutionary process. The network parameters are optimized by minimizing the following loss function:
	
	\begin{equation}
		L_{\mathrm{NN}}
		=
		\frac{1}{|\mathcal{D}|}
		\sum_{(\mathbf{u}_k^{(a)},\mathbf{v}_k^*) \in \mathcal{D}}
		\left\|
		NN(\mathbf{u}_k^{(a)};\theta)
		-
		\mathbf{v}_k^*
		\right\|^2.
		\label{eq:nn_loss}
	\end{equation}
	
	\noindent
	As the training dataset grows, the surrogate model can progressively improve its approximation of the lower-level optimal response function. Multilayer feedforward networks are universal approximators \cite{HORNIK1989359}; with sufficient data coverage and effective training, the surrogate can fit the response mapping over the visited region of the decision space. Consequently, the surrogate-generated solutions can increasingly approach the corresponding CHC-derived optima, reducing the frequency of expensive lower-level optimization.
	
	Finally, we consider upper-level convergence. At each generation, candidate solutions are sampled according to \eqref{eq:34} and updated through \eqref{eq:35}--\eqref{eq:cma_c_update}. Convergence properties of CMA-ES have been established for broad classes of continuous objective functions \cite{toure2023global}. As the surrogate-assisted lower-level solver progressively improves the approximation accuracy of the lower-level optimal response, the resulting fitness evaluations can approach the true bilevel objective values. Therefore, the CMA-ES search distribution is iteratively refined toward promising regions of the upper-level search space, providing a practical tendency toward stable, high-quality solutions. Combining the above, the lower-level optimization can attain a stable locally optimal or stationary solution, while the surrogate model progressively approximates the associated optimal response function. Consequently, the upper-level CMA-ES is guided by increasingly accurate fitness evaluations and is expected to approach a stable high-quality solution. Thus, the proposed surrogate-assisted bilevel optimization framework provides a practical solution procedure for $\boldsymbol{\mathcal{P}}_1^{(g)}$ rather than a guarantee of global optimality.
	\section{Additional Experimental Results}
	\label{sec:table}
	
	\begin{table}[htbp]
		\centering
		\footnotesize
		\caption{Unlearning strategy comparison on \textit{Set\#1}, 6 rounds}
		\label{tab:ablation_unlearn_10w}
		\begin{tabular}{lcccc}
			\hline
			Strategy & Acc & MgU & WkU & Time (s) \\
			\hline
			\multicolumn{5}{c}{\textit{AGNews}} \\
			\hline
			HermesHFL   & .883 & $-$0.7 & \textbf{7.0} & 1760 \\
			Rejoin (GA) & .887 & $-$0.7 & \textbf{7.0} & 1611 \\
			NoRejoin    & .883 & 1.3 & 6.0  & 1710 \\
			Retrain     & \textbf{.892} & \textbf{1.8} & 3.4  & 4665 \\
			FedOSD & .881 & 1.3 & 6.0 & \textbf{1610} \\
			\hline
			\multicolumn{5}{c}{\textit{Banking77}} \\
			\hline
			HermesHFL   & .937 & $-$0.7 & \textbf{7.0} & 1714 \\
			Rejoin (GA) & .939 & $-$0.7 & \textbf{7.0} & 1666 \\
			NoRejoin    & \textbf{.945} & 1.3 & 6.0  & \textbf{1645} \\
			Retrain     & .941 & \textbf{1.8} & 3.4  & 4347 \\
			FedOSD & \textbf{.945} & 1.3 & 6.0 & \textbf{1645} \\
			\hline
		\end{tabular}
	\end{table}
	
	FedOSD augments NoRejoin with a recovery projection that activates only when the OSD direction admits re-learning. Under \textit{Set\#1}, the recovery projection was never activated, so FedOSD matches NoRejoin (accuracy within 0.3 pp); at \textit{Set\#2} it never activated on AGNews and matches NoRejoin exactly, while on Banking77 it activated once at $g{=}3$, improving accuracy by 0.6 pp over NoRejoin (Table \ref{tab:ablation_unlearn_20w}).
	
	\begin{table}[htbp]
		\centering
		\footnotesize
		\caption{Unlearning strategy comparison on \textit{Set\#2}, 6 rounds}
		\label{tab:ablation_unlearn_20w}
		\begin{tabular}{lcccc}
			\hline
			Strategy  & Acc   & MgU   & WkU  & Time (s) \\
			\hline
			\multicolumn{5}{c}{\textit{AGNews}} \\
			\hline
			HermesHFL & .844  & 14.0  & \textbf{7.8} & 1543 \\
			NoRejoin  & \textbf{.859}  & 8.1   & \textbf{7.8} & \textbf{1397} \\
			Retrain   & .856  & \textbf{15.8}  & 3.2 & 4915 \\
			FedOSD    & \textbf{.859}  & 8.1   & \textbf{7.8} & 1409 \\
			\hline
			\multicolumn{5}{c}{\textit{Banking77}} \\
			\hline
			HermesHFL & \textbf{.917}  & 14.0  & \textbf{7.8} & 1451 \\
			NoRejoin  & .901  & 8.1   & \textbf{7.8} & \textbf{1415} \\
			Retrain   & .914  & \textbf{15.8}  & 3.2 & 4791 \\
			FedOSD    & .907  & 8.1   & \textbf{7.8} & 1430 \\
			\hline
		\end{tabular}
	\end{table}
	
	Rejoin (GA) is omitted at \textit{Set\#2}: it serves as an operator ablation, and a valid OSD--GA comparison requires the same request and schedule. Its independent Neogen optimization selected a different worker set, so the unlearning request targeted W14 (200 samples) rather than W10 (100 samples), which the other strategies replay; plotting it would mix the change of the unlearning operator with that of the target, its data volume, and the participant schedule, so the observed differences could not be attributed to any single factor. Its strategy-level comparison under the same request is retained at \textit{Set\#1} (Table \ref{tab:ablation_unlearn_10w}), and the same-target directional analysis at \textit{Set\#2} is reported in Sec.~\ref{sec:evaluation}.
	
	\subsection{Backdoor-Based Unlearning Verification}
	The KLD check in Sec.~\ref{subsec:unlearning_strategy} verifies that unlearning shifts the model's output distribution on the target worker's data. As a complementary, directly observable verification of deletion, we additionally run a backdoor canary on Banking77 \textit{Set\#1}. Worker W4's training set (200 samples) is poisoned: all of its samples carry a trigger (\texttt{"cf"} appended 16 times) and are relabeled to class 0. The unlearning request targets W4 at $g = 2$. All unlearning strategies share the poisoned worker data, and the backdoor run uses a checkpoint directory separate from the clean runs. The \textit{Keep} control receives the same request but ignores it.
	
	\begin{table}[htbp]
		\centering
		\footnotesize
		\caption{Backdoor attack success rate (ASR) per global round on Banking77 (\textit{Set\#1}); unlearning triggers at $g = 2$.}
		\label{tab:backdoor_asr}
		\begin{tabular}{lcccccc}
			\hline
			Strategy & $g{=}0$ & $g{=}1$ & $g{=}2$ & $g{=}3$ & $g{=}4$ & $g{=}5$ \\
			\hline
			HermesHFL   & .931 & .904 & .000 & .003 & .000 & .000 \\
			Rejoin (GA) & 1.00 & .973 & .000 & .000 & .000 & .000 \\
			NoRejoin    & 1.00 & .973 & .000 & .003 & .008 & .005 \\
			Retrain     & 1.00 & .973 & .000 & .003 & .000 & .000 \\
			FedOSD      & 1.00 & .973 & .000 & .003 & .008 & .005 \\
			Keep        & 1.00 & .973 & .909 & .941 & 1.00 & 1.00 \\
			\hline
		\end{tabular}
	\end{table}
	
	\begin{table}[htbp]
		\centering
		\footnotesize
		\caption{Optimization mechanism comparison on \textit{Set\#1}, 6 rounds}
		\label{tab:ablation_mechanism}
		\begin{tabular}{lcccc}
			\hline
			Method & Acc & MgU & WkU & Feval/r \\
			\hline
			\multicolumn{5}{c}{\textit{AGNews}} \\
			\hline
			Neogen      & \textbf{.898} & \textbf{3.8}  & 4.0  & 6000 \\
			EAOnly      & .742 & $-$8.1 & 16.0 & 8000 \\
			GenSA       & .877 & $-$2.5 & \textbf{19.1} & 200  \\
			GdyRatio    & .880 & 3.5  & 15.9 & 10   \\
			RandAssign  & .873 & 0.6  & 17.9 & \textbf{1}    \\
			\hline
			\multicolumn{5}{c}{\textit{Banking77}} \\
			\hline
			Neogen      & .935 & \textbf{3.8}  & 4.0  & 6000 \\
			EAOnly      & .833 & $-$8.1 & 16.0 & 8000 \\
			GenSA       & .931 & $-$2.5 & \textbf{19.1} & 200  \\
			GdyRatio    & .937 & 3.5  & 15.9 & 10   \\
			RandAssign  & \textbf{.947} & 0.6  & 17.9 & \textbf{1}    \\
			\hline
		\end{tabular}
	\end{table}
	
	\begin{table}[htbp]
		\centering
		\footnotesize
		\caption{Optimization mechanism comparison on \textit{Set\#2}, 6 rounds}
		\label{tab:ablation_20w4m}
		\begin{tabular}{lcccc}
			\hline
			Method     & Acc    & MgU     & WkU  & Feval/r \\
			\hline
			\multicolumn{5}{c}{\textit{AGNews}} \\
			\hline
			Neogen     & .883  & \textbf{13.9}   & 5.9  & 6333     \\
			EAOnly     & .867  & $-$20.5& \textbf{47.0} & 8000     \\
			GenSA      & \textbf{.909}  & 6.9   & 43.8 & 200      \\
			GdyRatio   & .902  & 8.3   & 41.4 & 20       \\
			RandAssign & .893  & 10.5   & 36.4 & \textbf{1}        \\
			\hline
			\multicolumn{5}{c}{\textit{Banking77}} \\
			\hline
			Neogen     & .938  & \textbf{13.9}   & 5.9  & 6333     \\
			EAOnly     & .919  & $-$20.5& \textbf{47.0} & 8000     \\
			GenSA      & .956  & 8.5   & 36.4 & 200      \\
			GdyRatio   & .956  & 8.5   & 36.4 & 20       \\
			RandAssign & \textbf{.960}  & 10.3   & 34.9 & \textbf{1}        \\
			\hline
		\end{tabular}
	\end{table}
	
\end{document}